\documentclass[sn-mathphys]{sn-jnl}

\usepackage[normalem]{ulem}

\jyear{2022}%

\theoremstyle{thmstyleone}%
\theoremstyle{thmstyletwo}%

\theoremstyle{thmstylethree}%

\begin{document}

\title[Article Title]{Physical Models for Solar Cycle Predictions}


\author*[1]{\fnm{Prantika} \sur{Bhowmik}}\email{prantika.bhowmik@durham.ac.uk}

\author[2,3]{\fnm{Jie} \sur{Jiang}}\email{jiejiang@buaa.edu.cn}

\author[4]{\fnm{Lisa} \sur{Upton}}\email{lisa.upton@swri.org}

\author[5,6]{\fnm{Alexandre} \sur{Lemerle}}\email{alexandre.lemerle@umontreal.ca}

\author[7,8]{\fnm{Dibyendu} \sur{Nandy}}\email{dnandi@iiserkol.ac.in}

\affil*[1]{\orgdiv{Department of Mathematical Sciences}, \orgname{Durham University}, \orgaddress{\street{ Stockton Road}, \city{Durham}, \postcode{DH1 3LE}, \state{} \country{The United Kingdom}}}

\affil[2]{\orgdiv{School of Space and Environment}, \orgname{Beihang University}, \orgaddress{\street{} \city{Beijing}, \postcode{} \state{} \country{China}}}

\affil[3]{\orgdiv{Key Laboratory of Space Environment Monitoring and Information Processing of MIIT}, \orgname{}\orgaddress{\street{} \city{Beijing}, \postcode{} \state{} \country{China}}}

\affil[4]{\orgdiv{Department of Solar and Heliospheric Physics}, \orgname{Southwest Research Institute}, \orgaddress{\street{} \city{Boulder}, \postcode{80302}, \state{Colorado}, \country{The United States of America}}}

\affil[5]{\orgdiv{D{\'e}partement de Physique}, \orgname{Coll{\`e}ge de Bois-de-Boulogne}, \orgaddress{\street{} \city{ Montr{\'e}al}, \state{Qu\'ebec}, \country{Canada}}}

\affil[6]{\orgdiv{D{\'e}partement de Physique}, \orgname{Universit{\'e} de Montr{\'e}al}, \orgaddress{\street{} \city{ Montr{\'e}al}, \state{Qu\'ebec}, \country{Canada}}}

\affil[7]{\orgdiv{Center of Excellence in Space Sciences India}, \orgname{Indian Institute of Science Education and Research Kolkata}, \orgaddress{\street{} \city{Mohanpur}, \postcode{741246}, \state{West Bengal}, \country{India}}}

\affil[8]{\orgdiv{Department of Physical Sciences}, \orgname{Indian Institute of Science Education and Research Kolkata}, \orgaddress{\street{} \city{Mohanpur}, \postcode{741246}, \state{West Bengal}, \country{India}}}


\abstract{

The dynamic activity of stars such as the Sun influences (exo)planetary space environments through modulation of stellar radiation, plasma wind, particle and magnetic fluxes. Energetic stellar phenomena such as flares and coronal mass ejections act as transient perturbations giving rise to hazardous space weather. Magnetic fields -- the primary driver of stellar activity -- are created via a magnetohydrodynamic dynamo mechanism within stellar convection zones. The dynamo mechanism in our host star -- the Sun -- is manifest in the cyclic appearance of magnetized sunspots on the solar surface. While sunspots have been directly observed for over four centuries, and theories of the origin of solar-stellar magnetism have been explored for over half a century, the inability to converge on the exact mechanism(s) governing cycle to cycle fluctuations and inconsistent predictions for the strength of future sunspot cycles have been challenges for models of solar cycle forecasts. This review discusses observational constraints on the solar magnetic cycle with a focus on those relevant for cycle forecasting, elucidates recent physical insights which aid in understanding solar cycle variability, and presents advances in solar cycle predictions achieved via data-driven, physics-based models. The most successful prediction approaches support the Babcock-Leighton solar dynamo mechanism as the primary driver of solar cycle variability and reinforces the flux transport paradigm as a useful tool for modelling solar-stellar magnetism.}


\maketitle

\section{Introduction}\label{sec1}

The Sun's magnetic field is the primary determinant of the electromagnetic and particulate environment around our planet as well as the heliosphere. Solar magnetic field variability is manifested through different spatial and temporal scales: from long-term decadal-scale variations in open magnetic flux, 10.7 cm radio flux, and total solar irradiance to short-term sporadic energetic events such as flares and coronal mass ejections (CMEs). While longer-term variations depend on the distribution and evolution of the large-scale global magnetic field of the Sun, short-term perturbations originate from magnetic structures of smaller spatial scales. Ultimately, solar magnetic fields therefore are responsible for shaping space weather and space climate. 

High energy radiation and particle fluxes originating from extreme space weather events (flares and CMEs) can damage satellites orbiting the Earth and are hazardous to astronaut health. The impact of such events can harm critical infrastructures on the ground, resulting in direct or cascading failures across vital services such as communications and navigational networks, electric power grids, water supply, healthcare, transportation services etc. \cite{Schrijver2015AdSpR}. 

Flares and CMEs are linked to the complex magnetic field distribution on the solar surface, which is dictated by the emergence of sunspots and the subsequent evolution of the active region associated magnetic flux. Thus the frequency of short-lived energetic events depends on the number of sunspots emerging within a solar cycle. Simultaneously, the slower and longer-term evolution of the magnetic field of sunspots determines the amplitude of open magnetic flux and the speed and structure of solar wind emanating from the Sun, in effect, defining the space climate. It has direct consequences on the forcing of planetary atmospheres, the life span of orbiting satellites and planning of future space missions. Thus understanding and predicting phenomena which governs space weather and space climate is a scientific pursuit with immense societal relevance -- in which solar cycle predictions occupy a central role \citep{NAP13060,NSWSAP2019,NSWSAP2022,UNOOSA2017,Schrijver2015AdSpR}.

The methodologies for predicting different aspects of space weather and space climate are diverse but broadly relies upon observations, empirical methods, computational models and consideration of the physics of the system. Here we focus primarily on the last theme, i.e., developing our understanding of solar variability using the laws of physics towards attaining the goal of solar cycle predictions. Now physics-based prediction on different time scales itself is an extensive topic, and a complete narrative is beyond the scope of this chapter. Instead, we limit ourselves to decadal-centennial scale variability associated with the sunspot cycle. We emphasize that physical understanding gleaned from successful solar cycle prediction models also apply to other Sun-like stars with similar dynamo mechanisms.

Sunspots are understood to be the product of a dynamo mechanism operating within the Sun's convection zone (SCZ, hereafter) dictated by the laws of magnetohydrodynamics \cite{Charbonneau2020LRSP}. In the SCZ, the kinetic energy stored in the ionised plasma converts to the magnetic energy primarily stored in the toroidal component of the magnetic field. The toroidal field, following significant amplification, rises through the SCZ due to magnetic buoyancy \citep{Parker1955ApJ} and emerges on the solar surface as strong localised magnetic field concentrations, forming Bipolar Magnetic Regions (BMRs, primarily), of which the larger ones are optically identified as sunspots. 
One of the mechanisms that contribute to poloidal field generation is the mean-field $\alpha$-effect which relies on helical turbulent convection twisting rising toroidal fields whose net impact is to produce a finite poloidal component. On the surface, observed small and large-scale plasma motion redistributes the magnetic field associated with the BMRs, resulting in the reversal and growth of the existing global magnetic dipole moment (poloidal component) of the Sun. This process -- termed as the Babcock-Leighton (B-L, hereafter) mechanism \citep{Babcock1961ApJ, Leighton1969ApJ} -- is another means to generate the poloidal component.

The strength of the magnetic dipole at the end of a solar cycle is found to be one of the best precursors for predicting the amplitude of the following cycle. This in itself is related to the stretching of the poloidal field through the deterministic process of differential rotation. 
However, observations, analytic theory and data driven models of decadal-centennial scale variability in the solar dynamo mechanism indicates that the B-L mechanism is the primary source of variability in the poloidal field and hence in the sunspot cycle \citep{Cameron2015Sci, Bhowmik2018NatCo}. 

Any physics-based model aiming for solar cycle predictions must be dictated by the laws of magnetohydrodynamics, contain the essence of the dynamo mechanism, and be constrained by observed plasma and magnetic field properties in the solar surface and the interior.  Some recent studies \citep{Petrovay2020LRSP, Nandy2021SoPh, Jiang2023JASTP} have explored the diversity of methods employed for sunspot cycle predictions and their credibility. Compared to these studies, this review will primarily focus on physics-based predictions of the sunspot cycle. 

In the following sections, we begin with a brief account of the observed distribution of magnetic field (primarily on the solar surface) and plasma flows which serve as building blocks and constraints for computational models (Section \ref{sec2}). This is followed by a short description of the computational models of magnetic field evolution on the solar surface and interior which has shown great promise as predictive models of the solar cycle (Section \ref{sec3}).
Physical insights on sources of irregularities in the strength of the solar cycle and amplitude modulation mechanisms -- which are gleaned from simulations and attempts to match observations -- are discussed in Section \ref{sec4}. In Section \ref{sec5} we present a review of physics-based solar cycle predictions limiting ourselves to data-driven modelling approaches. We conclude in Section \ref{sec6} with a discussion on the relevance of solar cycle predictions models for theories of solar-stellar magnetism and end with some remarks on future prospects in the field of solar cycle predictability.


\section{Constraints from solar observation}\label{sec2}

Solar magnetic field observations are primarily limited to the visible layer of the Sun, i.e., the photosphere. Plasma flows are observed both on the surface as well as in the interior through inference using tools of helioseismology. 
Computational models utilize these information for purposes of constraining and calibrating the models. The main goals driving this synergy between observations and models are to achieve a better understanding of the physical processes ongoing in the Sun and develop predictive models of solar activity. In this section, we focus on observations which are relevant to surface surface flux transport (SFT) and dynamo models of the solar magnetic field. For a detailed account of solar observations, see \citep{Chapter2} (Chapter 2 of this book).

\subsection{Sunspot Number} \label{secSSN}

\begin{figure}
 \noindent\includegraphics[width=\textwidth]{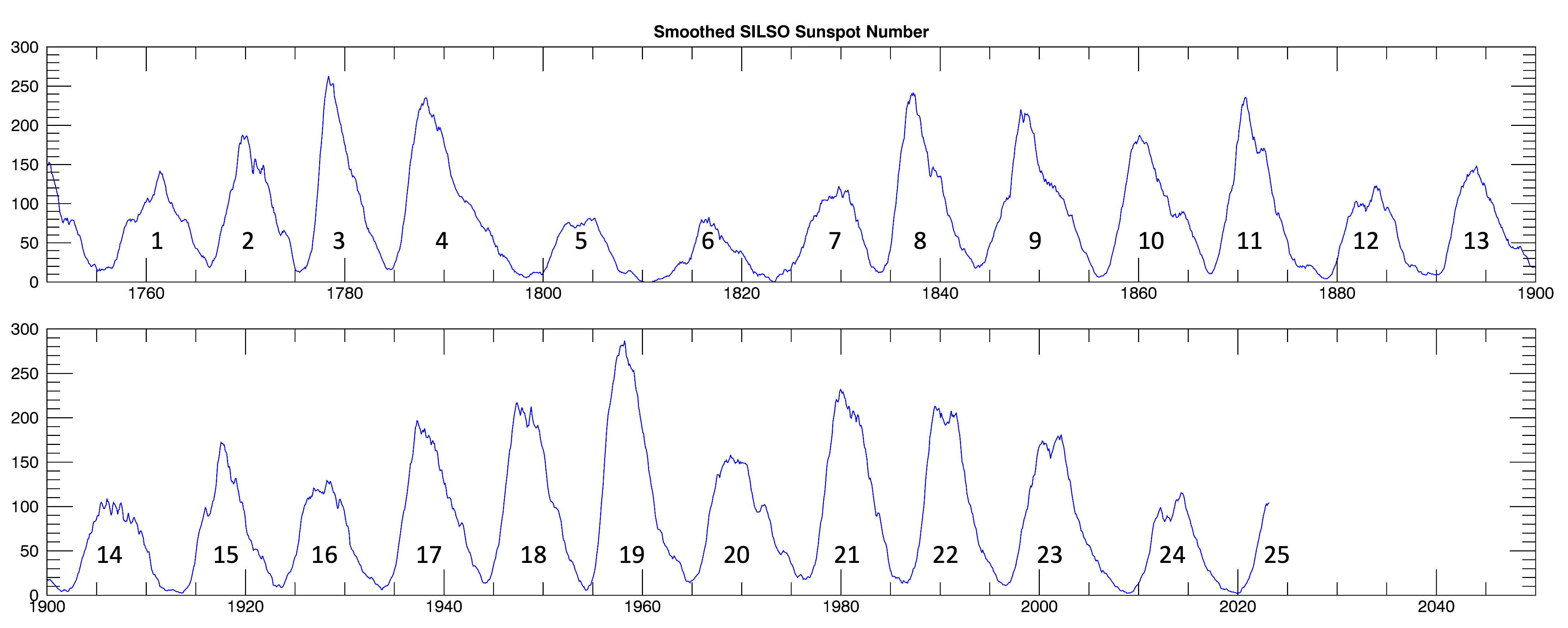}
\caption{The SILSO Sunspot Number. The smoothed SILSO Monthly mean total sunspot number (v2.0), smoothed illustrates the rise and fall of solar activity from 1750 to the present (marked with solar cycle numbers).}
\label{fig:SSN}
\end{figure}

Although sunspots have been observed systematically through telescopes from the early 1600s, in the early 1800's, solar astronomer Samuel Heinrich Schwabe began plotting the number of sunspots as a function of time and discovered that their appearance was cyclic with a period of about eleven years \citep{1844Schwabe}. As new phenomena (e.g., flares, CMEs, etc.) on the Sun was discovered, it was found that they too varied along with the sunspot number. Solar activity is now characterized by the Monthly Sunspot Number, a count of the number of Sunspots or Active Regions observed each month as a function of time. The official count is maintained by the Sunspot Index and Long-term Solar Observations (SILSO) at the Royal Observatory of Belgium, Brussels \footnote{\url{https://www.sidc.be/silso/home}}. In 2015, the sunspot number data were revised to version 2.0, to account for changes in observers and provide a more consistent data series throughout the historical record \citep{2015Clette_etal}. Version 2.0 of the SILSO Monthly mean total sunspot number, smoothed with a 13 month Gaussian filter, is illustrated in Figure \ref{fig:SSN}.

The SILSO data series now shows nearly 25 solar cycles. Each solar cycle has a period of about 11 years and an average amplitude (v2.0) of 180, with a range of about 80 (e.g., Solar Cycles 5) and 280 (e.g., Solar Cycle 19). The length of the cycle correlates with the amplitude of the cycle such that bigger cycles tend to be shorter in duration and weaker cycles tend to be longer. The shape of the solar cycle typically appears as an asymmetric Gaussian function, with a rapid rising phase and a longer decaying phase. Shorter term variability in the cycle, on the order of about 2 years, causes many cycles to have two or more peaks, which are often more pronounced in weaker cycles. Sunspot cycles have other irregularities too, for example, the Sun entered into a prolonged near-minimum state during 1645--1715 (this was pointed out by G. Sp\"{o}rer and E. W. Maunder in the 1890s). This phase, known as Maunder Minimum \citep{Eddy1976Sci}, is a period where the solar activity cycle was operating at an exceptionally weak state for several decades. 

\subsection{Magnetic field observations relevant for solar cycle models} \label{secMagBfly}

 \begin{figure}
 \noindent\includegraphics[width=\textwidth,trim={10 0 70 0},clip]{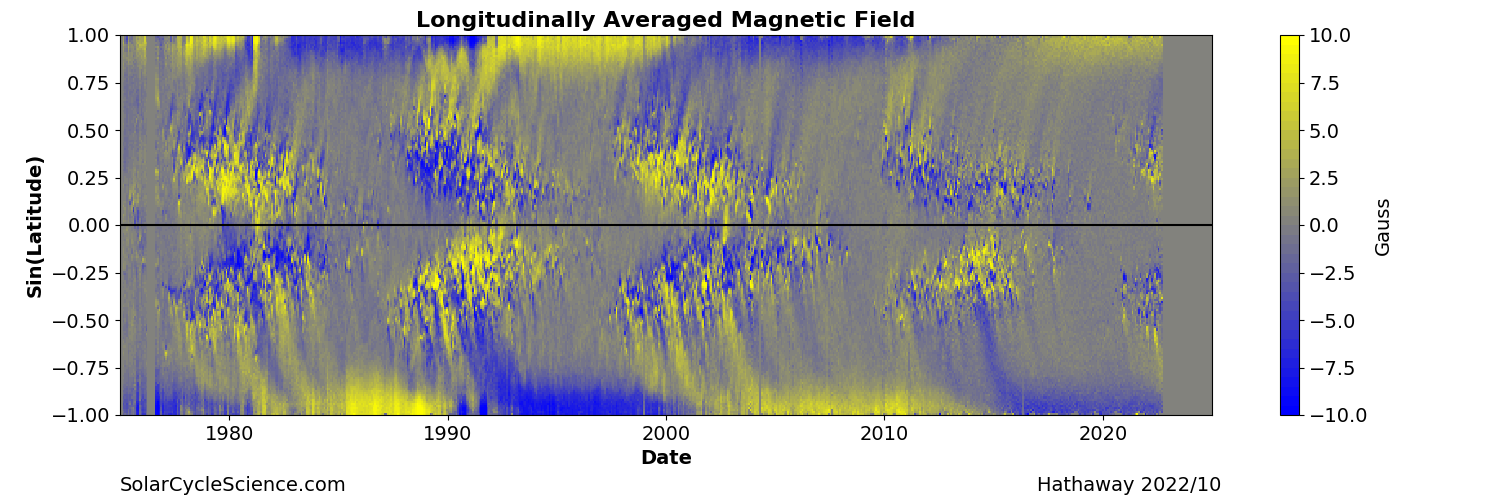}
\caption{The magnetic butterfly diagram illustrates the evolution of magnetic flux on the Sun over several cycles. The magnetic field averaged over a Carrington Rotation is plotted as a function of latitude and time, with the positive (negative) polarity shown in yellow (blue). This figure illustrates many observational constraints that govern models used to make solar cycle predictions.}
\label{fig:magbfly}
\end{figure}

Perhaps one of the most significant solar discoveries in the twentieth century, was the realization that sunspots are magnetic in nature \citep{1908Hale_a,1908Hale_b}. Sunspots are now known to host very strong magnetic fields on the order of 1000 G and often appear as a collection of spots known as an active region. An active region, in general, appears as a pair of spots (with opposite polarities) which are referred to as  Bipolar Magnetic Regions (BMR). Sunspots are the optical counterparts of active regions. Besides the spots, the background surface field has a strength of only a few Gauss. The dynamics of the Sun's surface magnetic field is captured by the `Magnetic Butterfly Diagram', see Figure~\ref{fig:magbfly}. This figure is created by plotting the longitudinally average radial magnetic field on the Sun as a function of latitude and time. This figure illustrates several properties that serve as constraints for the SFT and dynamo models and which are tests for predictive models. The properties include,
\begin{itemize}
  \item Sporer's Law: Active Regions begin emerging at mid-latitudes ($\sim$ 30 degrees). As the cycle progresses, active regions form bands of flux that moves equator-ward \citep{1858Carrington}. 
  \item Joy's Law for Tilt Angles: Active Regions tend to have a characteristic tilt such that the angle between the local parallel of latitude and the line joining the leading polarity spot (which appears closer to the equator) and the following polarity spot increases with increasing latitude \citep{1919Hale_etal}.
  \item Hale's Polarity Law: The relative polarity of Active Regions (between the leading and following spots) is opposite across the equator, and this polarity changes sign from one cycle to the next \citep{1919Hale_etal}. Thus a complete magnetic cycle has a periodicity of 22 years.
  \item Polar Fields: In addition to the flux in emerging Active Regions, the Sun possesses unipolar concentrations of magnetic flux near both of the poles. The polarity is opposite across hemispheres, reverses polarity at about the time of solar cycle maximum \citep{1958BabcockLivingston} and attains its maximum amplitude around cycle minimum. This large-scale field is known as the `Polar field', and its evolution plays an important role in solar cycle predictions. However, due to projection effects, polar field measurements suffer from erroneous values, thus prompting the need for off-ecliptic space missions focusing on the Sun's poles \citep{Nandy2023arXiv}.
 \item Magnetic Field Surges: Streams of weak polarity flux are carried from the active latitude to the poles in about 2-3 years. These streams are responsible for the build-up and reversal of the polar fields \citep{Babcock1961ApJ}. The strength of these surges reaching the poles can vary significantly based on the associated magnetic flux content of the group of active regions or a single `rogue region' (e.g., with a large area and high tilt angle \citep{Nagy2017}). The emergence latitudes, tilt and flux of the active region, and frequency of active region emergence are all important factors in determining how much a given active region will contribute to the polar field evolution.
\end{itemize}

The redistribution of the active regions' magnetic flux across the solar surface and the interior convection zone happens through the collective effect of small and large-scale plasma motions which provide additional constraints on models.

\subsection{Plasma flows} \label{secFlows}

Plasma flows in the solar convection zone may be divided into three categories based on the physical role they play in the solar dynamo mechanism: convective flows, differential rotation, and meridional circulation. The thermal flux through the solar convection zone and consequent temperature gradient causes the plasma within the solar convection zone to rise to upper layers, transfer or radiate their energy away and sink back down after cooling. As a result, convective cells with a spectrum of different scales \citep{2015Hathaway_etal} are formed ranging from granules (radius $\sim 1$ Mm) with lifetimes of minutes to hours, to supergranules (radius $\sim 30$ Mm) with lifetimes of days, and to the largest convective structures (radius $\sim 100$ Mm) with lifetimes of months. These convective motions are turbulent in nature and effectively distribute the magnetic field over the entire solar surface, similar to a diffusive process.

The Sun rotates differentially which was first found by tracking sunspots on the solar surface \citep{Adams1911,Belopolsky1933, Howard1984ARA&A}. This differential rotation at the surface is such that the large-scale plasma flow speed along the direction of solar rotation varies latitudinally with a faster-rotating equator than the poles. Later, helioseismology \citep{Schou1998ApJ,Basu2016LRSP} was utilized to obtain the structure and spatial variation of rotation rate inside the solar convection zone. The radiative zone rotates as a solid body resulting in a strong radial shear within the tachocline which is thought to encompass the (stable) overshoot layer at the base of the convection zone. The differential rotation plays a crucial role in the generation and amplification of the toroidal component of the Sun's large-scale magnetic field (see Section \ref{sec3}). 

Another large-scale subsurface plasma flow known as the meridional circulation \citep{Hanasoge2022LRSP} carries plasma from the equatorial region to the poles (in both hemispheres) with a varying speed dependent on latitude. The flow speed becomes zero at the equator and the poles, and the circulation attains its peak speed (10\,–\,20 m\,s$^{-1}$, about 1\% of the mean solar rotation rate) near mid-latitude. The law of mass conservation dictates an equator-ward return flow of plasma deeper inside the solar convection zone, which, however, remained hard to map using helioseismic observations due to its small amplitude. While some recent studies \citep{Rajaguru2015ApJ, Liang2018A&A} have suggested that meridional circulation is a single-cell flow where the return flow is at the depth of the solar convection zone (depth $<$ 0.8 $R_{\odot}$), others \citep{2012Hathaway, 2013Zhao_etal, 2022Hathaway_etal} suggest that it may be multi-cellular in depth and or latitude. The shape of the meridional profile in latitude and radius is crucial in determining the various properties of the sunspot cycles, including cycle duration. Early flux transport dynamo models suggest that a deep single-cell meridional flow threading the convection zone is necessary to match solar cycle observations \citep{Nandy2002Sci}.

Note that both small-scale and large-scale plasma flows are not static. Helioseismic observation shows that the Sun's differential rotation varies with time in an oscillatory fashion and is correlated with the solar activity cycle -- it is known as the solar torsional oscillation \citep{Zhao2004torsional, Howe2009LRSP}. Meridional circulation also exhibits cycle-dependent temporal variation in its peak speed, with reduced amplitude during cycle maximum compared to the minimum \citep{Hathaway2010Sci, Hathaway2014JGRAH, 2022Hathaway_etal}. However, for both large-scale plasma flows, such variations constitute less than 20\% of the average profiles. Thus, computational models with time-independent plasma flow profiles can reproduce the majority of the observed magnetic field variability. For a detailed account of the plasma flows in the Sun, see Chapter 10 of this book \citep{Chapter10}.

\begin{figure}[ht!]
 \noindent\includegraphics[width=\textwidth]{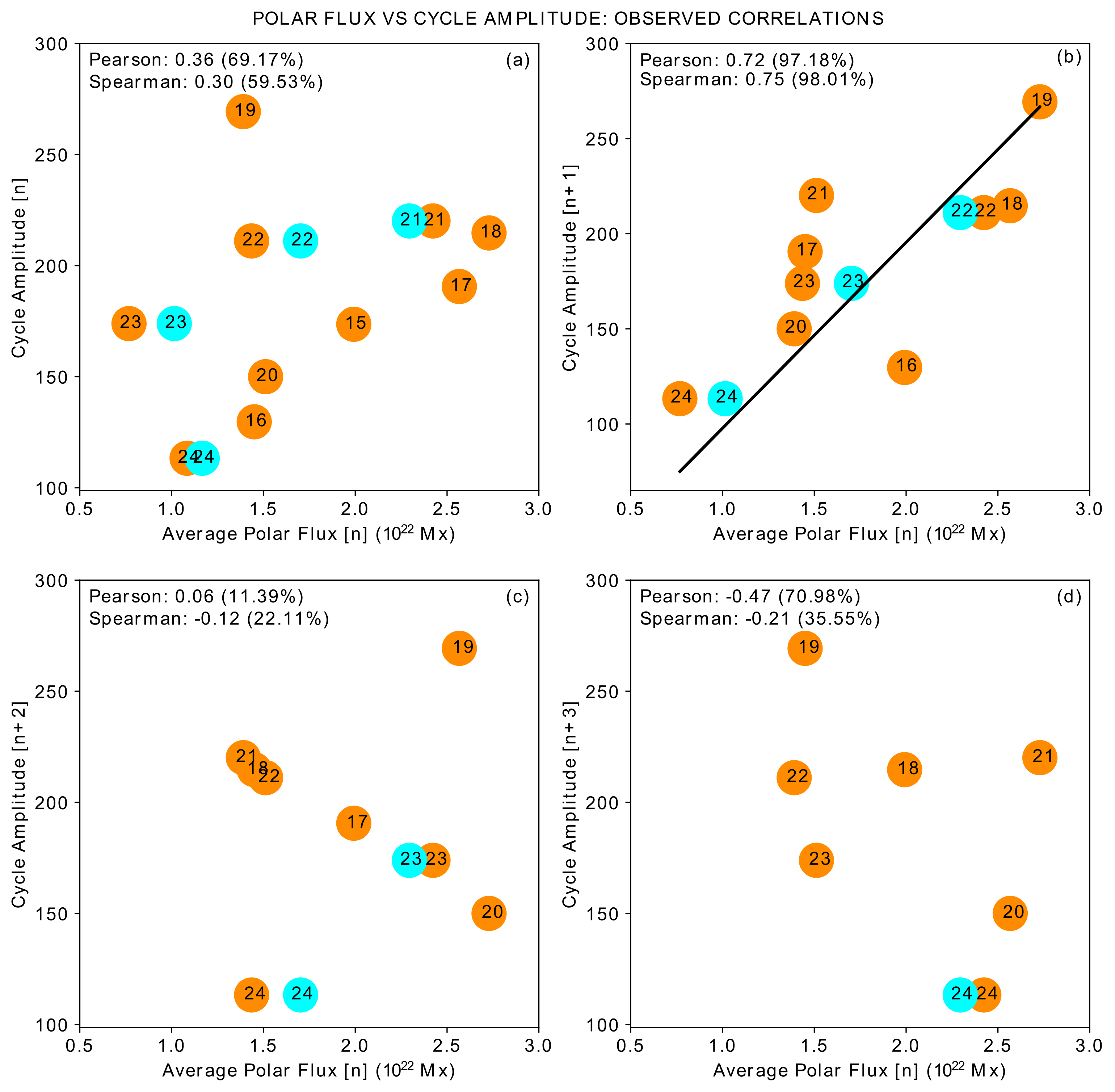}
\caption{Observed cycle-to-cycle correlations between the polar flux at cycle minima (say, [n]) and the cycle amplitude of different cycles, namely (a) cycle [n], (b) cycle [n+1], (c) cycle [n+2], and (d) cycle [n+3]. The numbers inside the circles indicate the associated solar cycle numbers. The colors of the circles differ based on the source of polar flux data, orange: averaged polar flux obtained from polar faculae count, cyan: the average dipole moment (scaled appropriately to place them in the figure). Image reproduced with permission from Nandy et al. \citep{Nandy2021SoPh} copyright by Springer Link.}
\label{fig:polar_corr}
\end{figure}

\subsection{Polar fields as precursors of the strength of sunspot cycles} \label{secPolar}

The temporal evolution of the averaged polar field has a $\pi/2$ phase difference with the sunspot cycle. As mentioned earlier, the average polar field strength at cycle minimum serves as an important element in predicting sunspot cycles. Although direct observation of the polar fields became available only in the 1970s, indirect measures of polar flux exist based on proxies. Polar flux evolution derived from polar faculae observations \citep{Munoz2012ApJ} cover a period of 100 years (during cycles 14\,-\,24). Note that the average polar flux during cycle minimum is a close representation of the Sun's magnetic dipole (axial) moment -- which acts as a seed to generate the following solar cycle. In fact, the average polar flux at the n$^{\mathrm{th}}$ cycle minimum has the maximum positive correlation with the amplitude of the n+1$^{\mathrm{th}}$ cycle [see, Figure \ref{fig:polar_corr}]. The correlation decreases drastically for the amplitude of cycles n$^{\mathrm{th}}$, n+2$^{\mathrm{th}}$ and n+3$^{\mathrm{th}}$ as depicted in Figure \ref{fig:polar_corr}. Figure \ref{fig:polar_corr} reflects on two crucial aspects of solar cycle predictability: first, a strong solar cycle does not result in a strong polar field generation at that cycle minimum [see, Figure \ref{fig:polar_corr}(a)] and the memory of the polar field (of n$^{\mathrm{th}}$ cycle) diminishes beyond the next (n+1$^{\mathrm{th}}$) cycle [see, Figure \ref{fig:polar_corr}(c) and (d)]. It is important to note here that these empirical observational evidences were preceded by flux transport dynamo models exploring the memory issue which predicted that the sunspot cycle memory is limited primarily to one cycle alone in certain parameter regimes related to flux transport timescales \citep{Yeates2008, Karak2012ApJ}.









\section{Physical Modeling Approaches}\label{sec3}

In an astrophysical magnetised plasma system like our Sun, we expect the following properties of the plasma will be satisfied: the velocity is non-relativistic, the collisional mean free path of the atomic or molecular constituents of the plasma is much shorter than competing plasma length scales, and the plasma is electrically neutral and non-degenerate. In such a system, the evolution of the magnetic field is dictated by the magnetohydrodynamic (MHD) induction equation, which is a combination of Ohm's law, Amp{\`e}re's law and Maxwell's equations:

\begin{equation}
    \frac{\partial\mathbf{B}}{\partial t} = \nabla \times (\mathbf{u} \times \mathbf{B} - \eta \nabla \times \mathbf{B}).
    \label{eq_induction}
\end{equation}
\noindent Here {\bf u} and {\bf B} are the plasma velocity and magnetic fields, respectively, and $\eta = 1/\mu_0\sigma$ is the magnetic diffusivity, with $\mu_0$ the magnetic permeability and $\sigma$ the electric conductivity. Additionally, the magnetic field satisfies the divergence-free condition, $\nabla\cdot\mathbf{B} = 0$. The spatio-temporal evolution of the plasma flow is dictated by Navier–Stokes equation,

\begin{equation}
    \frac{\partial \mathbf{u}}{\partial t} + (\mathbf{u}\cdot\nabla)\mathbf{u} = - \frac{1}{\rho} \nabla \mathrm{P} + \mathbf{g} + \frac{1}{\rho} (\mathbf{J} \times \mathbf{B})+ \nu \nabla^2 \mathbf{u},
    \label{eq_NS}
\end{equation}

\noindent where $\rho$ is plasma density, P is plasma pressure, \textbf{g} is the gravitational acceleration, $\mathbf{J} = \nabla \times \mathbf{B}$ is the electric current, and $\nu$ is kinematic viscosity. Additionally, the plasma flow obeys mass conservation through the continuity equation. Along with equations (\ref{eq_induction}) and (\ref{eq_NS}), one should take into account the conservation of energy and equation of states involving the pressure and plasma density. In an isolated system, where we can ignore the Poynting flux, the mechanical energy stored in the flow (\textbf{u}) acting in the opposite direction of the Lorentz force ($\mathbf{J} \times \mathbf{B}$) is converted to magnetic energy. This magnetic energy can decay through the dissipation of the electrical currents supporting the magnetic field. 

Thus the sustainability of a dynamo depends on the relative strength between the induction effect controlled by the velocity field (the first term on the R.H.S in equation (\ref{eq_induction}) and the Ohmic dissipation (the second term on the R.H.S in equation (\ref{eq_induction}). The ratio of these two terms is known as the magnetic Reynolds number, $\mathrm{R}_\mathrm{m} = \mathrm{u L}/ \eta$, where L is the length scale determining whether inductive effect overcomes the dissipative processes. In most astrophysical systems, a very large L ensures a very high $\mathrm{R}_\mathrm{m}$, which is crucial for the survival and growth of the dynamo.

In an ideal scenario, solving the complete set of MHD equations associated with the conservation of mass, momentum, energy, and magnetic flux including the magnetic induction equation in the SCZ should provide the Sun-like spatio-temporal evolution of the velocity and magnetic field with the given Sun-like plasma properties. However, this requires the numerical models to be capable of comprising a wide range of spatial and temporal scales characterizing fluid turbulence at high viscous and magnetic Reynolds number medium -- which is quite challenging from the computational point of view. While with increasing computational power and improved algorithms, full MHD models are becoming more realistic, the parameter regimes are still nowhere near the real solar convection zone. Moreover, all the existing MHD models operate with enhanced dissipation, much stronger than the characteristic dissipation in the solar interior. A comprehensive account of the MHD simulations of solar dynamos is presented in Chapter 15 of this book \citep{Chapter15}, thus, we restrain ourselves from going into further details.

The scope of the growth of the dynamo is encapsulated within the advective part of the induction equation [$\nabla \times (\mathbf{u} \times \mathbf{B})]$ in equation (\ref{eq_induction}), where any pre-existing magnetic field (\textbf{B}) is amplified by the plasma flow through the shearing term [$\mathbf{B}\cdot\nabla(\mathbf{u})$], compression and rarefication [$\mathbf{B}(\nabla\cdot\mathbf{u})$], and advection [$(\mathbf{u}\cdot\nabla)\mathbf{B}$]. While any positive gradient in the plasma flow ensures growth of \textbf{B}, the dynamo-generated magnetic field should have the following observed characteristics (see Section \ref{sec2}) in the solar context:
\begin{itemize}
\item The large-scale magnetic field (the dipole component) should reverse on a decadal scale.
\item The sunspot generating field component should have a $\pi/2$ phase difference with the dipole component, should exhibit an equator-ward migration, and the associated polarity should be anti-symmetric across the equator.
\item On the solar surface, the dynamo model is expected to reproduce observed features of sunspots and the associated flux evolution, which include pole-ward migration of the diffused field and generation of observed polar field.
\item Moreover, the solar dynamo models should result in amplitude fluctuations in both the sunspot-generating component and the large-scale dipole component, along with observed empirical patterns and correlations between them. 
\end{itemize}
Reproducing all these intricate details of the observed solar magnetic field and the  velocity field while solving the full set of MHD equations in the turbulent convection zone indeed becomes a challenging problem. Thus one major and very successful alternative approach in the dynamo community has been to focus on the evolution of the magnetic field only by solving the induction equation (\ref{eq_induction}) while utilizing prescribed plasma flow derived from observation \citep{Charbonneau2020LRSP}. These are often termed as kinematic or flux transport dynamo models. Another modelling approach, namely Surface Flux Transport (SFT) models, simulate only one half of the solar cycle, namely, the evolution of magnetic fields which have emerged on the solar surface mediated via prescribed flow parameters which are constrained by observations.
We discuss them briefly below. 

\subsection{Solar Surface Flux Transport Models as Tools for Polar Field Predictions}\label{sec3.1}

The genesis of solar surface magnetic field evolution models, as well as the Babcock-Leighton mechanism for polar field generation can be traced to the heuristic ideas first proposed by Babcock \citep{Babcock1961ApJ}. Babcock attempted to explain the behavior of the large-scale solar magnetic fields through a phenomenological description of the topology of the Sun's magnetic field and its evolution which was related to the emergence of systematically tilted BMRs and the subsequent diffusion of their flux, cross-equatorial cancellation and migration to the poles -- culminating in the large-scale dipolar field reversal. This process was envisaged to be complemented by the stretching of the submerged large-scale flux systems by differential rotation to produce the sunspot forming toroidal field.  Later, R. B. Leighton put these ideas on a firmer theoretical foundation \citep{Leighton1964}. He suggested that the radial magnetic field at the surface of the Sun is advected and diffused kinematically like a passive scalar field. 

The computational models capturing the evolution  of this radial magnetic field [$B_r\,(\theta, \phi, R_{\odot})$] associated with the active regions are known as Surface Flux Transport (SFT) models. The temporal evolution of the longitudinal averaged radial field obtained from such simulations should have the distinct features observed in the magnetic butterfly diagram (Figure \ref{fig:magbfly}). The SFT mechanism may also be coherently derived from the MHD induction equation (\ref{eq_induction}) as the time evolution of the radial component of the magnetic field $B_{r}$, evaluated at $r=R_{\odot}$, as:

\begin{align}
\frac{\partial B_{r}}{\partial t} &= -\frac{1}{r\sin \theta} \frac{\partial}{\partial \theta}\left(u_{\theta}B_{r} \sin\theta\right) -\frac{1}{r\sin \theta} \frac{\partial}{\partial \phi}\left(u_{\phi}B_{r}\right) + \eta_{T} \nabla ^{2} B_{r}.
\label{eq:SFT2}
\end{align}

\noindent Here, $u_{\theta}$ and $u_{\phi}$ denote two large-scale plasma flows on the solar surface: meridional circulation and differential rotation, respectively. The diffusivity $\eta_{T}$ represents a simplification of the effect of turbulent convective motions of the plasma on $B_r$. To the linear formulation above, a source term must be added to account for the additional influx of magnetic field associated with the emergence of active regions, which are primarily Bipolar Magnetic Regions (BMR). For a detailed description of SFT models and their theoretical considerations, refer to \citep{Chapter7} (Chapter 7 of this book) or \citep{2014Jiang_etal, 2023Yeates_etal}. 

\subsubsection{Genesis of Surface Flux Transport Simulations}

Following the pioneering work by Babcock \citep{Babcock1961ApJ} and Leighton \citep{Leighton1964} describing the evolution of $B_r$ on the solar surface,  DeVore et al. \citep{DeVore1984SoPh} created the first SFT model of the Sun. Their SFT model was originally used to constrain meridional flow at the surface, which was difficult to measure and very uncertain at that time. To mimic the emergence of active regions on the solar surface, Sheeley et al. \citep{Sheeley1985SoPh} included bipolar active region sources based on observed statistics. Wang et al. \citep{Wang1989Sci} explored the role of surface flux transport and dissipation processes such as differential rotation, meridional flow, and diffusion to investigate their role in the reversal and build-up of the polar fields. They found that a) differential rotation was essential for separating the leading and following polarity flux in bipolar active regions, b) diffusion played a crucial role in cross-equatorial flux cancellation of the leading polarities and b) meridional flow was essential for transporting the following polarity flux to the poles aiding in polar field reversal and build-up. The primary physical ingredients of the surface processes resulting in the observed solar cycle associated polar field dynamics were now in place. 

\subsubsection{Evolution towards Data Driven Surface Flux Transport Models}

More evolved SFT models now have the ability to incorporate the observed flows (static and time-evolving) and assimilate data for realistic simulations of solar surface field evolution and polar field predictions. These models are also paving the way for realistic coronal field and heliospheric modeling by providing time-dependent, data assimilated boundary conditions at the solar photosphere.

While many modern SFT models continue to parameterize the small-scale convective motions with a diffusivity coefficient, a novel class of magnetoconvective SFT (mSFT) models have been developed which emulate not only the spreading motions of convection outflows, but also the concentration of the magnetic network formed at the boundaries of convective structures. The first attempt at this was achieved by introducing a random attractor matrix \citep{2000WordenHarvey} to replace the diffusivity. The attractor method was later adapted by the Air Force Data-Assimilative Photospheric Flux Transport model (ADAPT) \citep{2010Arge_etal, 2015Hickmann_etal}. Another SFT model invoked a collision and fragmentation algorithm \citep{2001Schrijver}. An alternative approach known as the Advective Flux Transport (AFT) \citep{2014UptonHathaway_a} model has been developed which mimics surface convection through spherical harmonics to generate an evolving velocity flow field that reproduces the size, velocities, and lifetimes of the observed convective spectrum \citep{2015Hathaway_etal}. 

Another major advancement in SFT models, brought about by the space-based Doppler-Magnetographs, is the availability of high cadence and high-resolution magnetograms. Schrijver \& DeRosa (2003) \citep{2003SchrijverDeRosa} were one of the firsts to directly assimilate magnetogram data into the SFT model by incorporating SHOS/MDI magnetic field observations within $60^{\circ}$ of the disk center. Using their SFT maps as an inner boundary condition to a PFSS model, they were able to create an accurate reconstruction of the interplanetary magnetic field (IMF). More formal data assimilation processes (e.g., Kalman filtering) require that the observed data be merged with the simulated data in a way that accounts for the uncertainties. ADAPT \citep{2015Hickmann_etal} and AFT \citep{2014UptonHathaway_a} type SFT models employ Kalman filtering. The ADAPT model is used in conjunction with WSA-ENLIL model to aid in Space Weather Predictions. Since the surface field distribution drives the coronal magnetic field, SFT (and AFT) models have shown great capabilities for coronal field simulations and predictions \citep{Nandy2018ApJ,Dash2020ApJ,2018Mikic_etal, 2022MackayUpton,Yeates2022ApJ}.

Studies are illuminating the influence that flow variations have on polar field dynamics. Upton \& Hathaway \citep{2014UptonHathaway_b} found that variations in the meridional flow had a significant impact ($\sim$20\%) on the polar field strength. Cross equatorial flow \citep{2022Komm} significantly influenced the residual magnetic flux transported to the poles. These simulations clearly show that despite being relatively weak, the shape and amplitude of the medicinal circulation is a crucial element shaping the solar cycle dynamics.

Incorporating the observed sunspots statistics \citep{Jiang2011} on the solar surface is crucial -- where the active region's emergence latitude and associated tilt angle and magnetic flux become major deciding factors to the final contribution to the dipole moment evolution \citep{Petrovay2020JSWSC, Wang2021}. It may appear that the final dipole moment at the end of a cycle (which acts as a seed to the following cycle) will then be deterministic - a strong cycle producing a strong poloidal field at cycle minimum. However, observation suggests otherwise [see Figure \ref{fig:polar_corr}(a)]: saturation in the final dipole moment, which is linked with two factors: tilt-quenching \citep{Dasi2010, Jiang2011} and latitude-quenching \citep{Jiang2020} (see Section \ref{sec4.3.2} for more details). Moreover, scatter in the active region tilt angles (in addition to the systematic tilt according to Joy's law) introduces substantial uncertainty in the final dipole moment at cycle minimum \citep{Jiang2014, Bhowmik2019A&A}. For example, a few big rouge active regions emerging at low latitudes with a “wrong” (i.e., opposite to the majority for this cycle) tilt angles can reduce the dipole moment amplitude, thus weakening the seed for the following cycle \citep{Jiang2015,Nagy2017}.

Observationally constrained and flux calibrated SFT models can now match decadal to centennial-scale solar surface magnetic field dynamics and are being used for predicting the polar field amplitude based on synthetic data inputs of the declining phase of the cycle -- with a reasonably good degree of success \citep{Cameron2010ApJ, 2014UptonHathaway_a, Cameron2016ApJ, Bhowmik2018NatCo, Wang2002ApJ}. These models have become useful tools for understanding solar surface flux transport dynamics, exploring the nuances of the Babcock-Leighton mechanism for solar poloidal field generation, and are being coupled to data driven dynamo models for predicting the strength of the sunspot cycle.

\subsection{Flux Transport Dynamo Models as a Tool for Sunspot Cycle Predictions}\label{sec}\label{sec3.2}

Kinematic or flux transport dynamo models have shown exceptional fidelity for being used as tools for solar cycle predictions. The utility of these models are due to the possibility of prescribing observationally constrained, or theoretically ``expected'' velocity profiles (\textbf{u}) and assimilating observations of the poloidal field to obtain the spatio-temporal evolution of the solar magnetic field (\textbf{B}) by using the (\ref{eq_induction}). These models use two large-scale time-independent velocity profiles to incorporate the observed differential rotation and meridional circulation (see Section \ref{sec2}). 

Based on the observed properties of the surface field, the large-scale solar magnetic field at cycle minimum can be reasonably approximated to be axisymmetric (independent of $\phi$) and antisymmetric across the equatorial plane. This simplifies the kinematic dynamo problem further. Thus in spherical polar coordinates ($r, \theta, \phi$), the magnetic field (\textbf{B}) can be expressed as,
\begin{equation}
    \mathbf{B}(r,\theta, t) = \nabla \times \mathcal{A}(r, \theta, t)\, \hat{\mathbf{e}}_{\phi} + \mathcal{B}(r, \theta, t)\, \hat{\mathbf{e}}_{\phi}.
\end{equation}

\noindent The first term in the R.H.S. of the above equation is the poloidal component ($\mathbf{B_P}$, hereafter) in the meridional plane expressed through a vector potential ($\mathcal{A}$) and the second term ($\mathcal{B}$) corresponds to the toroidal component ($\mathbf{B_T}$, hereafter). The velocity can also be expressed similarly as a combination of the poloidal (meridional circulation) and toroidal (differential rotation) components. All these simplifications lead us to two separate but coupled equations for $\mathbf{B_P}$ and $\mathbf{B_T}$, where the first corresponds to the axial dipole moment (or averaged polar field), and the latter is related to the sunspot-generating strong magnetic field. 

\begin{figure}[h]%
	\centering
	\includegraphics[width=0.95\textwidth]{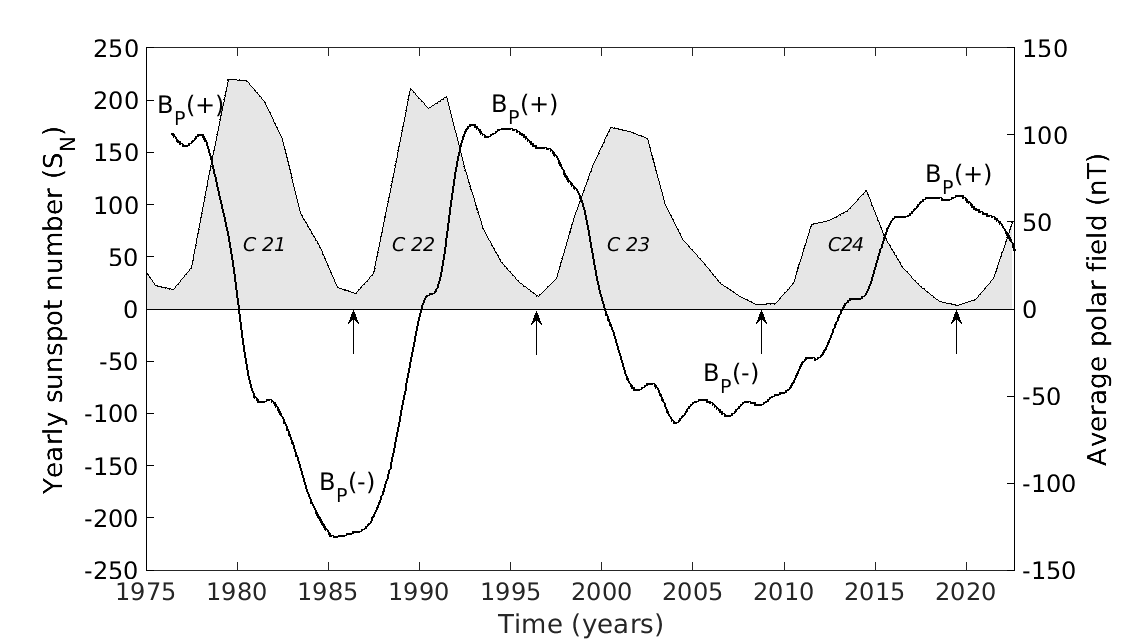}
	\caption{The temporal evolution of yearly sunspot number (source: WDC-SILSO, Royal Observatory of Belgium, Brussels) and average polar field (source: Wilcox Solar Observatory) during sunspot cycles: 21--24. The arrows denote the epochs of cycle minima approximately. The sign of the average polar field corresponds to the sign of  $\mathrm{B_P(+/-)}$, and the sunspot number is related to the amplitude of $\mathbf{B_T}$.}
 \label{fig:Bt_Bp}
\end{figure}

The solution to the set of equations produces $\mathbf{B_P}$ and $\mathbf{B_T}$ with a $\pi/2$ phase difference, both having roughly decadal-scale periodicity (considering amplitude only). It is reflected in Figure \ref{fig:Bt_Bp} showing the observed evolution of averaged polar field and sunspot cycles over four decades. A cycle begins with the strongest $\mathbf{B_P}$ and the weakest $\mathbf{B_T}$. In the rising phase of a cycle, with increasing $\mathbf{B_T}$ (i.e., more sunspots of opposite polarity), $\mathbf{B_P}$ (i.e., average polar field) weakens gradually through the B-L mechanism. $\mathbf{B_P}$ changes its polarity during the cycle maximum and progresses towards a new maximum value (with opposite polarity) during the declining phase of the cycle while $\mathbf{B_T}$ continues to decrease till the cycle minimum. Polarity-wise, $\mathbf{B_P}$ and $\mathbf{B_T}$ have a 22-year-long periodicity, which is also evinced through Hale's polarity law (as discussed in Section \ref{sec2}). In the following section, we describe how the generation process of the two components of the Sun's magnetic field rely on each other.

\subsubsection{Poloidal to Toroidal Field}

The induction equation for the toroidal component ($\mathbf{B_T}$) includes a source term originating from the differential rotation [$\Omega (r, \theta)$] in the SCZ, compared to which the sink term due to diffusive decay is negligible. Thus any pre-existing $\mathbf{B_P}$ will be amplified through shearing along the azimuthal direction ($\phi$) by $\mathbf{B_P}\cdot\nabla\Omega$ and generate new $\mathbf{B_T}$. The differential rotation in the solar convection zone and the stable overshoot layer at its base (coinciding with the tachocline where turbulent diffusivity is suppressed) plays important roles in the amplification and storage of $\mathbf{B_T}$ \citep{munoz2009}.

Following sufficient amplification by differential rotation, $\mathbf{B_T}$ satisfies the magnetic buoyancy condition \citep{Jouve2018ApJ,Fan2021}. Any perturbed part of the toroidal field rises as a flux rope through the SCZ, where it encounters the Coriolis force and turbulent diffusivity in the medium \citep{Weber2011}. The flux tube which eventually emerges through the solar surface creates a pair of spots (in general, Hale-spot) with a certain latitude-dependent tilt angle (following Joy's law), a somewhat fragmented structure and a reduced strength compared to its initial amplitude. A more detailed account of active region emergence has been discussed in a separate Chapter 6 of this book \citep{Chapter6}.

\subsubsection{Toroidal to Poloidal Field generation}

Only axisymmetric flows and fields cannot sustain the dynamo process (Cowling's theorem). Thus to sustain a dynamo, a non-axisymmetric process must be invoked. Elaborate studies on kinematic dynamo models have utilized different mechanisms to convert $\mathbf{B_T}$ to $\mathbf{B_P}$ \citep{Cameron2017SSRv, Charbonneau2020LRSP} by utilizing intrinsically non-axisymmetric processes which are parameterized in the dynamo equations. We present below a very brief narrative of such approaches.

\paragraph{Turbulence and Mean-field Electrodynamics Approach}

The thermally driven environment in the SCZ results in turbulent motion of the plasma, which therefore has a mean large-scale flow along with a fluctuating component [$\mathbf{u} = \langle\mathbf{u}\rangle + \mathbf{u}^{\prime}$]. While the mean component, $\langle\mathbf{u}\rangle$, corresponds to the standard axisymmetric large-scale plasma velocity (differential rotation and meridional circulation), the fluctuating component, $\mathbf{u}^{\prime}$, vanishes when averaged in the azimuthal direction. The magnetic field can be decomposed in a similar fashion: $\mathbf{B} = \langle\mathbf{B}\rangle + \mathbf{B}^{\prime}$. Although the fluctuating parts of the velocity and the magnetic field vanish individually when averaged azimuthally, their product, $\mathcal{E} = \langle\mathbf{u^{\prime}} \times \mathbf{B^{\prime}}\rangle$ will sustain and overcome the restriction set by Cowling's theorem. $\mathcal{E}$ is known as the mean turbulent electromotive force, and part of it serves as a source term (the $\alpha$-effect) in the induction equation of $\mathbf{B_P}$. From a physical point of view, it can be linked to the helical twisting of the toroidal field component ($\mathbf{B_T}$) by helical turbulent convection. For a thorough description, please refer to \cite{hazraISSIchapter2003}. 

\paragraph{The Babcock-Leighton Mechanism}

The magnetic axis connecting the opposite polarities of active regions has a certain tilt with respect to the east-west (toroidal) direction which arises due to the action of the Coriolis force on buoyantly rising toroidal flux tubes (an inherently non-axisymmetric process). Thus, all active regions have non-zero components of magnetic moments along the north-south (poloidal) direction -- which collectively contributes to the axial dipole moment generation and evolution \citep{Petrovay2020JSWSC}. Section \ref{sec3.1} describes how the magnetic flux initially concentrated within tilted active regions decay and redistributes across the solar surface to generate the large-scale magnetic field. Thus is the so called Babcock-Leighton mechanism which converts $\mathbf{B_T}$ to $\mathbf{B_P}$. Observational evidence not only strongly supports the the B-L mechanism, they also help constrain data driven predictive SFT and dynamo models \citep{Passos2014AA,Cameron2015Sci,Bhowmik2018NatCo, Bhowmik2019A&A}. For a more detailed account readers may consult Chapter 13 of this book \citep{Chapter13}.

One critical aspect of B-L type dynamos is the spatial dissociation between the source regions of $\mathbf{B_P}$ (on the solar surface) and $\mathbf{B_T}$ (in the deep SCZ). For the B-L dynamo to function effectively, the spatially segregated layers must be connected to complete the dynamo loop. The transport of $\mathbf{B_P}$ generated on the solar surface to the deeper layers of SCZ can occur through various processes. These include meridional circulation \citep{Choudhuri1995A&A,Nandy2002Sci} which acts as a conveyor belt connecting surface layers to the deep convection zone, turbulent diffusion \citep{Yeates2008} as well as turbulent pumping \cite{HazraS2016ApJ}. All these processes are an integral part of any flux-transport dynamo model, irrespective of whether the dominant poloidal field generation process is the mean field or the B-L mechanism.

A new approach towards predictive solar cycle modeling is the coupling of a 2D SFT model to an internal 2D dynamo model -- where the output from the first model serves as an upper boundary condition of the second one \citep{Lemerle2017ApJ}. Subsequently, the internal distribution of the toroidal magnetic field ($\mathbf{B_T}$) in the dynamo model generates synthetic sunspots emerging in the SFT model. This model, therefore, has the advantage of incorporating a full non-axisymmetric representation of the solar surface at a much lower numerical cost than the 3D models. The primary weakness of this 2\,$\times$\,2D model is its obligation to tackle the different spatial resolutions in the SFT and the dynamo components.

Presenting an elaborate account of all important works on SFT and solar dynamo modelling approaches is beyond the scope of this review; instead we have elaborated only on the primary physical mechanisms that are at the heart of the solar dynamo mechanism. We now turn our focus to processes that are at the basis of solar cycle fluctuations, understanding which is important from the perspective of solar cycle predictions.

\section{Physical processes influencing solar cycle predictability}\label{sec4}

As shown in Section \ref{sec2}, apart from its about 11-year periodicity, a prominent property of the solar activity record is the strong variability of the cycle amplitudes, including extended intervals of very low activity, e.g., Maunder minimum, or particularly high activity, e.g., modern maximum \citep{Usoskin2017}. Stochastic perturbations inherent in the turbulent solar-stellar convection zones and nonlinearities are two viable candidates for explaining solar cycle fluctuations. Understanding what drives these fluctuations and our ability to account for them either through first principles or observational data assimilation paves the way towards physics-based solar cycle predictions. 

\subsection{Numerical weather forecasts and nonlinear time series analysis of solar activity proxies}

Insights into the development of numerical weather or climate forecasting models over half a century serve as an useful analogy for physics-based solar cycle predictions \citep{Wiin-Nielsen1991} and could inspire the progress in physics-based solar cycle predictions. Numerical weather forecasts correspond to applying physical laws to the atmosphere, solving mathematical equations associated with these laws, and generating reliable forecasts within a certain timescale. The breakthrough from the 1930s to the 1950s can be classified into two groups. One is the physics of atmospheric dynamics. Vilhelm Bjerknes formulated the atmospheric prediction problem. C.~G. Rossby \cite{Rossby1939} derived the barotropic vorticity equation and proposed the first theory of the atmospheric long waves, i.e., Rossby waves. J. Charney \cite{Charney1948} developed the quasi-geostrophic theory for calculating the large-scale motions of planetary-scale waves. The second is the genesis of numerical calculation methods and the application of the computational method led by Lewis Fry Richardson, John von Neumann. From 1955 onwards, numerical forecasts generated by computers were issued regularly. People mainly concentrate on four domains to increase the performance of predictions \citep{Kalnay2003}; improve the representation of small-scale physical processes, utilize more comprehensive (spatial and temporal) observational data, use more accurate methods of data assimilation, and utilize more and more powerful supercomputers. 

Edward Lorenz \cite{Lorenz1963} opened the doors of  physics based weather forecasting by establishing the importance of nonlinear dynamics in the context of convecting systems and meteorology. In the subsequent remarkable papers \citep{Lorenz1965, Lorenz1969}, Lorenz made a fundamental discovery related to the predictability of weather arguing that nonlinearity leads to chaotic dynamics making long-range forecasts impossible. We now know that the chaotic nature of the atmosphere imposes a limit of about two weeks in weather forecasts even with ideal models and perfect observations. 

Advances in time-series analysis of non-linear dynamics since the 1980s have made it possible to distinguish between stochastic behavior and deterministic chaos in principle. Strange attractor reconstruction based on correlation integral and embedding dimension \citep{Takens1981, Grassberger1983} and the method of surrogate data \citep{Theiler1992, Paluvs1999} are the most widely used methods to look for chaotic behavior. Numerous attempts in this field have been invoked in the literature by analyzing different time series of solar activity proxies, e.g., sunspot number data \citep{Price1992}, sunspot area data \citep{Carbonell1994}, cosmogenic data \citep{Hanslmeier2010}, polar faculae \citep{Deng2016}, and so on. However, these studies show highly diverging results. For example, some studies \citep{Mundt1991,Rozelot1995,Hanslmeier2010,Deng2016} report evidence for the presence of low-dimensional deterministic chaos in solar cycles. On the other hand, others \cite{Price1992, Carbonell1994, Mininni2000} find no evidence that sunspots are generated by a low-dimensional chaotic process. Even in studies showing evidence of chaos, divergent values of the system parameters (e.g., maximum Lyapunov exponent) were estimated -- indicating divergent prediction time scales. It is suggested that results claiming the existence of chaos were derived from short scaling regions obtained using very low time delays in the computations for the correlation dimension \cite{Carbonell1994}. Furthermore, the ways to filter or smooth solar activity data also strongly impact the results \citep{Price1992, Petrovay2020LRSP}.  

In brief, despite the intensive investigation of solar activity data, there is no consensus on whether chaos or inherent stochastic perturbations or noise, or a combination of these processes drive solar cycle fluctuations. The insufficient length of the solar activity data, that is sparsity of data in phase space, compromises statistical sampling. Clearly distinguishing between stochastic modulation and deterministic chaos remains an outstanding issue. However, driving predictive physical models with observational data of the poloidal component of Sun's magnetic field (which is primarily subject to stochasticity and nonlinear effects) provides a way out of this conundrum. 

\subsection{Low-order models of the solar cycle}
\label{sec:lowOrderModel}

Model building aims to use our understanding of a physical system by establishing dynamical equations explaining a physical phenomena and furthermore, aid in the interpretation of observational data. One such approach -- low-order dynamo models -- usually approximate physical processes that occur in a dynamo through truncated equations. Such models have the advantage of exploring a wide variety of solar behavior that is governed by the same underlying mathematical structure, without studying the dynamo process in detail, or making other modeling assumptions. Sections 3.5 and 3.6 of Petrovay (2020) \cite{Petrovay2020LRSP} give an overview of this topic classified by two types of low-order models: truncated models and generic normal-form models. See also the review by Lopes et al. \cite{Lopes2014}. Here we present the progress in this field by classifying them based on different physical processes influencing solar cycle variability and predictability. Although there is no conclusive evidence of the presence or absence of chaos, most recent studies suggest that the irregular component in the variation of solar activity is dominated by stochastic mechanisms. 

\subsubsection{Deterministic chaos subject to weak stochastic perturbations}

\label{sec:chaos}
Such studies assume the non-linear solar dynamo is a chaotic oscillator, subject only to weak stochastic perturbations. The generic normal-form equations are investigated utilizing the theory of non-linear dynamics by bifurcation analysis. The bifurcation sequences are robust. Although the approach has no actual predictive power, they provide an understanding of generic properties and explain the origin of assumed chaotic behavior. Tobias et al. \cite{Tobias1995} used a Poincare-Birkhoff normal form for a saddle-node or Hopf bifurcation. Their results show that stellar dynamos are governed by equations that possess the bifurcation structure. Modulation of the basic cycle and chaos are found to be a natural consequence of the successive transitions from a non-magnetic state to periodic cyclic activity and then to periodically modulated cyclic activity followed by chaotically modulated cycles.  This behaviour can be further complicated by symmetry-breaking bifurcations that lead to mixed-mode modulation of cyclic behaviour \citep{Knobloch1998}.
Trajectories in the phase space spend most of the time in the dipole subspace, displaying modulated cyclic activity, but occasionally flip during a deep grand minimum. Wilmot-Smith et al. \cite{Wilmot-Smith2005} extended the model of Tobias et al. \cite{Tobias1995} to include an axisymmetry-breaking term. The model is able to reproduce some properties found in observations of solar-type stars. Their solution also exhibits clustering of minima, together with periods of reduced and enhanced magnetic activity.

There are also some studies using truncated dynamo models to investigate chaotic behaviour. For example, some argue that mixed modes of symmetry can only appear as a result of symmetry-breaking bifurcations in the nonlinear domain based on a constructed minimal nonlinear $\alpha$-$\Omega$ dynamo  \cite{Jennings1991}. Tobias et al. \cite{Tobias1996} show that grand minima naturally occur in their low-order non-linear $\alpha$-$\Omega$  dynamo if the magnetic Prandtl number is small. The pattern of magnetic activity during grand minima can be contrasted both with sunspot observations and with the cosmogenic record.

H. Yoshimura \cite{Yoshimura1978} suggested that a time-delay mechanism is intrinsic to the feedback action of a magnetic field on the dynamo process. Wilmot-Smith et al. \cite{Wilmot-Smith2006} constructed a truncated dynamo model to mimic the generation of field components in spatially segregated layers and their communication was mimicked through the use of time delays in a dynamo model involving delay differential equations. A variety of dynamic behaviors including periodic and aperiodic oscillations similar to solar variability arise as a direct consequence of the introduction of time delays in the system. Hazra et al. \cite{HazraS2014ApJ} extended the model of \cite{Wilmot-Smith2006} by introducing stochastic fluctuations to investigate the solar behaviour during a grand minimum. Recently Tripathy at al. \citep{Tripathi2021} apply the model with an additive noise to understand the breakdown of stellar gyrochronology relations at about the age of the Sun \citep{vanSaders2016}. The one-dimensional iterative map is an effective and classical method to investigate the dynamics of a system. Using this method, studies \cite{Durney2000, Charbonneau2001} explored the dynamical consequences of the time delay in the B-L type dynamo. As the dynamo number increases beyond criticality, the system exhibits a classical transition to chaos through successive period doubling bifurcations. The odd-even pattern in sunspot cycle peak amplitudes is also reproduced when low amplitude stochastic fluctuations are introduced.

\subsubsection{Weakly nonlinear limit cycle affected by random noise}

\label{sec:randomNoise}

Since the non-stationary nature of solar convection is an intimate part of the solar dynamo, a rich body of literature regards that solar variability is largely governed by stochastic perturbations.  Random noise has been used to fully mimic the behaviour of the solar cycle \cite{Barnes1980} while others describe the global behavior of the solar cycle in terms of a Van der Pol oscillator \cite{Mininni2001}; a stochastic parameter corresponding to a stochastic mechanism in the dynamo process was introduced in the Van der Pol equations to model irregularities in the solar cycle were modeled.  The mean values and deviations obtained for the periods, rise times, and peak values, were in good agreement with the values obtained from the sunspot time series. Another example is a low-order dynamo model with a stochastic $\alpha$-effect \cite{Passos2011} in which grand minima episodes manifested; this model is characterized by a non-linear oscillator whose coefficients retain most of the physics behind dynamo theory.

While most low-order models have a loose connection with observations, Cameron \& Sch{\"u}ssler \cite{Cameron2017} developed a generic normal-form model, whose parameters are all constrained by observations. They introduce multiplicative noise to the generic normal-form model of a weakly nonlinear system near a Hopf bifurcation. Their model reproduces the characteristics of the solar cycle variability on timescales between decades and millennia, including the properties of grand minima, which suggest that the variability of the solar cycle can be understood in terms of a weakly nonlinear limit cycle affected by random noise. In addition, they argue that no intrinsic periodicities apart from the 11-year cycle are required to understand the variability.

\subsection{Babcock-Leighton-type kinematic dynamo models}

Over the past two decades -- supported by advances in flux transport models and observational evidence -- Babcock-Leighton type solar dynamo models have become the mainstream approach to solar cycle modeling. The B-L mechanism imbibes the processes of emergence of toroidal fields through the convection zone and their subsequent decay and transport by supergranular diffusion and large-scale surface flow fields over the surface, i.e., the SFT processes discussed in Section \ref{sec3.1}. Since the SFT processes are directly observed, these act as a source of data assimilation in the B-L dynamo paving the way towards data-driven predictions -- akin to what has been achieved in weather models.

\subsubsection{Effects of the meridional flow and the time delay}

The meridional flow plays an essential role in the B-L type flux transport dynamo (FTD). The flow strength can modulate not only the cycle strength but also the cycle period. There exists a rich literature describing the effects of the meridional flow on modulation of solar cycles based on FTD models \citep{Yeates2008, Karak2010, Nandy2011, Karak2013}. Bushby \& Tobias \cite{Bushby2007} introduced weak stochastic perturbations in the penetration depth of the meridional flow and the results showed significant modulation in the activity cycle; while they argue that this modulation leads to a loss of predictability. Nandy \cite{Nandy2021SoPh} provides counter arguments pointing out short-term prediction up to one cycle is possible due to the inherent one-cycle memory in the sunspot cycle; see also \cite{Yeates2008,Karak2012ApJ,HazraS2020A&A}. We note that recently Zhang \& Jiang \cite{Zhang2022} develop a B-L type dynamo working in the bulk of the convection zone. The model has a much weaker dependence on the flow. Only the flow at the solar surface plays a key role in the polar field generation as in the SFT models implying that the flux transport paradigm is not hostage to meridional circulation being the primary transporter of magnetic flux within the SCZ -- as also argued in Hazra \& Nandy \cite{HazraS2016ApJ}.

\begin{figure}[h]%
	\centering
	\includegraphics[width=0.5\textwidth]{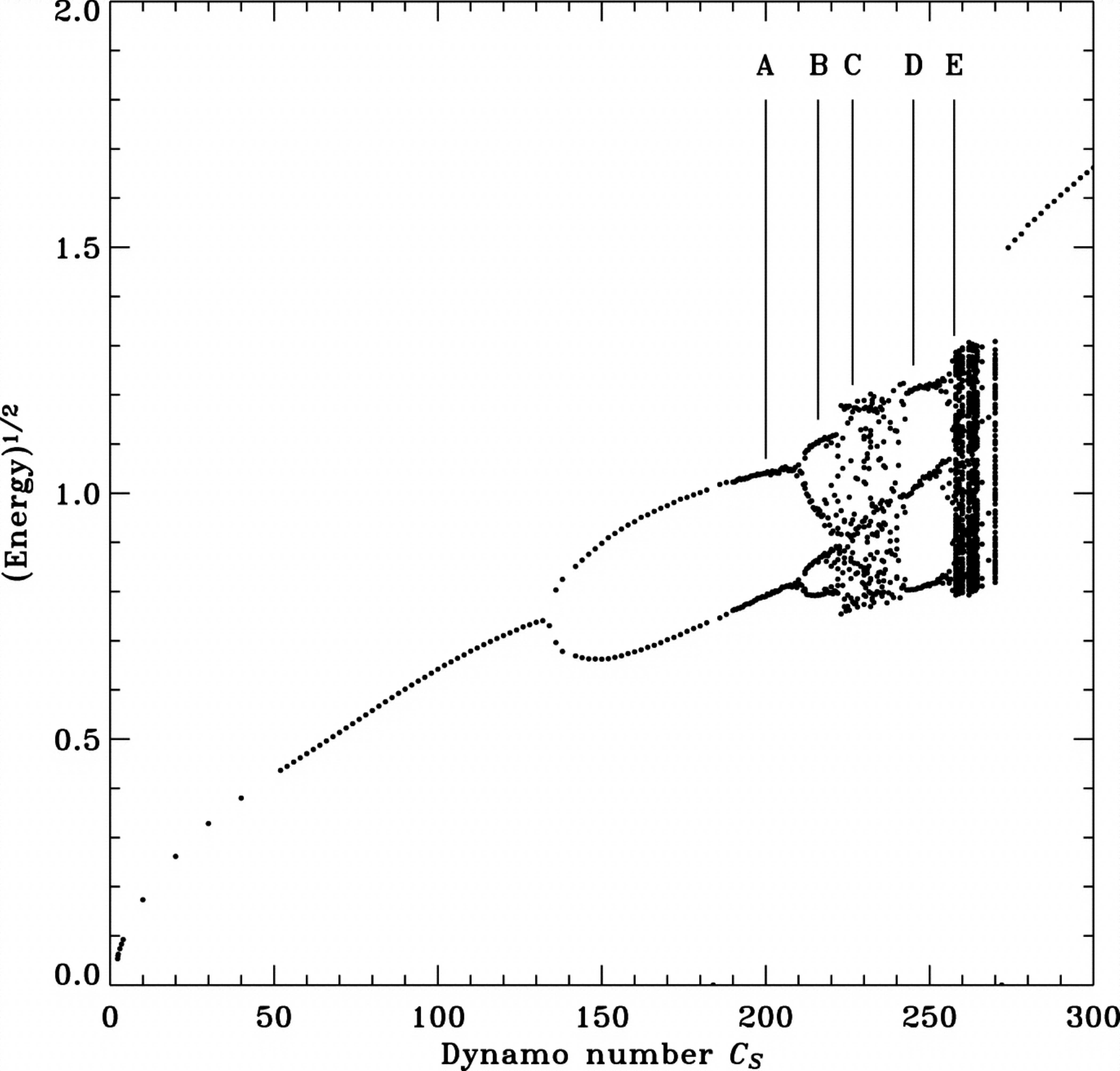}
	\caption{Bifurcation diagram reconstructed from a sequence of numerical 2D B-L type dynamo solutions with increasing dynamo numbers $C_S$. Vertical lines labeled `A' to `E' correspond to representative 2-periodic, 4-periodic, 5-periodic, 3-periodic, and chaotic solutions, respectively. Image reproduced with permission from \cite{Charbonneau2005}, copyright by the American Astronomical Society.}\label{fig1_Charbonneau2005}
\end{figure}

In Section \ref{sec:lowOrderModel} we have shown numerous attempts at the analysis of the dynamical system using low-order models. For the first time, Charbonneau et al. \cite{Charbonneau2005} presented a series of numerical simulations of the PDE-based 2D B-L type dynamo model incorporating amplitude-limiting quenching nonlinearity. The solutions show a well-defined transition to chaos via a sequence of period-doubling bifurcations as the dynamo numbers $C_S=s_0 R_\odot/\eta_t$ ($s_0$: strength of the source term, $\eta_t$ is the turbulent magnetic diffusivity in the Sun’s convective envelope) is increased. The results are presented in Figure \ref{fig1_Charbonneau2005}. Hence they suggest that the time delay inherent to the B L type dynamo process, acting in conjunction with a simple amplitude-quenching algebraic-type nonlinearity could naturally lead to the observed fluctuations in the amplitude of the solar cycle. The time delay was regarded as the third class of fluctuation mechanisms by Charbonneau et al. \cite{Charbonneau2005}. The method was further extended \cite{Charbonneau2007} to investigate the odd-even pattern in sunspot cycle peak amplitudes. Indeed, it is now being recognized that time delays introduced into the dynamo system due to the finite time necessary for flux transport processes to bridge the source layers of the poloidal and toroidal fields across the convection zone introduces a memory into the system which makes solar cycle predictions a realistic possibility \cite{Nandy2021SoPh}.

\subsubsection{Observable nonlinear and stochastic mechanisms in the source term}\label{sec4.3.2}

Within the framework of the B-L type dynamo, the emergence and decay of tilted bipolar sunspots give rise to the poloidal field. The amount of poloidal field depends on the sunspot properties, e.g., the tilt angle of the bipolar sunspots, which show both the systematic property resulting from Joy's law and the stochastic one due to the tilt scatter. Hence, the B-L mechanism has an inherent randomness. Studies like \cite{Charbonneau2000,Choudhuri2012,Passos2014AA,Karak2017,Kitchatinov2018,Hazra2019,saha2022} took the stochastic fluctuation in the poloidal field source term as a free parameter and investigated their possible effects on the cycle modulation. Based on a B-L type solar dynamo with an additional mean-field $\alpha$-effect, Sanchez et al.  \cite{Sanchez2014} quantified the intrinsic limit of predictability, i.e., $e$-folding time $\tau$, which is the equivalent of the two weeks for the weather forecast. As expected, the $e$-folding time is shown to decrease corresponding to a short forecast horizon, with the increase of the $\alpha$-effect.

Studies \cite{Kitchatinov2011,Olemskoy2013} attempted to estimate the parameters of the B-L mechanism and their fluctuations using historical sunspot data. Jiang et al. \cite{Jiang2014} measured the tilt-angle scatter using the observed tilt angle data and quantified the effects of this scatter on the evolution of the solar surface field using SFT simulations with flux input based upon the recorded sunspot groups. The result showed that the effect of the scatter on the evolution of the large-scale magnetic field at the solar surface reaches a level of over 30\%.  When a BMR with area $A$ and tilt angle $\alpha$ emerges, it has the (initial) axial dipole field strength $D_i\propto A^{1.5}\sin\alpha$. We define the final contribution of a BMR to the axial dipole field as the final axial dipole field strength $D_f$. Jiang et al. \cite{Jiang2014} show that $D_f$ has the Gaussian latitudinal dependence. The result was confirmed by others \cite{Nagy2017, Whitbread2018, Petrovay2020b}, that is 
\begin{equation}
	D_f=D_i\exp(-\lambda^{2}/\lambda_R^2),
	\label{eqn:D_f}
\end{equation}
where $\lambda_R$ is determined by the ratio of equatorial flow divergence to diffusivity. Wang et al. \cite{Wang2021} further generalized the result to ARs with realistic configuration, which usually show large differences in evolution from the idealized BMR approximation for $\delta$-type ARs \citep{Jiang2019,Yeates2020}. Hence big ARs emerging close to the equator could have big effects on the polar field evolution, and on the subsequent cycle evolution based on the correlation between the polar field at cycle minimum and the subsequent cycle strength. These are referred to as rogue ARs by Nagy et al. \cite{Nagy2017}. Jiang et al. \cite{Jiang2015} demonstrated that these low-latitude regions with abnormal polarity could indeed be the cause of the weak polar field at the end of Cycle 23, hence the low amplitude of Cycle 24. Simulations by Nagy et al. \cite{Nagy2017} indicate that in the most extreme case, such an event could lead to a grand minimum; they argue that the emergence of rogue ARs in the late phase of a cycle may limit the scope of predicting the dipole moment (and polar field amplitude) at the minimum of a cycle. However, it is likely that the impact of such rogue regions may be estimated through the ensemble prediction approach \cite{Bhowmik2018NatCo}.

The stochastic properties of sunspot group emergence mentioned above provide the observable stochastic mechanisms in solar cycle modulation. The systematic properties of sunspot group emergence have recently been suggested to be observable nonlinearities. Historical data show that the cycle amplitude has an anti-correlation with the mean tilt angle of BMRs \citep{Dasi2010,Jiao2021} and a positive correlation with the mean latitudes \citep{Solanki2008,Jiang2011}. J. Jiang \cite{Jiang2020} investigated the effects of the latitude and tilt's properties on the solar cycle, which are referred to as latitudinal quenching and tilt quenching, respectively. They defined the final total dipole moment, which is the total dipole moment generated by all sunspot emergence during a whole cycle. Both forms of quenching lead to the expected final total dipolar moment being enhanced for weak cycles and saturated to a nearly constant value for normal and strong cycles. This naturally explains observed long-term solar cycle variability, e.g., the odd-even rule. B.~B. Karak \cite{Karak2020} verified that latitudinal quenching is a potential mechanism for limiting the magnetic field growth in the Sun using a three-dimensional B-L type dynamo model. Talafha et al. \cite{Talafha2022} systematically explored the relative importance played by these two forms of quenching in the solar dynamo showing that this is governed by $\lambda_R$.

\begin{figure}[h]%
	\centering
	\includegraphics[width=0.5\textwidth]{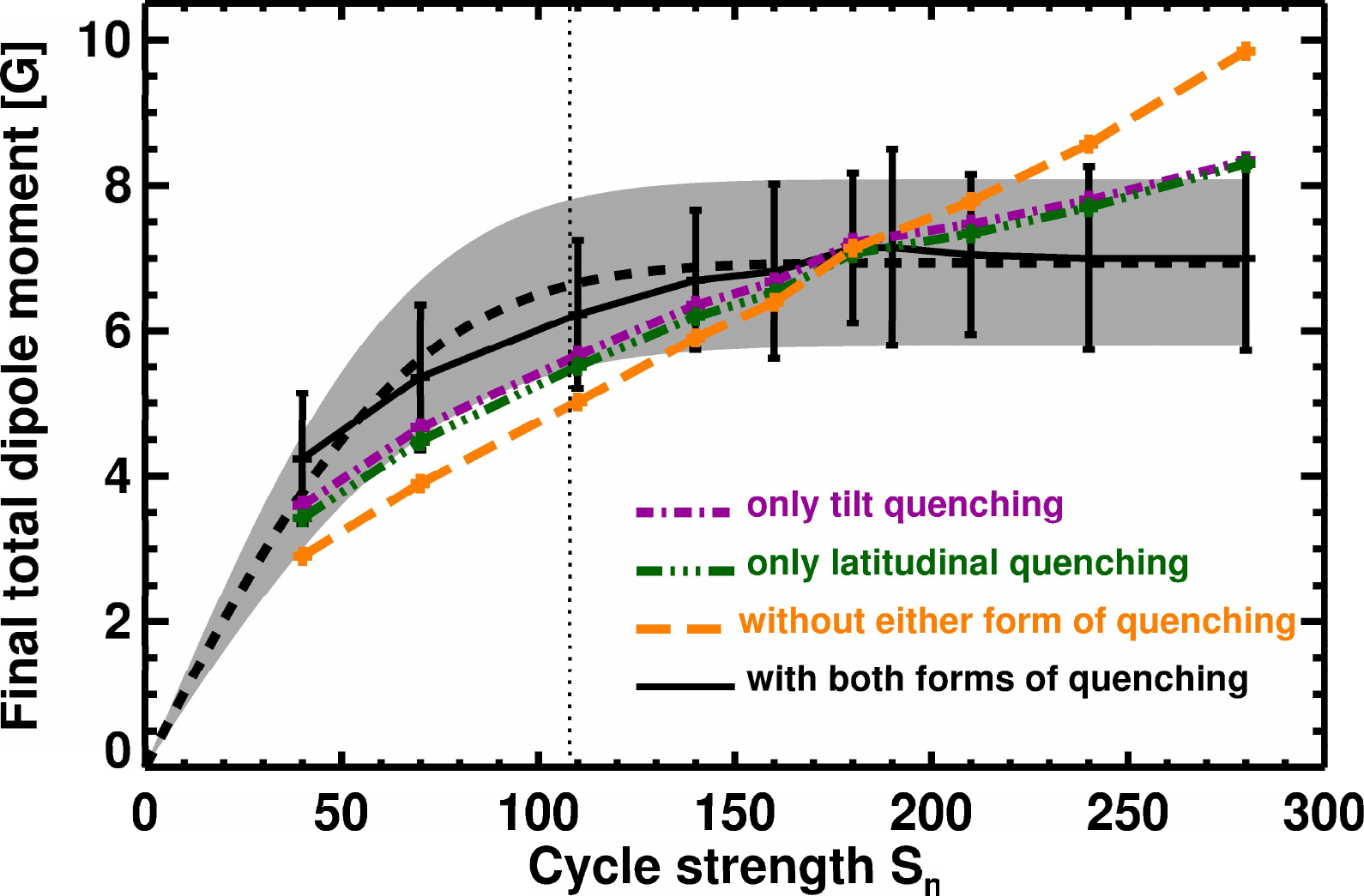}
	\caption{Effects of observable nonlinear and stochastic mechanisms in the source term on the poloidal field generation. The black solid curve indicates the expected values from 100 SFT simulations using random sunspot group realizations including latitudinal and tilt quenching. Error bars correspond to the 1$\sigma$ standard deviation, caused by the randomness in the properties of sunspot groups. Green dashed–triple-dotted and purple dashed–dotted curves show the expected values for SFT simulations with only the latitudinal and tilt quenching, respectively. The orange dashed curve shows the expected value of SFT simulations without latitudinal or tilt quenching. Image reproduced with permission from \cite{Jiang2020}, copyright by the American Astronomical Society.}
	\label{fig2_Jiang2020}
\end{figure}

\section{Physics-based Solar Cycle Predictions}\label{sec5}

The importance of the dipole moment (or the average polar field) in solar cycle predictions is established through observation and dynamo theory. The most successful empirical method for solar cycle predictions based on the polar field precursor \cite{schatten1978} in fact predated the solar dynamo model based predictions. Thus, physics-based predictions of the solar cycle, in general, are either based on SFT simulations aiming to estimate the dipole moment (related to polar flux) at the cycle minimum or involve dynamo simulations with modified poloidal field source term according to the observed (or simulated) dipole moment at the cycle minimum. Nandy \cite{nandy2002predict} first alluded to the possibility of developing data driven predictive solar dynamo models by utilizing the observed poloidal field as inputs; although this particular branch of solar cycle predictions is relatively new, significant progress has been achieved through contributions from multiple works predicting the past cycle 24 and present cycle 25 using physics-based models. 

\subsection{Role of SFT Models in Solar Cycle Predictions} 
\label{sec5.1} 

Despite the dissimilarities among different SFT models regarding their treatments of the emerged sunspot statistics and observed transport processes on the photosphere, SFT models have played a major role in physics-based solar cycle predictions, especially of cycle 25. The idea lies in the fact that the Sun's magnetic axial dipole moment at the end of a solar cycle is strongly correlated with the following cycle's peak amplitude. This positive correlation is found from observation spanning multiple cycles \citep{Munoz2013ApJ} and is also supported by the principles of the solar dynamo mechanism (see Section \ref{sec3.2} for more details). 

Using SFT simulations, Cameron et al. \cite{Cameron2016ApJ} presented the first prediction of cycle 25 about four years before the cycle minimum (which occurred at the end of 2019). Their simulation started with the observed synoptic magnetogram at the end of 2015. The sunspot emergence statistics in the declining phase (years: 2016\,-\,2020) of cycle 24 were generated using 50 randomizations which included uncertainties associated with the emergence timing, latitude-longitude position, tilt angle and magnetic flux of the sunspots. They provided a possible range of the axial dipole moment at the end of cycle 24. Based on the positive correlation between dipole moment and the following cycle amplitude, they predicted a moderately strong sunspot cycle 25. Hathaway \& Upton \cite{Hathaway2016JGRA,Upton2018GeoRL} took a similar approach to estimate the axial dipole moment at the end of cycle 24 using their AFT simulations. However, the sunspot statistics corresponding to the declining phase of cycle 24 were taken from the declining phase of solar cycle 14. The uncertainties in predicting the dipole moment were realised by considering stochastic variations in the convective motion details, sunspot tilt angles, and changes in the meridional flow profile. Their predicted dipole moment amplitude suggested that cycle 25 would be a weak to moderate cycle. Iijima et al. \cite{Iijima2017A&A} argued that the axial dipole moment does not vary significantly in the last three years of the declining phase of any sunspot cycle. Thus to predict the dipole moment at the end of cycle 24, they initiated their SFT simulation in 2017 with an observed synoptic magnetogram and continued till the end of 2019 without assimilating any sunspots. Their prediction suggested a weak solar cycle 25. The importance of correctly simulating the surface magnetic field distribution and their consecutive inclusion in an interior dynamo model was extensively utilized by other studies \citep{Labonville2019SoPh, Bhowmik2018NatCo} for cycle predictions. 

\subsection{Dynamo-based Solar Cycle Predictions with Data Assimilation}\label{sec5.2}

The B-L type 2D flux transport dynamo models were utilized for the first time to predict cycle 24 during the mid-2000s \citep{Dikpati2006GeoRL, Choudhuri2007PhRvL}. However, the only two physics-based predictions of cycle 24 diverged significantly from each other (with a difference of $\sim$ 100 sunspots during the maximum). Despite using similar dynamo models, such divergence can arise from two aspects: differences in the dominating flux transport processes \cite{Yeates2008} and how the B-L source term is designed according to the observed magnetic field. Exploring these two points is crucial for understanding the physics behind sunspot cycle predictions as well as providing realistic forecasts.

All kinematic flux transport dynamo models consider the following transport processes at the least: differential rotation, meridional circulation and magnetic diffusion. They use analytical functions corresponding to the observed differential rotation \citep{Howe2009LRSP}. The observed meridional flow on the solar surface sets constraints for the meridional circulation profile used within the SCZ. The exact structuring of this flow at different layers of the SCZ still requires further observational support \cite{Hanasoge2022LRSP}, but recent helioseismic studies suggest a one-cell meridional circulation \citep{Rajaguru2015ApJ,Liang2018A&A}. Nonetheless, these models assume an equatorward branch of the meridional flow near the tachocline which ensures observed latitudinal propagation of sunspot activity belts in both hemispheres and an appropriate cycle duration \citep{Choudhuri1995A&A, Dikpati1999ApJ, Nandy2002Sci, Chatterjee2004A&A, HazraS2014ApJ, Hazra2019MNRAS}. Furthermore, the amplitude and profiles of magnetic diffusivity are also based on analytical functions, which only vary with the depth of the SCZ \citep{Andres2011ApJ}. 

However, based on the strength of diffusivity, flux transport dynamo models behave differently and can be categorized into two major classes: advection dominated (diffusivity order, $\eta \sim 10^{10}$ cm$^2$s$^{-1}$, see \cite{Dikpati1999ApJ}) and diffusion dominated ($\eta \sim 10^{12}$ cm$^2$s$^{-1}$, see \cite{Choudhuri2007PhRvL}). The strength of the diffusivity decides which transport mechanism between meridional circulation and magnetic diffusivity will be more effective for convecting $\mathbf{B_P}$ to the deeper layers of the SCZ \cite{Yeates2008}. It also determines whether $\mathbf{B_P}$ associated with multiple past solar cycles can survive in the SCZ at the prescribed diffusivity and contribute simultaneously to the generation of new $\mathbf{B_T}$ of the following cycle \citep{Yeates2008}. However, the inclusion of turbulent pumping \citep{Karak2012ApJ, HazraS2016ApJ} as an additional transport process in flux transport dynamo diminishes the difference between the advection-dominated and diffusion-dominated regimes. All these results are crucial for estimating the dynamical memory of the solar dynamo models and their ability to accurately predict the future solar cycle amplitude. Dynamical memory is a measure of determining the range of temporal association of the poloidal field ($\mathbf{B_P}$) of a certain sunspot cycle (say, n$^\mathrm{th}$) with the toroidal field ($\mathbf{B_T}$) of following cycles (say, n+1$^\mathrm{th}$, n+2$^\mathrm{th}$, n+3$^\mathrm{th}$, etc.). Note that for advection-dominated dynamo models, the dynamical memory is about two solar cycles, whereas it's about half a solar cycle for diffusion-dominated dynamo models (or in models with turbulent pumping, \citep{Yeates2008, Karak2012ApJ}). 

Besides the transport parameters, how we assimilate observational data to model the poloidal field ($\mathbf{B_P}$) source will influence the successive generation of the toroidal field ($\mathbf{B_T}$), thus is crucial for solar cycle predictions. Below, we discuss this aspect of `data-driven' dynamo models in the context of solar cycle prediction.

As mentioned in Section \ref{sec3.2}, surface flux transport processes acting on emerged active regions produces the large-scale photospheric field and serves as a reliable means for poloidal field generation. Multiple efforts have been made to assimilate the observed surface data in flux transport dynamo models. Dikpati et al. \cite{Dikpati2006ApJ} included observed sunspot group areas to formulate the $\mathbf{B_P}$ source while assuming all spots of any solar cycle are distributed in the same latitudinal belt (between 5$^\circ$ and 35$^\circ$) and have similar tilt angles. However, their data-driven model failed to correctly predict the solar cycle 24 peak \citep{Dikpati2006GeoRL}. The primary reasons for this disparity were that the idealized realization of the sunspots (fixed latitude and tilt angle) results in a poloidal source at the minimum (of n$^\mathrm{th}$ cycle) directly proportional to the preceding cycle (n$^\mathrm{th}$) amplitude 
and that the low magnetic diffusivity in their flux transport dynamo model increased the dynamical memory to more than two solar cycles. Thus according to their model, not only does a strong solar cycle (n$^\mathrm{th}$) produce a strong poloidal source, its strength influences several following solar cycles (n+1$^\mathrm{th}$, n+2$^\mathrm{th}$ and n+3$^\mathrm{th}$).

In contrast, Choudhuri et al. \citep{Choudhuri2007PhRvL} and Jiang et al. \citep{Jiang2007MNRAS} used a diffusion-dominated flux transport dynamo model to predict sunspot cycle 24. For modeling the $\mathbf{B_P}$ source, they relied on observed large-scale surface magnetic field distribution (for example, the axial dipole moment) during the solar cycle 23 minimum. Their prediction was a good match to the observed peak of cycle 24. The positive correlation between $\mathbf{B_P}\mathrm{(n)}$ and the $\mathbf{B_T}\mathrm{(n+1)}$ using a diffusion-dominated dynamo model where the dynamic memory is half a solar cycle ensured the success of their prediction. Guo et al. \citep{Guo2021SoPh} took a similar approach by combining observed axial dipole moment to a diffusion-dominated dynamo model to predict cycle 25. In a recent review, Nandy \cite{Nandy2021SoPh} discusses in details how observations and stochastically forced dynamo simulations support only a short half- to one-cycle memory in the solar cycle, suggesting that the latter class of dynamo models are the right approach to take for solar cycle predictions.

Recently, Bhowmik \& Nandy \citep{Bhowmik2018NatCo} assimilated a series of surface magnetic field distribution at eight successive solar minima (cycles 16\,-\,23 minima) in a flux transport dynamo model (diffusion-dominated). The surface maps were obtained from their calibrated century-scale SFT simulation, which assimilates the observed statistics of emerging bipolar sunspot pairs during that period. Their coupled SFT-dynamo simulations reproduced past solar cycles (17\,-\,23) with reasonable accuracy (except cycle 19). The same methodology was utilized to provide an ensemble forecast for cycle 25 while assimilating the predicted surface field distributions at cycle 24 minimum from their SFT simulations. Their prediction indicates a weak cycle 25 (with a peak SSN of 118 and a range from 109 - 139), similar to, or slightly stronger than cycle 24. Note that the upper bound of the Bhowmik \& Nandy \citep{Bhowmik2018NatCo} prediction as reported in Nandy \citep{Nandy2021SoPh} was misreported as 155 but should have been 139. The 2\,$\times$\,2D SFT-dynamo model by Lemerle et al. \citep{Lemerle2017ApJ} is another example of coupling the surface magnetic field to the internal dynamo, which occurs more intimately in their model. Labonville et al. \citep{Labonville2019SoPh} utilized the same model to assimilate the series of BMRs observed during cycles 23 and 24 (from \citep{Yeatesetal2007}) into the SFT part of the simulations. They first calibrated the model by including cycle 23 data to produce an ensemble forecast for cycle 24 and subsequently comparing it with observation. They then assimilated the BMRs series for cycles 23 and 24 to present an ensemble forecast for cycle 25, including its amplitude (weaker than cycle 24), rising and declining phases and northern and southern asymmetry. 

\begin{figure}[h!]
	\centering
	\includegraphics[width=0.9\textwidth]{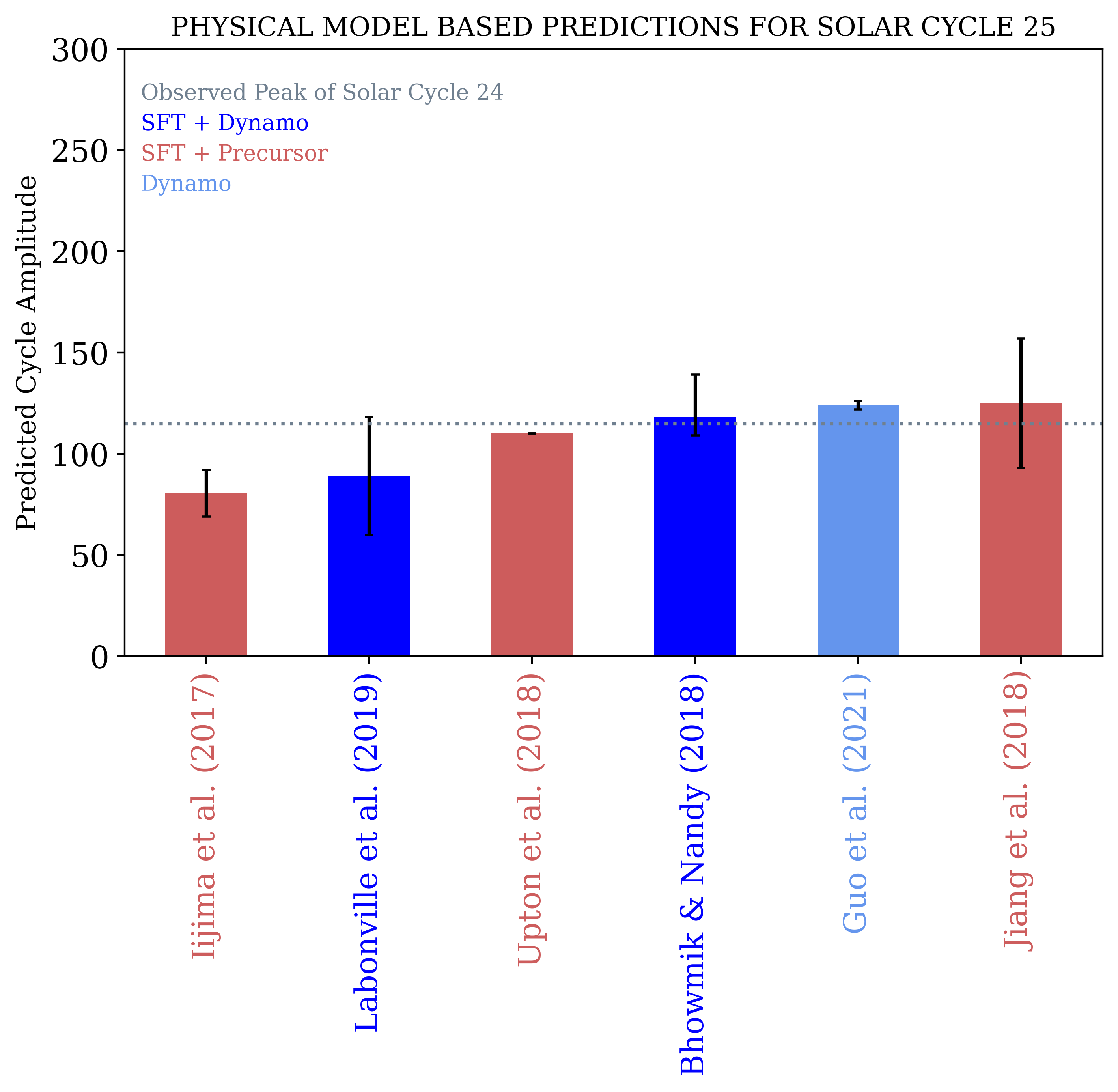}
	\caption{Comparison of physical model-based predictions of solar cycle 25 peak amplitude. The average of the six predictions is 107.75 sunspots (with $\pm 1\sigma = 17.15$). Details on each of these predictions are described in Sections \ref{sec5.1} and \ref{sec5.2}.}\label{fig_prediction}
\end{figure}

All physics-based predictions for cycle 25 are depicted in Figure \ref{fig_prediction}. 

\subsection{Comparison with other prediction methods}\label{sec5.3}

The methodologies of predictions are not restricted to physical modeling only. They can be based on (a) precursor technique-based forecasts, (b) non-linear mathematical modeling, (c) statistical forecasts, (d) machine learning and neural network, (e) spectral methods etc. \citep{Petrovay2020LRSP, Nandy2021SoPh}. Note that most of the precursor-based forecasts consider the physics of solar dynamo and cycle predictability to some extent without performing any computational modeling. For example, based on semi-empirical and semi-physical approaches Hazra et al. \citep{HazraG2019ApJ} and Kumar et al. \citep{Kumar2022MNRAS} predicted solar cycle 25 amplitude, where the latter group claimed that polar field evolution after the polarity reversal exhibits similar features like the Waldmeier effect and can be utilized for cycle predictions. Nonetheless, the forecasts based on diverging techniques show a wide variation in predicted cycle amplitudes: with peak sunspot numbers ranging between 50 and 250 for solar cycle 25 (see Figure 3 of Nandy \citep{Nandy2021SoPh}). In that regard, physics-based predictions of cycle 25 have reached a consensus with an average of 107.75 sunspots (with $\pm 1\sigma = 17.15$). In contrast, for cycle 24 physics-based predictions, the average was 179.44 (with $\pm 1\sigma = 63.44$).

\section{Summary and Future Prospects}\label{sec6}

It is noteworthy that while the only two physics-based predictions of solar cycle 24 diverged significantly from each other, physics-based predictions of cycle 25 show significantly more convergence. This is possibly an indication of increasing understanding of the solar dynamo process -- as argued in Nandy (2002) \cite{nandy2002predict} -- and advances in assimilating observational data in the computational models used for predictions. However, there are significant improvements that are still necessary in the adopted modeling approaches.

While all physical models of solar cycle predictions have been 2D in nature, global 3D dynamo models have the promise of capturing the surface flux transport dynamics and internal magnetic field evolution self-consistently. Some recent works have solved the induction equation in three spatial dimensions within a dynamo framework thus going beyond the 2D axisymmetric models \citep{Yeates2013MNRAS, Miesch2014ApJ, Hazra2017ApJ, Karak2017, Kumar2019A&A, Whitbread2019A&A}. These are B-L type dynamos working in a kinematic mode with modules to incorporate realistic sunspot emergence and decay of the associated flux. These models provide the opportunity of further development towards dynamical models, imbibing in-built axisymmetric and non-axisymmetric feedback mechanisms \citep{Nagyetal2020}, thus slowly closing the gap between these phenomenological B-L type dynamo models and the full MHD dynamo models. However, such models still have two significant discrepancies: the polar field generated on the surface is much stronger than the observed order of magnitude, and sunspot emergence at higher latitudes is prevented artificially. Nonetheless, these models hold the promise of self-consistently imbibing elements of the surface flux transport dynamics leading to polar field reversal and build-up as well as solar internal magnetic field evolution, which 2D models cannot.

Finally, we end by posing the following provocative questions. From a purely utilitarian view, are dynamo models at all necessary for solar cycle predictions and can they provide higher accuracy or advantage compared to empirical polar field precursor based techniques? There is no doubt that physical approaches based on surface flux transport models and dynamo models have significantly advanced our understanding of solar cycle predictability; however, the resolution of the above questions are fundamental to sustained growth prospects of this field of research. 

\backmatter

\bmhead{Acknowledgments}

D.N. acknowledges financial support for the Center of Excellence in Space Sciences India from the Ministry of Education, Government of India. P.B. acknowledges support from the project ST/W00108X/1 funded by the Science and Technology Facilities Council (STFC), United Kingdom. J.J. was supported by the National Natural Science Foundation of China grant Nos. 12173005 and 11873023. L.U. was supported by NASA Heliophysics Living With a Star grants NNH16ZDA010N-LWS and NNH18ZDA001N-LWS and by NASA grant NNH18ZDA001N-DRIVE to the COFFIES DRIVE Center managed by Stanford University. A.L. acknowledges support from the Fonds de Recherche du Qu\'ebec – Nature et Technologie (Programme de recherche coll\'egiale). All authors acknowledge support from the International Space Science Institute (Bern) for facilitating enriching interactions which led to the planning of this review.

\section*{Declarations}

D.N. coordinated and planned the writing of this review in consultation with all authors. P.B. managed and led the manuscript writing process to which all authors contributed. The authors declare no conflicts of interest.

\bibliography{main}


\begin{thebibliography}{176}
\ifx \bisbn   \undefined \def \bisbn  #1{ISBN #1}\fi
\ifx \binits  \undefined \def \binits#1{#1}\fi
\ifx \bauthor  \undefined \def \bauthor#1{#1}\fi
\ifx \batitle  \undefined \def \batitle#1{#1}\fi
\ifx \bjtitle  \undefined \def \bjtitle#1{#1}\fi
\ifx \bvolume  \undefined \def \bvolume#1{\textbf{#1}}\fi
\ifx \byear  \undefined \def \byear#1{#1}\fi
\ifx \bissue  \undefined \def \bissue#1{#1}\fi
\ifx \bfpage  \undefined \def \bfpage#1{#1}\fi
\ifx \blpage  \undefined \def \blpage #1{#1}\fi
\ifx \burl  \undefined \def \burl#1{\textsf{#1}}\fi
\ifx \doiurl  \undefined \def \doiurl#1{\url{https://doi.org/#1}}\fi
\ifx \betal  \undefined \def \betal{\textit{et al.}}\fi
\ifx \binstitute  \undefined \def \binstitute#1{#1}\fi
\ifx \binstitutionaled  \undefined \def \binstitutionaled#1{#1}\fi
\ifx \bctitle  \undefined \def \bctitle#1{#1}\fi
\ifx \beditor  \undefined \def \beditor#1{#1}\fi
\ifx \bpublisher  \undefined \def \bpublisher#1{#1}\fi
\ifx \bbtitle  \undefined \def \bbtitle#1{#1}\fi
\ifx \bedition  \undefined \def \bedition#1{#1}\fi
\ifx \bseriesno  \undefined \def \bseriesno#1{#1}\fi
\ifx \blocation  \undefined \def \blocation#1{#1}\fi
\ifx \bsertitle  \undefined \def \bsertitle#1{#1}\fi
\ifx \bsnm \undefined \def \bsnm#1{#1}\fi
\ifx \bsuffix \undefined \def \bsuffix#1{#1}\fi
\ifx \bparticle \undefined \def \bparticle#1{#1}\fi
\ifx \barticle \undefined \def \barticle#1{#1}\fi
\bibcommenthead
\ifx \bconfdate \undefined \def \bconfdate #1{#1}\fi
\ifx \botherref \undefined \def \botherref #1{#1}\fi
\ifx \url \undefined \def \url#1{\textsf{#1}}\fi
\ifx \bchapter \undefined \def \bchapter#1{#1}\fi
\ifx \bbook \undefined \def \bbook#1{#1}\fi
\ifx \bcomment \undefined \def \bcomment#1{#1}\fi
\ifx \oauthor \undefined \def \oauthor#1{#1}\fi
\ifx \citeauthoryear \undefined \def \citeauthoryear#1{#1}\fi
\ifx \endbibitem  \undefined \def \endbibitem {}\fi
\ifx \bconflocation  \undefined \def \bconflocation#1{#1}\fi
\ifx \arxivurl  \undefined \def \arxivurl#1{\textsf{#1}}\fi
\csname PreBibitemsHook\endcsname

\bibitem{Schrijver2015AdSpR}
\begin{barticle}
\bauthor{\bsnm{{Schrijver}}, \binits{C.J.}},
\bauthor{\bsnm{{Kauristie}}, \binits{K.}},
\bauthor{\bsnm{{Aylward}}, \binits{A.D.}},
\bauthor{\bsnm{{Denardini}}, \binits{C.M.}},
\bauthor{\bsnm{{Gibson}}, \binits{S.E.}},
\bauthor{\bsnm{{Glover}}, \binits{A.}},
\bauthor{\bsnm{{Gopalswamy}}, \binits{N.}},
\bauthor{\bsnm{{Grande}}, \binits{M.}},
\bauthor{\bsnm{{Hapgood}}, \binits{M.}},
\bauthor{\bsnm{{Heynderickx}}, \binits{D.}},
\bauthor{\bsnm{{Jakowski}}, \binits{N.}},
\bauthor{\bsnm{{Kalegaev}}, \binits{V.V.}},
\bauthor{\bsnm{{Lapenta}}, \binits{G.}},
\bauthor{\bsnm{{Linker}}, \binits{J.A.}},
\bauthor{\bsnm{{Liu}}, \binits{S.}},
\bauthor{\bsnm{{Mandrini}}, \binits{C.H.}},
\bauthor{\bsnm{{Mann}}, \binits{I.R.}},
\bauthor{\bsnm{{Nagatsuma}}, \binits{T.}},
\bauthor{\bsnm{{Nandy}}, \binits{D.}},
\bauthor{\bsnm{{Obara}}, \binits{T.}},
\bauthor{\bsnm{{Paul O'Brien}}, \binits{T.}},
\bauthor{\bsnm{{Onsager}}, \binits{T.}},
\bauthor{\bsnm{{Opgenoorth}}, \binits{H.J.}},
\bauthor{\bsnm{{Terkildsen}}, \binits{M.}},
\bauthor{\bsnm{{Valladares}}, \binits{C.E.}},
\bauthor{\bsnm{{Vilmer}}, \binits{N.}}:
\batitle{{Understanding space weather to shield society: A global road map for
  2015-2025 commissioned by COSPAR and ILWS}}.
\bjtitle{Advances in Space Research}
\bvolume{55}(\bissue{12}),
\bfpage{2745}--\blpage{2807}
(\byear{2015})
{\href{https://arxiv.org/abs/1503.06135}{{arXiv:1503.06135}}}
{[physics.space-ph]}.
\doiurl{10.1016/j.asr.2015.03.023}
\end{barticle}
\endbibitem

\bibitem{NAP13060}
\begin{bbook}
\bauthor{\bsnm{Council}, \binits{N.R.}}:
\bbtitle{Solar and Space Physics: A Science for a Technological Society}.
\bpublisher{The National Academies Press},
\blocation{Washington, DC}
(\byear{2013}).
\doiurl{10.17226/13060}.
\burl{https://nap.nationalacademies.org/catalog/13060/solar-and-space-physics-a-science-for-a-technological-society}
\end{bbook}
\endbibitem

\bibitem{NSWSAP2019}
\begin{botherref}
National Space Weather Strategy and Action Plan.
The National Academies Press,
Washington, DC
(2019).
\url{https://trumpwhitehouse.archives.gov/wp-content/uploads/2019/03/National-Space-Weather-Strategy-and-Action-Plan-2019.pdf}
\end{botherref}
\endbibitem

\bibitem{NSWSAP2022}
\begin{botherref}
Space Weather Research-to-Operations and Operations-to-Research Framework.
The National Academies Press,
Washington, DC
(2022).
\url{https://www.whitehouse.gov/wp-content/uploads/2022/03/03-2022-Space-Weather-R2O2R-Framework.pdf}
\end{botherref}
\endbibitem

\bibitem{UNOOSA2017}
\begin{botherref}
UNOOSA Space Weather: 2017, Special Report of the Inter-agency Meeting on Outer
  Space Activities on Developments Within the United Nations System Related to
  Space Weather.
United Nations,
New York, US
(2017).
\url{http://www.unoosa.org/oosa/oosadoc/data/documents/2017/aac.105/aac.1051146
  0.html}
\end{botherref}
\endbibitem

\bibitem{Charbonneau2020LRSP}
\begin{barticle}
\bauthor{\bsnm{{Charbonneau}}, \binits{P.}}:
\batitle{{Dynamo models of the solar cycle}}.
\bjtitle{Living Reviews in Solar Physics}
\bvolume{17}(\bissue{1}),
\bfpage{4}
(\byear{2020}).
\doiurl{10.1007/s41116-020-00025-6}
\end{barticle}
\endbibitem

\bibitem{Parker1955ApJ}
\begin{barticle}
\bauthor{\bsnm{{Parker}}, \binits{E.N.}}:
\batitle{{The Formation of Sunspots from the Solar Toroidal Field.}}
\bjtitle{The Astrophysical Journal}
\bvolume{121},
\bfpage{491}
(\byear{1955}).
\doiurl{10.1086/146010}
\end{barticle}
\endbibitem

\bibitem{Babcock1961ApJ}
\begin{barticle}
\bauthor{\bsnm{{Babcock}}, \binits{H.W.}}:
\batitle{{The Topology of the Sun's Magnetic Field and the 22-YEAR Cycle.}}
\bjtitle{The Astrophysical Journal}
\bvolume{133},
\bfpage{572}
(\byear{1961}).
\doiurl{10.1086/147060}
\end{barticle}
\endbibitem

\bibitem{Leighton1969ApJ}
\begin{barticle}
\bauthor{\bsnm{{Leighton}}, \binits{R.B.}}:
\batitle{{A Magneto-Kinematic Model of the Solar Cycle}}.
\bjtitle{The Astrophysical Journal}
\bvolume{156},
\bfpage{1}
(\byear{1969}).
\doiurl{10.1086/149943}
\end{barticle}
\endbibitem

\bibitem{Cameron2015Sci}
\begin{barticle}
\bauthor{\bsnm{{Cameron}}, \binits{R.}},
\bauthor{\bsnm{{Sch{\"u}ssler}}, \binits{M.}}:
\batitle{{The crucial role of surface magnetic fields for the solar dynamo}}.
\bjtitle{Science}
\bvolume{347}(\bissue{6228}),
\bfpage{1333}--\blpage{1335}
(\byear{2015})
{\href{https://arxiv.org/abs/1503.08469}{{arXiv:1503.08469}}}
{[astro-ph.SR]}.
\doiurl{10.1126/science.1261470}
\end{barticle}
\endbibitem

\bibitem{Bhowmik2018NatCo}
\begin{barticle}
\bauthor{\bsnm{{Bhowmik}}, \binits{P.}},
\bauthor{\bsnm{{Nandy}}, \binits{D.}}:
\batitle{{Prediction of the strength and timing of sunspot cycle 25 reveal
  decadal-scale space environmental conditions}}.
\bjtitle{Nature Communications}
\bvolume{9},
\bfpage{5209}
(\byear{2018})
{\href{https://arxiv.org/abs/1909.04537}{{arXiv:1909.04537}}}
{[astro-ph.SR]}.
\doiurl{10.1038/s41467-018-07690-0}
\end{barticle}
\endbibitem

\bibitem{Petrovay2020LRSP}
\begin{barticle}
\bauthor{\bsnm{{Petrovay}}, \binits{K.}}:
\batitle{{Solar cycle prediction}}.
\bjtitle{Living Reviews in Solar Physics}
\bvolume{17}(\bissue{1}),
\bfpage{2}
(\byear{2020})
{\href{https://arxiv.org/abs/1907.02107}{{arXiv:1907.02107}}}
{[astro-ph.SR]}.
\doiurl{10.1007/s41116-020-0022-z}
\end{barticle}
\endbibitem

\bibitem{Nandy2021SoPh}
\begin{barticle}
\bauthor{\bsnm{{Nandy}}, \binits{D.}}:
\batitle{{Progress in Solar Cycle Predictions: Sunspot Cycles 24-25 in
  Perspective}}.
\bjtitle{Solar Physics}
\bvolume{296}(\bissue{3}),
\bfpage{54}
(\byear{2021})
{\href{https://arxiv.org/abs/2009.01908}{{arXiv:2009.01908}}}
{[astro-ph.SR]}.
\doiurl{10.1007/s11207-021-01797-2}
\end{barticle}
\endbibitem

\bibitem{Jiang2023JASTP}
\begin{barticle}
\bauthor{\bsnm{{Jiang}}, \binits{J.}},
\bauthor{\bsnm{{Zhang}}, \binits{Z.}},
\bauthor{\bsnm{{Petrovay}}, \binits{K.}}:
\batitle{{Comparison of physics-based prediction models of solar cycle 25}}.
\bjtitle{Journal of Atmospheric and Solar-Terrestrial Physics}
\bvolume{243},
\bfpage{106018}
(\byear{2023})
{\href{https://arxiv.org/abs/2212.01158}{{arXiv:2212.01158}}}
{[astro-ph.SR]}.
\doiurl{10.1016/j.jastp.2023.106018}
\end{barticle}
\endbibitem

\bibitem{Chapter2}
\begin{botherref}
\oauthor{\bsnm{{Norton}}, \binits{A.}},
\oauthor{\bsnm{{Howe}}, \binits{R.}},
\oauthor{\bsnm{{Upton}}, \binits{L.}},
\oauthor{\bsnm{{Usoskin}}, \binits{I.}}:
{Solar Cycle Observations}.
Submitted to Space Science Reviews
(2023)
\end{botherref}
\endbibitem

\bibitem{1844Schwabe}
\begin{barticle}
\bauthor{\bsnm{{Schwabe}}, \binits{H.}}:
\batitle{{Sonnenbeobachtungen im Jahre 1843. Von Herrn Hofrath Schwabe in
  Dessau}}.
\bjtitle{Astronomische Nachrichten}
\bvolume{21}(\bissue{15}),
\bfpage{233}
(\byear{1844}).
\doiurl{10.1002/asna.18440211505}
\end{barticle}
\endbibitem

\bibitem{2015Clette_etal}
\begin{bchapter}
\bauthor{\bsnm{{Clette}}, \binits{F.}},
\bauthor{\bsnm{{Svalgaard}}, \binits{L.}},
\bauthor{\bsnm{{Vaquero}}, \binits{J.M.}},
\bauthor{\bsnm{{Cliver}}, \binits{E.W.}}:
\bctitle{{Revisiting the Sunspot Number}}.
In: \beditor{\bsnm{{Balogh}}, \binits{A.}},
\beditor{\bsnm{{Hudson}}, \binits{H.}},
\beditor{\bsnm{{Petrovay}}, \binits{K.}},
\beditor{\bsnm{{von Steiger}}, \binits{R.}} (eds.)
\bbtitle{The Solar Activity Cycle}
vol. \bseriesno{53},
p. \bfpage{35}
(\byear{2015}).
\doiurl{10.1007/978-1-4939-2584-1_3}
\end{bchapter}
\endbibitem

\bibitem{Eddy1976Sci}
\begin{barticle}
\bauthor{\bsnm{{Eddy}}, \binits{J.A.}}:
\batitle{{The Maunder Minimum}}.
\bjtitle{Science}
\bvolume{192},
\bfpage{1189}--\blpage{1202}
(\byear{1976}).
\doiurl{10.1126/science.192.4245.1189}
\end{barticle}
\endbibitem

\bibitem{1908Hale_a}
\begin{barticle}
\bauthor{\bsnm{{Hale}}, \binits{G.E.}}:
\batitle{{The Zeeman Effect in the Sun}}.
\bjtitle{The Astronomical Society of the Pacific}
\bvolume{20}(\bissue{123}),
\bfpage{287}
(\byear{1908}).
\doiurl{10.1086/121847}
\end{barticle}
\endbibitem

\bibitem{1908Hale_b}
\begin{barticle}
\bauthor{\bsnm{{Hale}}, \binits{G.E.}}:
\batitle{{On the Probable Existence of a Magnetic Field in Sun-Spots}}.
\bjtitle{The Astrophysical Journal}
\bvolume{28},
\bfpage{315}
(\byear{1908}).
\doiurl{10.1086/141602}
\end{barticle}
\endbibitem

\bibitem{1858Carrington}
\begin{barticle}
\bauthor{\bsnm{{Carrington}}, \binits{R.C.}}:
\batitle{{On the Distribution of the Solar Spots in Latitudes since the
  Beginning of the Year 1854, with a Map}}.
\bjtitle{Monthly Notices of the Royal Astronomical Society}
\bvolume{19},
\bfpage{1}--\blpage{3}
(\byear{1858}).
\doiurl{10.1093/mnras/19.1.1}
\end{barticle}
\endbibitem

\bibitem{1919Hale_etal}
\begin{barticle}
\bauthor{\bsnm{{Hale}}, \binits{G.E.}},
\bauthor{\bsnm{{Ellerman}}, \binits{F.}},
\bauthor{\bsnm{{Nicholson}}, \binits{S.B.}},
\bauthor{\bsnm{{Joy}}, \binits{A.H.}}:
\batitle{{The Magnetic Polarity of Sun-Spots}}.
\bjtitle{The Astrophysical Journal}
\bvolume{49},
\bfpage{153}
(\byear{1919}).
\doiurl{10.1086/142452}
\end{barticle}
\endbibitem

\bibitem{1958BabcockLivingston}
\begin{barticle}
\bauthor{\bsnm{{Babcock}}, \binits{H.D.}},
\bauthor{\bsnm{{Livingston}}, \binits{W.C.}}:
\batitle{{Changes in the Sun's Polar Magnetic Field}}.
\bjtitle{Science}
\bvolume{127},
\bfpage{1058}
(\byear{1958})
\end{barticle}
\endbibitem

\bibitem{Nandy2023arXiv}
\begin{botherref}
\oauthor{\bsnm{{Nandy}}, \binits{D.}},
\oauthor{\bsnm{{Banerjee}}, \binits{D.}},
\oauthor{\bsnm{{Bhowmik}}, \binits{P.}},
\oauthor{\bsnm{{Brun}}, \binits{A.S.}},
\oauthor{\bsnm{{Cameron}}, \binits{R.H.}},
\oauthor{\bsnm{{Gibson}}, \binits{S.E.}},
\oauthor{\bsnm{{Hanasoge}}, \binits{S.}},
\oauthor{\bsnm{{Harra}}, \binits{L.}},
\oauthor{\bsnm{{Hassler}}, \binits{D.M.}},
\oauthor{\bsnm{{Jain}}, \binits{R.}},
\oauthor{\bsnm{{Jiang}}, \binits{J.}},
\oauthor{\bsnm{{Jouve}}, \binits{L.}},
\oauthor{\bsnm{{Mackay}}, \binits{D.H.}},
\oauthor{\bsnm{{Mahajan}}, \binits{S.S.}},
\oauthor{\bsnm{{Mandrini}}, \binits{C.H.}},
\oauthor{\bsnm{{Owens}}, \binits{M.}},
\oauthor{\bsnm{{Pal}}, \binits{S.}},
\oauthor{\bsnm{{Pinto}}, \binits{R.F.}},
\oauthor{\bsnm{{Saha}}, \binits{C.}},
\oauthor{\bsnm{{Sun}}, \binits{X.}},
\oauthor{\bsnm{{Tripathi}}, \binits{D.}},
\oauthor{\bsnm{{Usoskin}}, \binits{I.G.}}:
{Exploring the Solar Poles: The Last Great Frontier of the Sun}.
arXiv e-prints,
2301--00010
(2022)
{\href{https://arxiv.org/abs/2301.00010}{{arXiv:2301.00010}}}
{[astro-ph.IM]}.
\doiurl{10.48550/arXiv.2301.00010}
\end{botherref}
\endbibitem

\bibitem{Nagy2017}
\begin{barticle}
\bauthor{\bsnm{{Nagy}}, \binits{M.}},
\bauthor{\bsnm{{Lemerle}}, \binits{A.}},
\bauthor{\bsnm{{Labonville}}, \binits{F.}},
\bauthor{\bsnm{{Petrovay}}, \binits{K.}},
\bauthor{\bsnm{{Charbonneau}}, \binits{P.}}:
\batitle{{The Effect of ``Rogue'' Active Regions on the Solar Cycle}}.
\bjtitle{Solar Physics}
\bvolume{292}(\bissue{11}),
\bfpage{167}
(\byear{2017})
{\href{https://arxiv.org/abs/1712.02185}{{arXiv:1712.02185}}}
{[astro-ph.SR]}.
\doiurl{10.1007/s11207-017-1194-0}
\end{barticle}
\endbibitem

\bibitem{2015Hathaway_etal}
\begin{barticle}
\bauthor{\bsnm{{Hathaway}}, \binits{D.H.}},
\bauthor{\bsnm{{Teil}}, \binits{T.}},
\bauthor{\bsnm{{Norton}}, \binits{A.A.}},
\bauthor{\bsnm{{Kitiashvili}}, \binits{I.}}:
\batitle{{The Sun's Photospheric Convection Spectrum}}.
\bjtitle{The Astrophysical Journal}
\bvolume{811}(\bissue{2}),
\bfpage{105}
(\byear{2015})
{\href{https://arxiv.org/abs/1508.03022}{{arXiv:1508.03022}}}
{[astro-ph.SR]}.
\doiurl{10.1088/0004-637X/811/2/105}
\end{barticle}
\endbibitem

\bibitem{Adams1911}
\begin{barticle}
\bauthor{\bsnm{{Adams}}, \binits{W.S.}}:
\batitle{An investigation of the rotation period of the sun by spectroscopic
  methods}.
\bjtitle{Publication of Carnegie Institution of Washington}
\bvolume{138},
\bfpage{1}--\blpage{132}
(\byear{1911})
\end{barticle}
\endbibitem

\bibitem{Belopolsky1933}
\begin{barticle}
\bauthor{\bsnm{{Belopolsky}}, \binits{A.}}:
\batitle{{Bestimmung der Sonnenrotation auf spektroskopischem Wege in den
  Jahren 1931, 1932 und 1933 in Pulkovo. Mit 3 Abbildungen.}}
\bjtitle{Zeitschrift fur Astrophysik}
\bvolume{7},
\bfpage{357}
(\byear{1933})
\end{barticle}
\endbibitem

\bibitem{Howard1984ARA&A}
\begin{barticle}
\bauthor{\bsnm{{Howard}}, \binits{R.}}:
\batitle{{Solar Rotation}}.
\bjtitle{Ann. Rev. Astron. Astrophys.}
\bvolume{22},
\bfpage{131}--\blpage{155}
(\byear{1984}).
\doiurl{10.1146/annurev.aa.22.090184.001023}
\end{barticle}
\endbibitem

\bibitem{Schou1998ApJ}
\begin{barticle}
\bauthor{\bsnm{{Schou}}, \binits{J.}},
\bauthor{\bsnm{{Antia}}, \binits{H.M.}},
\bauthor{\bsnm{{Basu}}, \binits{S.}},
\bauthor{\bsnm{{Bogart}}, \binits{R.S.}},
\bauthor{\bsnm{{Bush}}, \binits{R.I.}},
\bauthor{\bsnm{{Chitre}}, \binits{S.M.}},
\bauthor{\bsnm{{Christensen-Dalsgaard}}, \binits{J.}},
\bauthor{\bsnm{{Di Mauro}}, \binits{M.P.}},
\bauthor{\bsnm{{Dziembowski}}, \binits{W.A.}},
\bauthor{\bsnm{{Eff-Darwich}}, \binits{A.}},
\bauthor{\bsnm{{Gough}}, \binits{D.O.}},
\bauthor{\bsnm{{Haber}}, \binits{D.A.}},
\bauthor{\bsnm{{Hoeksema}}, \binits{J.T.}},
\bauthor{\bsnm{{Howe}}, \binits{R.}},
\bauthor{\bsnm{{Korzennik}}, \binits{S.G.}},
\bauthor{\bsnm{{Kosovichev}}, \binits{A.G.}},
\bauthor{\bsnm{{Larsen}}, \binits{R.M.}},
\bauthor{\bsnm{{Pijpers}}, \binits{F.P.}},
\bauthor{\bsnm{{Scherrer}}, \binits{P.H.}},
\bauthor{\bsnm{{Sekii}}, \binits{T.}},
\bauthor{\bsnm{{Tarbell}}, \binits{T.D.}},
\bauthor{\bsnm{{Title}}, \binits{A.M.}},
\bauthor{\bsnm{{Thompson}}, \binits{M.J.}},
\bauthor{\bsnm{{Toomre}}, \binits{J.}}:
\batitle{{Helioseismic Studies of Differential Rotation in the Solar Envelope
  by the Solar Oscillations Investigation Using the Michelson Doppler Imager}}.
\bjtitle{The Astrophysical Journal}
\bvolume{505}(\bissue{1}),
\bfpage{390}--\blpage{417}
(\byear{1998}).
\doiurl{10.1086/306146}
\end{barticle}
\endbibitem

\bibitem{Basu2016LRSP}
\begin{barticle}
\bauthor{\bsnm{{Basu}}, \binits{S.}}:
\batitle{{Global seismology of the Sun}}.
\bjtitle{Living Reviews in Solar Physics}
\bvolume{13}(\bissue{1}),
\bfpage{2}
(\byear{2016})
{\href{https://arxiv.org/abs/1606.07071}{{arXiv:1606.07071}}}
{[astro-ph.SR]}.
\doiurl{10.1007/s41116-016-0003-4}
\end{barticle}
\endbibitem

\bibitem{Hanasoge2022LRSP}
\begin{barticle}
\bauthor{\bsnm{{Hanasoge}}, \binits{S.M.}}:
\batitle{{Surface and interior meridional circulation in the Sun}}.
\bjtitle{Living Reviews in Solar Physics}
\bvolume{19}(\bissue{1}),
\bfpage{3}
(\byear{2022}).
\doiurl{10.1007/s41116-022-00034-7}
\end{barticle}
\endbibitem

\bibitem{Rajaguru2015ApJ}
\begin{barticle}
\bauthor{\bsnm{{Rajaguru}}, \binits{S.P.}},
\bauthor{\bsnm{{Antia}}, \binits{H.M.}}:
\batitle{{Meridional Circulation in the Solar Convection Zone: Time-Distance
  Helioseismic Inferences from Four Years of HMI/SDO Observations}}.
\bjtitle{The Astrophysical Journal}
\bvolume{813}(\bissue{2}),
\bfpage{114}
(\byear{2015})
{\href{https://arxiv.org/abs/1510.01843}{{arXiv:1510.01843}}}
{[astro-ph.SR]}.
\doiurl{10.1088/0004-637X/813/2/114}
\end{barticle}
\endbibitem

\bibitem{Liang2018A&A}
\begin{barticle}
\bauthor{\bsnm{{Liang}}, \binits{Z.-C.}},
\bauthor{\bsnm{{Gizon}}, \binits{L.}},
\bauthor{\bsnm{{Birch}}, \binits{A.C.}},
\bauthor{\bsnm{{Duvall}}, \binits{T.L.}},
\bauthor{\bsnm{{Rajaguru}}, \binits{S.P.}}:
\batitle{{Solar meridional circulation from twenty-one years of SOHO/MDI and
  SDO/HMI observations. Helioseismic travel times and forward modeling in the
  ray approximation}}.
\bjtitle{Astronomy \& Astrophysics}
\bvolume{619},
\bfpage{99}
(\byear{2018})
{\href{https://arxiv.org/abs/1808.08874}{{arXiv:1808.08874}}}
{[astro-ph.SR]}.
\doiurl{10.1051/0004-6361/201833673}
\end{barticle}
\endbibitem

\bibitem{2012Hathaway}
\begin{barticle}
\bauthor{\bsnm{{Hathaway}}, \binits{D.H.}}:
\batitle{{Supergranules as Probes of the Sun's Meridional Circulation}}.
\bjtitle{The Astrophysical Journal}
\bvolume{760}(\bissue{1}),
\bfpage{84}
(\byear{2012})
{\href{https://arxiv.org/abs/1210.3343}{{arXiv:1210.3343}}}
{[astro-ph.SR]}.
\doiurl{10.1088/0004-637X/760/1/84}
\end{barticle}
\endbibitem

\bibitem{2013Zhao_etal}
\begin{barticle}
\bauthor{\bsnm{{Zhao}}, \binits{J.}},
\bauthor{\bsnm{{Bogart}}, \binits{R.S.}},
\bauthor{\bsnm{{Kosovichev}}, \binits{A.G.}},
\bauthor{\bsnm{{Duvall}}, \binits{J.} \bsuffix{T.~L.}},
\bauthor{\bsnm{{Hartlep}}, \binits{T.}}:
\batitle{{Detection of Equatorward Meridional Flow and Evidence of Double-cell
  Meridional Circulation inside the Sun}}.
\bjtitle{The Astrophysical Journal Letter}
\bvolume{774}(\bissue{2}),
\bfpage{29}
(\byear{2013})
{\href{https://arxiv.org/abs/1307.8422}{{arXiv:1307.8422}}}
{[astro-ph.SR]}.
\doiurl{10.1088/2041-8205/774/2/L29}
\end{barticle}
\endbibitem

\bibitem{2022Hathaway_etal}
\begin{barticle}
\bauthor{\bsnm{{Hathaway}}, \binits{D.H.}},
\bauthor{\bsnm{{Upton}}, \binits{L.A.}},
\bauthor{\bsnm{{Mahajan}}, \binits{S.S.}}:
\batitle{{Variations in differential rotation and meridional flow within the
  Sun's surface shear layer 1996{\textendash}2022}}.
\bjtitle{Frontiers in Astronomy and Space Sciences}
\bvolume{9},
\bfpage{1007290}
(\byear{2022})
{\href{https://arxiv.org/abs/2212.10619}{{arXiv:2212.10619}}}
{[astro-ph.SR]}.
\doiurl{10.3389/fspas.2022.1007290}
\end{barticle}
\endbibitem

\bibitem{Nandy2002Sci}
\begin{barticle}
\bauthor{\bsnm{{Nandy}}, \binits{D.}},
\bauthor{\bsnm{{Choudhuri}}, \binits{A.R.}}:
\batitle{{Explaining the Latitudinal Distribution of Sunspots with Deep
  Meridional Flow}}.
\bjtitle{Science}
\bvolume{296}(\bissue{5573}),
\bfpage{1671}--\blpage{1673}
(\byear{2002}).
\doiurl{10.1126/science.1070955}
\end{barticle}
\endbibitem

\bibitem{Zhao2004torsional}
\begin{barticle}
\bauthor{\bsnm{Zhao}, \binits{J.}},
\bauthor{\bsnm{Kosovichev}, \binits{A.G.}}:
\batitle{Torsional oscillation, meridional flows, and vorticity inferred in the
  upper convection zone of the sun by time-distance helioseismology}.
\bjtitle{The Astrophysical Journal}
\bvolume{603}(\bissue{2}),
\bfpage{776}
(\byear{2004})
\end{barticle}
\endbibitem

\bibitem{Howe2009LRSP}
\begin{barticle}
\bauthor{\bsnm{{Howe}}, \binits{R.}}:
\batitle{{Solar Interior Rotation and its Variation}}.
\bjtitle{Living Reviews in Solar Physics}
\bvolume{6}(\bissue{1}),
\bfpage{1}
(\byear{2009})
{\href{https://arxiv.org/abs/0902.2406}{{arXiv:0902.2406}}}
{[astro-ph.SR]}.
\doiurl{10.12942/lrsp-2009-1}
\end{barticle}
\endbibitem

\bibitem{Hathaway2010Sci}
\begin{barticle}
\bauthor{\bsnm{{Hathaway}}, \binits{D.H.}},
\bauthor{\bsnm{{Rightmire}}, \binits{L.}}:
\batitle{{Variations in the Sun{\textquoteright}s Meridional Flow over a Solar
  Cycle}}.
\bjtitle{Science}
\bvolume{327}(\bissue{5971}),
\bfpage{1350}
(\byear{2010}).
\doiurl{10.1126/science.1181990}
\end{barticle}
\endbibitem

\bibitem{Hathaway2014JGRAH}
\begin{barticle}
\bauthor{\bsnm{{Hathaway}}, \binits{D.H.}},
\bauthor{\bsnm{{Upton}}, \binits{L.}}:
\batitle{{The solar meridional circulation and sunspot cycle variability}}.
\bjtitle{Journal of Geophysical Research (Space Physics)}
\bvolume{119}(\bissue{5}),
\bfpage{3316}--\blpage{3324}
(\byear{2014})
{\href{https://arxiv.org/abs/1404.5893}{{arXiv:1404.5893}}}
{[astro-ph.SR]}.
\doiurl{10.1002/2013JA019432}
\end{barticle}
\endbibitem

\bibitem{Chapter10}
\begin{botherref}
\oauthor{\bsnm{{Hotta}}, \binits{H.}},
\oauthor{\bsnm{{Bekki}}, \binits{Y.}},
\oauthor{\bsnm{{Gizon}}, \binits{L.}},
\oauthor{\bsnm{{Noraz}}, \binits{Q.}},
\oauthor{\bsnm{{Rast}}, \binits{M.P.}}:
Large-scale flow dynamics.
Submitted to Space Science Reviews
(2023)
\end{botherref}
\endbibitem

\bibitem{Munoz2012ApJ}
\begin{barticle}
\bauthor{\bsnm{{Mu{\~n}oz-Jaramillo}}, \binits{A.}},
\bauthor{\bsnm{{Sheeley}}, \binits{N.R.}},
\bauthor{\bsnm{{Zhang}}, \binits{J.}},
\bauthor{\bsnm{{DeLuca}}, \binits{E.E.}}:
\batitle{{Calibrating 100 Years of Polar Faculae Measurements: Implications for
  the Evolution of the Heliospheric Magnetic Field}}.
\bjtitle{The Astrophysical Journal}
\bvolume{753}(\bissue{2}),
\bfpage{146}
(\byear{2012})
{\href{https://arxiv.org/abs/1303.0345}{{arXiv:1303.0345}}}
{[astro-ph.SR]}.
\doiurl{10.1088/0004-637X/753/2/146}
\end{barticle}
\endbibitem

\bibitem{Yeates2008}
\begin{barticle}
\bauthor{\bsnm{{Yeates}}, \binits{A.R.}},
\bauthor{\bsnm{{Nandy}}, \binits{D.}},
\bauthor{\bsnm{{Mackay}}, \binits{D.H.}}:
\batitle{{Exploring the Physical Basis of Solar Cycle Predictions: Flux
  Transport Dynamics and Persistence of Memory in Advection- versus
  Diffusion-dominated Solar Convection Zones}}.
\bjtitle{The Astrophysical Journal}
\bvolume{673}(\bissue{1}),
\bfpage{544}--\blpage{556}
(\byear{2008})
{\href{https://arxiv.org/abs/0709.1046}{{arXiv:0709.1046}}}
{[astro-ph]}.
\doiurl{10.1086/524352}
\end{barticle}
\endbibitem

\bibitem{Karak2012ApJ}
\begin{barticle}
\bauthor{\bsnm{{Karak}}, \binits{B.B.}},
\bauthor{\bsnm{{Nandy}}, \binits{D.}}:
\batitle{{Turbulent Pumping of Magnetic Flux Reduces Solar Cycle Memory and
  thus Impacts Predictability of the Sun's Activity}}.
\bjtitle{The Astrophysical Journal Letter}
\bvolume{761}(\bissue{1}),
\bfpage{13}
(\byear{2012})
{\href{https://arxiv.org/abs/1206.2106}{{arXiv:1206.2106}}}
{[astro-ph.SR]}.
\doiurl{10.1088/2041-8205/761/1/L13}
\end{barticle}
\endbibitem

\bibitem{Chapter15}
\begin{botherref}
\oauthor{\bsnm{{K{\"a}pyl{\"a}}}, \binits{P.J.}},
\oauthor{\bsnm{{Browning}}, \binits{M.K.M.}},
\oauthor{\bsnm{{Brun}}, \binits{A.S.}},
\oauthor{\bsnm{{Guerrero}}, \binits{G.}},
\oauthor{},
\oauthor{\bsnm{{Masada}}, \binits{Y.}},
\oauthor{\bsnm{{Warnecke}}, \binits{J.}}:
{Simulations of Solar and Stellar Dynamos and their Theoretical
  Interpretation}.
Submitted to Space Science Reviews
(2023)
\end{botherref}
\endbibitem

\bibitem{Leighton1964}
\begin{barticle}
\bauthor{\bsnm{{Leighton}}, \binits{R.B.}}:
\batitle{{Transport of Magnetic Fields on the Sun.}}
\bjtitle{The Astrophysical Journal}
\bvolume{140},
\bfpage{1547}
(\byear{1964}).
\doiurl{10.1086/148058}
\end{barticle}
\endbibitem

\bibitem{Chapter7}
\begin{botherref}
\oauthor{\bsnm{{Yeates}}, \binits{A.R.}},
\oauthor{\bsnm{{Cheung}}, \binits{M.C.M.}},
\oauthor{\bsnm{{Jiang}}, \binits{J.}},
\oauthor{\bsnm{{Petrovay}}, \binits{K.}},
\oauthor{\bsnm{{Wang}}, \binits{Y.M.}}:
{Surface Flux Transport}.
Submitted to Space Science Reviews
(2023)
\end{botherref}
\endbibitem

\bibitem{2014Jiang_etal}
\begin{barticle}
\bauthor{\bsnm{{Jiang}}, \binits{J.}},
\bauthor{\bsnm{{Hathaway}}, \binits{D.H.}},
\bauthor{\bsnm{{Cameron}}, \binits{R.H.}},
\bauthor{\bsnm{{Solanki}}, \binits{S.K.}},
\bauthor{\bsnm{{Gizon}}, \binits{L.}},
\bauthor{\bsnm{{Upton}}, \binits{L.}}:
\batitle{{Magnetic Flux Transport at the Solar Surface}}.
\bjtitle{Space Science Review}
\bvolume{186}(\bissue{1-4}),
\bfpage{491}--\blpage{523}
(\byear{2014})
{\href{https://arxiv.org/abs/1408.3186}{{arXiv:1408.3186}}}
{[astro-ph.SR]}.
\doiurl{10.1007/s11214-014-0083-1}
\end{barticle}
\endbibitem

\bibitem{2023Yeates_etal}
\begin{botherref}
\oauthor{\bsnm{{Yeates}}, \binits{A.R.}},
\oauthor{\bsnm{{Cheung}}, \binits{M.C.M.}},
\oauthor{\bsnm{{Jiang}}, \binits{J.}},
\oauthor{\bsnm{{Petrovay}}, \binits{K.}},
\oauthor{\bsnm{{Wang}}, \binits{Y.-M.}}:
{Surface Flux Transport}.
arXiv e-prints,
2303--01209
(2023)
{\href{https://arxiv.org/abs/2303.01209}{{arXiv:2303.01209}}}
{[astro-ph.SR]}.
\doiurl{10.48550/arXiv.2303.01209}
\end{botherref}
\endbibitem

\bibitem{DeVore1984SoPh}
\begin{barticle}
\bauthor{\bsnm{{DeVore}}, \binits{C.R.}},
\bauthor{\bsnm{{Boris}}, \binits{J.P.}},
\bauthor{\bsnm{{Sheeley}}, \binits{J.} \bsuffix{N.~R.}}:
\batitle{{The concentration of the large-scale solar magnetic field by a
  meridional surface flow}}.
\bjtitle{Solar Physics}
\bvolume{92}(\bissue{1-2}),
\bfpage{1}--\blpage{14}
(\byear{1984}).
\doiurl{10.1007/BF00157230}
\end{barticle}
\endbibitem

\bibitem{Sheeley1985SoPh}
\begin{barticle}
\bauthor{\bsnm{{Sheeley}}, \binits{J.} \bsuffix{N.~R.}},
\bauthor{\bsnm{{DeVore}}, \binits{C.R.}},
\bauthor{\bsnm{{Boris}}, \binits{J.P.}}:
\batitle{{Simulations of the Mean Solar Magnetic Field during Sunspot
  CYCLE-21}}.
\bjtitle{Solar Physics}
\bvolume{98}(\bissue{2}),
\bfpage{219}--\blpage{239}
(\byear{1985}).
\doiurl{10.1007/BF00152457}
\end{barticle}
\endbibitem

\bibitem{Wang1989Sci}
\begin{barticle}
\bauthor{\bsnm{{Wang}}, \binits{Y.-M.}},
\bauthor{\bsnm{{Nash}}, \binits{A.G.}},
\bauthor{\bsnm{{Sheeley}}, \binits{J.} \bsuffix{N.~R.}}:
\batitle{{Magnetic Flux Transport on the Sun}}.
\bjtitle{Science}
\bvolume{245}(\bissue{4919}),
\bfpage{712}--\blpage{718}
(\byear{1989}).
\doiurl{10.1126/science.245.4919.712}
\end{barticle}
\endbibitem

\bibitem{2000WordenHarvey}
\begin{barticle}
\bauthor{\bsnm{{Worden}}, \binits{J.}},
\bauthor{\bsnm{{Harvey}}, \binits{J.}}:
\batitle{{An Evolving Synoptic Magnetic Flux map and Implications for the
  Distribution of Photospheric Magnetic Flux}}.
\bjtitle{Solar Physics}
\bvolume{195}(\bissue{2}),
\bfpage{247}--\blpage{268}
(\byear{2000}).
\doiurl{10.1023/A:1005272502885}
\end{barticle}
\endbibitem

\bibitem{2010Arge_etal}
\begin{bchapter}
\bauthor{\bsnm{{Arge}}, \binits{C.N.}},
\bauthor{\bsnm{{Henney}}, \binits{C.J.}},
\bauthor{\bsnm{{Koller}}, \binits{J.}},
\bauthor{\bsnm{{Compeau}}, \binits{C.R.}},
\bauthor{\bsnm{{Young}}, \binits{S.}},
\bauthor{\bsnm{{MacKenzie}}, \binits{D.}},
\bauthor{\bsnm{{Fay}}, \binits{A.}},
\bauthor{\bsnm{{Harvey}}, \binits{J.W.}}:
\bctitle{{Air Force Data Assimilative Photospheric Flux Transport (ADAPT)
  Model}}.
In: \beditor{\bsnm{{Maksimovic}}, \binits{M.}},
\beditor{\bsnm{{Issautier}}, \binits{K.}},
\beditor{\bsnm{{Meyer-Vernet}}, \binits{N.}},
\beditor{\bsnm{{Moncuquet}}, \binits{M.}},
\beditor{\bsnm{{Pantellini}}, \binits{F.}} (eds.)
\bbtitle{Twelfth International Solar Wind Conference}.
\bsertitle{American Institute of Physics Conference Series},
vol. \bseriesno{1216},
pp. \bfpage{343}--\blpage{346}
(\byear{2010}).
\doiurl{10.1063/1.3395870}
\end{bchapter}
\endbibitem

\bibitem{2015Hickmann_etal}
\begin{barticle}
\bauthor{\bsnm{{Hickmann}}, \binits{K.S.}},
\bauthor{\bsnm{{Godinez}}, \binits{H.C.}},
\bauthor{\bsnm{{Henney}}, \binits{C.J.}},
\bauthor{\bsnm{{Arge}}, \binits{C.N.}}:
\batitle{{Data Assimilation in the ADAPT Photospheric Flux Transport Model}}.
\bjtitle{Solar Physics}
\bvolume{290}(\bissue{4}),
\bfpage{1105}--\blpage{1118}
(\byear{2015})
{\href{https://arxiv.org/abs/1410.6185}{{arXiv:1410.6185}}}
{[math-ph]}.
\doiurl{10.1007/s11207-015-0666-3}
\end{barticle}
\endbibitem

\bibitem{2001Schrijver}
\begin{barticle}
\bauthor{\bsnm{{Schrijver}}, \binits{C.J.}}:
\batitle{{Simulations of the Photospheric Magnetic Activity and Outer
  Atmospheric Radiative Losses of Cool Stars Based on Characteristics of the
  Solar Magnetic Field}}.
\bjtitle{The Astrophysical Journal}
\bvolume{547}(\bissue{1}),
\bfpage{475}--\blpage{490}
(\byear{2001}).
\doiurl{10.1086/318333}
\end{barticle}
\endbibitem

\bibitem{2014UptonHathaway_a}
\begin{barticle}
\bauthor{\bsnm{{Upton}}, \binits{L.}},
\bauthor{\bsnm{{Hathaway}}, \binits{D.H.}}:
\batitle{{Predicting the Sun's Polar Magnetic Fields with a Surface Flux
  Transport Model}}.
\bjtitle{The Astrophysical Journal}
\bvolume{780}(\bissue{1}),
\bfpage{5}
(\byear{2014})
{\href{https://arxiv.org/abs/1311.0844}{{arXiv:1311.0844}}}
{[astro-ph.SR]}.
\doiurl{10.1088/0004-637X/780/1/5}
\end{barticle}
\endbibitem

\bibitem{2003SchrijverDeRosa}
\begin{barticle}
\bauthor{\bsnm{{Schrijver}}, \binits{C.J.}},
\bauthor{\bsnm{{De Rosa}}, \binits{M.L.}}:
\batitle{{Photospheric and heliospheric magnetic fields}}.
\bjtitle{Solar Physics}
\bvolume{212}(\bissue{1}),
\bfpage{165}--\blpage{200}
(\byear{2003}).
\doiurl{10.1023/A:1022908504100}
\end{barticle}
\endbibitem

\bibitem{Nandy2018ApJ}
\begin{barticle}
\bauthor{\bsnm{{Nandy}}, \binits{D.}},
\bauthor{\bsnm{{Bhowmik}}, \binits{P.}},
\bauthor{\bsnm{{Yeates}}, \binits{A.R.}},
\bauthor{\bsnm{{Panda}}, \binits{S.}},
\bauthor{\bsnm{{Tarafder}}, \binits{R.}},
\bauthor{\bsnm{{Dash}}, \binits{S.}}:
\batitle{{The Large-scale Coronal Structure of the 2017 August 21 Great
  American Eclipse: An Assessment of Solar Surface Flux Transport Model Enabled
  Predictions and Observations}}.
\bjtitle{The Astrophysical Journal}
\bvolume{853}(\bissue{1}),
\bfpage{72}
(\byear{2018}).
\doiurl{10.3847/1538-4357/aaa1eb}
\end{barticle}
\endbibitem

\bibitem{Dash2020ApJ}
\begin{barticle}
\bauthor{\bsnm{{Dash}}, \binits{S.}},
\bauthor{\bsnm{{Bhowmik}}, \binits{P.}},
\bauthor{\bsnm{{Athira}}, \binits{B.S.}},
\bauthor{\bsnm{{Ghosh}}, \binits{N.}},
\bauthor{\bsnm{{Nandy}}, \binits{D.}}:
\batitle{{Prediction of the Sun's Coronal Magnetic Field and Forward-modeled
  Polarization Characteristics for the 2019 July 2 Total Solar Eclipse}}.
\bjtitle{The Astrophysical Journal}
\bvolume{890}(\bissue{1}),
\bfpage{37}
(\byear{2020})
{\href{https://arxiv.org/abs/1906.10201}{{arXiv:1906.10201}}}
{[astro-ph.SR]}.
\doiurl{10.3847/1538-4357/ab6a91}
\end{barticle}
\endbibitem

\bibitem{2018Mikic_etal}
\begin{barticle}
\bauthor{\bsnm{{Miki{\'c}}}},
\bauthor{\bsnm{{}}, \binits{Z.}},
\bauthor{\bsnm{{Downs}}, \binits{C.}},
\bauthor{\bsnm{{Linker}}, \binits{J.A.}},
\bauthor{\bsnm{{Caplan}}, \binits{R.M.}},
\bauthor{\bsnm{{Mackay}}, \binits{D.H.}},
\bauthor{\bsnm{{Upton}}, \binits{L.A.}},
\bauthor{\bsnm{{Riley}}, \binits{P.}},
\bauthor{\bsnm{{Lionello}}, \binits{R.}},
\bauthor{\bsnm{{T{\"o}r{\"o}k}}, \binits{T.}},
\bauthor{\bsnm{{Titov}}, \binits{V.S.}},
\bauthor{\bsnm{{Wijaya}}, \binits{J.}},
\bauthor{\bsnm{{Druckm{\"u}ller}}, \binits{M.}},
\bauthor{\bsnm{{Pasachoff}}, \binits{J.M.}},
\bauthor{\bsnm{{Carlos}}, \binits{W.}}:
\batitle{{Predicting the corona for the 21 August 2017 total solar eclipse}}.
\bjtitle{Nature Astronomy}
\bvolume{2},
\bfpage{913}--\blpage{921}
(\byear{2018}).
\doiurl{10.1038/s41550-018-0562-5}
\end{barticle}
\endbibitem

\bibitem{2022MackayUpton}
\begin{barticle}
\bauthor{\bsnm{{Mackay}}, \binits{D.H.}},
\bauthor{\bsnm{{Upton}}, \binits{L.A.}}:
\batitle{{A Comparison of Global Magnetofrictional Simulations of the 2015
  March 20 Solar Eclipse}}.
\bjtitle{The Astrophysical Journal}
\bvolume{939}(\bissue{1}),
\bfpage{9}
(\byear{2022}).
\doiurl{10.3847/1538-4357/ac94c7}
\end{barticle}
\endbibitem

\bibitem{Yeates2022ApJ}
\begin{barticle}
\bauthor{\bsnm{{Yeates}}, \binits{A.R.}},
\bauthor{\bsnm{{Bhowmik}}, \binits{P.}}:
\batitle{{Automated Driving for Global Nonpotential Simulations of the Solar
  Corona}}.
\bjtitle{The Astrophysical Journal}
\bvolume{935}(\bissue{1}),
\bfpage{13}
(\byear{2022}).
\doiurl{10.3847/1538-4357/ac7de4}
\end{barticle}
\endbibitem

\bibitem{2014UptonHathaway_b}
\begin{barticle}
\bauthor{\bsnm{{Hathaway}}, \binits{D.H.}},
\bauthor{\bsnm{{Upton}}, \binits{L.}}:
\batitle{{The solar meridional circulation and sunspot cycle variability}}.
\bjtitle{Journal of Geophysical Research (Space Physics)}
\bvolume{119}(\bissue{5}),
\bfpage{3316}--\blpage{3324}
(\byear{2014})
{\href{https://arxiv.org/abs/1404.5893}{{arXiv:1404.5893}}}
{[astro-ph.SR]}.
\doiurl{10.1002/2013JA019432}
\end{barticle}
\endbibitem

\bibitem{2022Komm}
\begin{barticle}
\bauthor{\bsnm{{Komm}}, \binits{R.}}:
\batitle{{Is the Subsurface Meridional Flow Zero at the Equator?}}
\bjtitle{Solar Physics}
\bvolume{297}(\bissue{7}),
\bfpage{99}
(\byear{2022}).
\doiurl{10.1007/s11207-022-02027-z}
\end{barticle}
\endbibitem

\bibitem{Jiang2011}
\begin{barticle}
\bauthor{\bsnm{{Jiang}}, \binits{J.}},
\bauthor{\bsnm{{Cameron}}, \binits{R.H.}},
\bauthor{\bsnm{{Schmitt}}, \binits{D.}},
\bauthor{\bsnm{{Sch{\"u}ssler}}, \binits{M.}}:
\batitle{{The solar magnetic field since 1700. I. Characteristics of sunspot
  group emergence and reconstruction of the butterfly diagram}}.
\bjtitle{Astronomy \& Astrophysics}
\bvolume{528},
\bfpage{82}
(\byear{2011})
{\href{https://arxiv.org/abs/1102.1266}{{arXiv:1102.1266}}}
{[astro-ph.SR]}.
\doiurl{10.1051/0004-6361/201016167}
\end{barticle}
\endbibitem

\bibitem{Petrovay2020JSWSC}
\begin{barticle}
\bauthor{\bsnm{{Petrovay}}, \binits{K.}},
\bauthor{\bsnm{{Nagy}}, \binits{M.}},
\bauthor{\bsnm{{Yeates}}, \binits{A.R.}}:
\batitle{{Towards an algebraic method of solar cycle prediction. I. Calculating
  the ultimate dipole contributions of individual active regions}}.
\bjtitle{Journal of Space Weather and Space Climate}
\bvolume{10},
\bfpage{50}
(\byear{2020})
{\href{https://arxiv.org/abs/2009.02299}{{arXiv:2009.02299}}}
{[astro-ph.SR]}.
\doiurl{10.1051/swsc/2020050}
\end{barticle}
\endbibitem

\bibitem{Wang2021}
\begin{barticle}
\bauthor{\bsnm{{Wang}}, \binits{Z.-F.}},
\bauthor{\bsnm{{Jiang}}, \binits{J.}},
\bauthor{\bsnm{{Wang}}, \binits{J.-X.}}:
\batitle{{Algebraic quantification of an active region contribution to the
  solar cycle}}.
\bjtitle{Astronomy \& Astrophysics}
\bvolume{650},
\bfpage{87}
(\byear{2021})
{\href{https://arxiv.org/abs/2104.04307}{{arXiv:2104.04307}}}
{[astro-ph.SR]}.
\doiurl{10.1051/0004-6361/202140407}
\end{barticle}
\endbibitem

\bibitem{Dasi2010}
\begin{barticle}
\bauthor{\bsnm{{Dasi-Espuig}}, \binits{M.}},
\bauthor{\bsnm{{Solanki}}, \binits{S.K.}},
\bauthor{\bsnm{{Krivova}}, \binits{N.A.}},
\bauthor{\bsnm{{Cameron}}, \binits{R.}},
\bauthor{\bsnm{{Pe{\~n}uela}}, \binits{T.}}:
\batitle{{Sunspot group tilt angles and the strength of the solar cycle}}.
\bjtitle{Astronomy \& Astrophysics}
\bvolume{518},
\bfpage{7}
(\byear{2010})
{\href{https://arxiv.org/abs/1005.1774}{{arXiv:1005.1774}}}
{[astro-ph.SR]}.
\doiurl{10.1051/0004-6361/201014301}
\end{barticle}
\endbibitem

\bibitem{Jiang2020}
\begin{barticle}
\bauthor{\bsnm{{Jiang}}, \binits{J.}}:
\batitle{{Nonlinear Mechanisms that Regulate the Solar Cycle Amplitude}}.
\bjtitle{The Astrophysical Journal}
\bvolume{900}(\bissue{1}),
\bfpage{19}
(\byear{2020})
{\href{https://arxiv.org/abs/2007.07069}{{arXiv:2007.07069}}}
{[astro-ph.SR]}.
\doiurl{10.3847/1538-4357/abaa4b}
\end{barticle}
\endbibitem

\bibitem{Jiang2014}
\begin{barticle}
\bauthor{\bsnm{{Jiang}}, \binits{J.}},
\bauthor{\bsnm{{Cameron}}, \binits{R.H.}},
\bauthor{\bsnm{{Sch{\"u}ssler}}, \binits{M.}}:
\batitle{{Effects of the Scatter in Sunspot Group Tilt Angles on the
  Large-scale Magnetic Field at the Solar Surface}}.
\bjtitle{The Astrophysical Journal}
\bvolume{791}(\bissue{1}),
\bfpage{5}
(\byear{2014})
{\href{https://arxiv.org/abs/1406.5564}{{arXiv:1406.5564}}}
{[astro-ph.SR]}.
\doiurl{10.1088/0004-637X/791/1/5}
\end{barticle}
\endbibitem

\bibitem{Bhowmik2019A&A}
\begin{barticle}
\bauthor{\bsnm{{Bhowmik}}, \binits{P.}}:
\batitle{{Polar flux imbalance at the sunspot cycle minimum governs hemispheric
  asymmetry in the following cycle}}.
\bjtitle{Astronomy \& Astrophysics}
\bvolume{632},
\bfpage{117}
(\byear{2019}).
\doiurl{10.1051/0004-6361/201834425}
\end{barticle}
\endbibitem

\bibitem{Jiang2015}
\begin{barticle}
\bauthor{\bsnm{{Jiang}}, \binits{J.}},
\bauthor{\bsnm{{Cameron}}, \binits{R.H.}},
\bauthor{\bsnm{{Sch{\"u}ssler}}, \binits{M.}}:
\batitle{{The Cause of the Weak Solar Cycle 24}}.
\bjtitle{The Astrophysical Journal Letter}
\bvolume{808}(\bissue{1}),
\bfpage{28}
(\byear{2015})
{\href{https://arxiv.org/abs/1507.01764}{{arXiv:1507.01764}}}
{[astro-ph.SR]}.
\doiurl{10.1088/2041-8205/808/1/L28}
\end{barticle}
\endbibitem

\bibitem{Cameron2010ApJ}
\begin{barticle}
\bauthor{\bsnm{{Cameron}}, \binits{R.H.}},
\bauthor{\bsnm{{Jiang}}, \binits{J.}},
\bauthor{\bsnm{{Schmitt}}, \binits{D.}},
\bauthor{\bsnm{{Sch{\"u}ssler}}, \binits{M.}}:
\batitle{{Surface Flux Transport Modeling for Solar Cycles 15-21: Effects of
  Cycle-Dependent Tilt Angles of Sunspot Groups}}.
\bjtitle{The Astrophysical Journal}
\bvolume{719}(\bissue{1}),
\bfpage{264}--\blpage{270}
(\byear{2010})
{\href{https://arxiv.org/abs/1006.3061}{{arXiv:1006.3061}}}
{[astro-ph.SR]}.
\doiurl{10.1088/0004-637X/719/1/264}
\end{barticle}
\endbibitem

\bibitem{Cameron2016ApJ}
\begin{barticle}
\bauthor{\bsnm{{Cameron}}, \binits{R.H.}},
\bauthor{\bsnm{{Jiang}}, \binits{J.}},
\bauthor{\bsnm{{Sch{\"u}ssler}}, \binits{M.}}:
\batitle{{Solar Cycle 25: Another Moderate Cycle?}}
\bjtitle{The Astrophysical Journal Letter}
\bvolume{823}(\bissue{2}),
\bfpage{22}
(\byear{2016})
{\href{https://arxiv.org/abs/1604.05405}{{arXiv:1604.05405}}}
{[astro-ph.SR]}.
\doiurl{10.3847/2041-8205/823/2/L22}
\end{barticle}
\endbibitem

\bibitem{Wang2002ApJ}
\begin{barticle}
\bauthor{\bsnm{{Wang}}, \binits{Y.-M.}},
\bauthor{\bsnm{{Lean}}, \binits{J.}},
\bauthor{\bsnm{{Sheeley}}, \binits{J.} \bsuffix{N.~R.}}:
\batitle{{Role of a Variable Meridional Flow in the Secular Evolution of the
  Sun's Polar Fields and Open Flux}}.
\bjtitle{The Astrophysical Journal Letter}
\bvolume{577}(\bissue{1}),
\bfpage{53}--\blpage{57}
(\byear{2002}).
\doiurl{10.1086/344196}
\end{barticle}
\endbibitem

\bibitem{munoz2009}
\begin{barticle}
\bauthor{\bsnm{{Mu{\~n}oz-Jaramillo}}, \binits{A.}},
\bauthor{\bsnm{{Nandy}}, \binits{D.}},
\bauthor{\bsnm{{Martens}}, \binits{P.C.H.}}:
\batitle{{Helioseismic Data Inclusion in Solar Dynamo Models}}.
\bjtitle{The Astrophysical Journal}
\bvolume{698}(\bissue{1}),
\bfpage{461}--\blpage{478}
(\byear{2009})
{\href{https://arxiv.org/abs/0811.3441}{{arXiv:0811.3441}}}
{[astro-ph]}.
\doiurl{10.1088/0004-637X/698/1/461}
\end{barticle}
\endbibitem

\bibitem{Jouve2018ApJ}
\begin{barticle}
\bauthor{\bsnm{{Jouve}}, \binits{L.}},
\bauthor{\bsnm{{Brun}}, \binits{A.S.}},
\bauthor{\bsnm{{Aulanier}}, \binits{G.}}:
\batitle{{Interactions of Twisted {\ensuremath{\Omega}}-loops in a Model Solar
  Convection Zone}}.
\bjtitle{The Astrophysical Journal}
\bvolume{857}(\bissue{2}),
\bfpage{83}
(\byear{2018})
{\href{https://arxiv.org/abs/1803.04709}{{arXiv:1803.04709}}}
{[astro-ph.SR]}.
\doiurl{10.3847/1538-4357/aab5b6}
\end{barticle}
\endbibitem

\bibitem{Fan2021}
\begin{barticle}
\bauthor{\bsnm{{Fan}}, \binits{Y.}}:
\batitle{{Magnetic fields in the solar convection zone}}.
\bjtitle{Living Reviews in Solar Physics}
\bvolume{18}(\bissue{1}),
\bfpage{5}
(\byear{2021}).
\doiurl{10.1007/s41116-021-00031-2}
\end{barticle}
\endbibitem

\bibitem{Weber2011}
\begin{barticle}
\bauthor{\bsnm{{Weber}}, \binits{M.A.}},
\bauthor{\bsnm{{Fan}}, \binits{Y.}},
\bauthor{\bsnm{{Miesch}}, \binits{M.S.}}:
\batitle{{The Rise of Active Region Flux Tubes in the Turbulent Solar
  Convective Envelope}}.
\bjtitle{The Astrophysical Journal}
\bvolume{741},
\bfpage{11}
(\byear{2011})
{\href{https://arxiv.org/abs/1109.0240}{{arXiv:1109.0240}}}
{[astro-ph.SR]}.
\doiurl{10.1088/0004-637X/741/1/11}
\end{barticle}
\endbibitem

\bibitem{Chapter6}
\begin{botherref}
\oauthor{\bsnm{{Weber}}, \binits{M.A.}},
\oauthor{\bsnm{{Schunker}}, \binits{H.}},
\oauthor{\bsnm{{Jouve}}, \binits{L.}},
\oauthor{\bsnm{{I{\c{s}}{\i}k}}, \binits{E.}}:
{Emergence of Coherent Magnetic Structures through Convection Zones}.
Submitted to Space Science Reviews
(2023)
\end{botherref}
\endbibitem

\bibitem{Cameron2017SSRv}
\begin{barticle}
\bauthor{\bsnm{{Cameron}}, \binits{R.H.}},
\bauthor{\bsnm{{Dikpati}}, \binits{M.}},
\bauthor{\bsnm{{Brandenburg}}, \binits{A.}}:
\batitle{{The Global Solar Dynamo}}.
\bjtitle{Space Science Review}
\bvolume{210}(\bissue{1-4}),
\bfpage{367}--\blpage{395}
(\byear{2017})
{\href{https://arxiv.org/abs/1602.01754}{{arXiv:1602.01754}}}
{[astro-ph.SR]}.
\doiurl{10.1007/s11214-015-0230-3}
\end{barticle}
\endbibitem

\bibitem{hazraISSIchapter2003}
\begin{botherref}
\oauthor{\bsnm{{Hazra}}, \binits{G.}},
\oauthor{\bsnm{{Nandy}}, \binits{D.}},
\oauthor{\bsnm{{Kitchatinov}}, \binits{L.}},
\oauthor{\bsnm{{Choudhuri}}, \binits{A.R.}}:
{Mean field models of flux transport dynamo and meridional circulation in the
  Sun and stars}.
arXiv e-prints,
2302--09390
(2023)
{\href{https://arxiv.org/abs/2302.09390}{{arXiv:2302.09390}}}
{[astro-ph.SR]}.
\doiurl{10.48550/arXiv.2302.09390}
\end{botherref}
\endbibitem

\bibitem{Passos2014AA}
\begin{barticle}
\bauthor{\bsnm{{Passos}}, \binits{D.}},
\bauthor{\bsnm{{Nandy}}, \binits{D.}},
\bauthor{\bsnm{{Hazra}}, \binits{S.}},
\bauthor{\bsnm{{Lopes}}, \binits{I.}}:
\batitle{{A solar dynamo model driven by mean-field alpha and Babcock-Leighton
  sources: fluctuations, grand-minima-maxima, and hemispheric asymmetry in
  sunspot cycles}}.
\bjtitle{Astronomy \& Astrophysics}
\bvolume{563},
\bfpage{18}
(\byear{2014})
{\href{https://arxiv.org/abs/1309.2186}{{arXiv:1309.2186}}}
{[astro-ph.SR]}.
\doiurl{10.1051/0004-6361/201322635}
\end{barticle}
\endbibitem

\bibitem{Chapter13}
\begin{botherref}
\oauthor{\bsnm{{Cameron}}, \binits{R.H.}},
\oauthor{\bsnm{{Sch{\"u}ssler}}, \binits{M.}}:
{Simplified Dynamo Models Based on the Babcock-Leighton Scenario}.
Submitted to Space Science Reviews
(2023)
\end{botherref}
\endbibitem

\bibitem{Choudhuri1995A&A}
\begin{barticle}
\bauthor{\bsnm{{Choudhuri}}, \binits{A.R.}},
\bauthor{\bsnm{{Schussler}}, \binits{M.}},
\bauthor{\bsnm{{Dikpati}}, \binits{M.}}:
\batitle{{The solar dynamo with meridional circulation.}}
\bjtitle{Astronomy \& Astrophysics}
\bvolume{303},
\bfpage{29}
(\byear{1995})
\end{barticle}
\endbibitem

\bibitem{HazraS2016ApJ}
\begin{barticle}
\bauthor{\bsnm{{Hazra}}, \binits{S.}},
\bauthor{\bsnm{{Nandy}}, \binits{D.}}:
\batitle{{A Proposed Paradigm for Solar Cycle Dynamics Mediated via Turbulent
  Pumping of Magnetic Flux in Babcock-Leighton-type Solar Dynamos}}.
\bjtitle{The Astrophysical Journal}
\bvolume{832}(\bissue{1}),
\bfpage{9}
(\byear{2016})
{\href{https://arxiv.org/abs/1608.08167}{{arXiv:1608.08167}}}
{[astro-ph.SR]}.
\doiurl{10.3847/0004-637X/832/1/9}
\end{barticle}
\endbibitem

\bibitem{Lemerle2017ApJ}
\begin{barticle}
\bauthor{\bsnm{{Lemerle}}, \binits{A.}},
\bauthor{\bsnm{{Charbonneau}}, \binits{P.}}:
\batitle{{A Coupled 2 {\texttimes} 2D Babcock-Leighton Solar Dynamo Model. II.
  Reference Dynamo Solutions}}.
\bjtitle{The Astrophysical Journal}
\bvolume{834}(\bissue{2}),
\bfpage{133}
(\byear{2017})
{\href{https://arxiv.org/abs/1606.07375}{{arXiv:1606.07375}}}
{[astro-ph.SR]}.
\doiurl{10.3847/1538-4357/834/2/133}
\end{barticle}
\endbibitem

\bibitem{Usoskin2017}
\begin{barticle}
\bauthor{\bsnm{{Usoskin}}, \binits{I.G.}}:
\batitle{{A history of solar activity over millennia}}.
\bjtitle{Living Reviews in Solar Physics}
\bvolume{14}(\bissue{1}),
\bfpage{3}
(\byear{2017}).
\doiurl{10.1007/s41116-017-0006-9}
\end{barticle}
\endbibitem

\bibitem{Wiin-Nielsen1991}
\begin{barticle}
\bauthor{\bsnm{{Wiin-Nielsen}}, \binits{A.}}:
\batitle{{The birth of numerical weather prediction}}.
\bjtitle{Tellus A: Dynamic Meteorology and Oceanography}
\bvolume{43}(\bissue{4}),
\bfpage{36}--\blpage{52}
(\byear{1991}).
\doiurl{10.3402/tellusa.v43i4.11937}
\end{barticle}
\endbibitem

\bibitem{Rossby1939}
\begin{barticle}
\bauthor{\bsnm{{Rossby}}, \binits{C.G.}}:
\batitle{{Relation between Variations in the Intensity of the Zonal Circulation
  of the Atmosphere and the Displacements of the Semi-Permanent Centers of
  Action}}.
\bjtitle{Journal of Marine Research}
\bvolume{2},
\bfpage{38}--\blpage{55}
(\byear{1939}).
\doiurl{10.1357/002224039806649023}
\end{barticle}
\endbibitem

\bibitem{Charney1948}
\begin{barticle}
\bauthor{\bsnm{{Charney}}, \binits{J.}}:
\batitle{{On the scale of atmospheric motions.}}
\bjtitle{Geofys. Publikasjoner}
\bvolume{17},
\bfpage{1}--\blpage{17}
(\byear{1948})
\end{barticle}
\endbibitem

\bibitem{Kalnay2003}
\begin{bbook}
\bauthor{\bsnm{{Kalnay}}, \binits{E.}}:
\bbtitle{{Atmospheric Modeling, Data Assimilation and Predictability.}},
(\byear{2003})
\end{bbook}
\endbibitem

\bibitem{Lorenz1963}
\begin{barticle}
\bauthor{\bsnm{{Lorenz}}, \binits{E.N.}}:
\batitle{{Deterministic Nonperiodic Flow.}}
\bjtitle{Journal of Atmospheric Sciences}
\bvolume{20}(\bissue{2}),
\bfpage{130}--\blpage{148}
(\byear{1963}).
\doiurl{10.1175/1520-0469(1963)020<0130:DNF>2.0.CO;2}
\end{barticle}
\endbibitem

\bibitem{Lorenz1965}
\begin{barticle}
\bauthor{\bsnm{Lorenz}, \binits{E.N.}}:
\batitle{A study of the predictability of a 28-variable atmospheric model}.
\bjtitle{Tellus}
\bvolume{17}(\bissue{3}),
\bfpage{321}--\blpage{333}
(\byear{1965})
{\href{https://arxiv.org/abs/https://onlinelibrary.wiley.com/doi/pdf/10.1111/j.2153-3490.1965.tb01424.x}{{https://onlinelibrary.wiley.com/doi/pdf/10.1111/j.2153-3490.1965.tb01424.x}}}.
\doiurl{10.1111/j.2153-3490.1965.tb01424.x}
\end{barticle}
\endbibitem

\bibitem{Lorenz1969}
\begin{barticle}
\bauthor{\bsnm{LORENZ}, \binits{E.N.}}:
\batitle{The predictability of a flow which possesses many scales of motion}.
\bjtitle{Tellus}
\bvolume{21}(\bissue{3}),
\bfpage{289}--\blpage{307}
(\byear{1969})
{\href{https://arxiv.org/abs/https://onlinelibrary.wiley.com/doi/pdf/10.1111/j.2153-3490.1969.tb00444.x}{{https://onlinelibrary.wiley.com/doi/pdf/10.1111/j.2153-3490.1969.tb00444.x}}}.
\doiurl{10.1111/j.2153-3490.1969.tb00444.x}
\end{barticle}
\endbibitem

\bibitem{Takens1981}
\begin{bchapter}
\bauthor{\bsnm{{Takens}}, \binits{F.}}:
\bctitle{{Detecting strange attractors in turbulence}}.
In: \bbtitle{Lecture Notes in Mathematics, Berlin Springer Verlag}
vol. \bseriesno{898},
p. \bfpage{366}
(\byear{1981}).
\doiurl{10.1007/BFb0091924}
\end{bchapter}
\endbibitem

\bibitem{Grassberger1983}
\begin{barticle}
\bauthor{\bsnm{{Grassberger}}, \binits{P.}},
\bauthor{\bsnm{{Procaccia}}, \binits{I.}}:
\batitle{{Characterization of strange attractors}}.
\bjtitle{Physical Review Letters}
\bvolume{50}(\bissue{5}),
\bfpage{346}--\blpage{349}
(\byear{1983}).
\doiurl{10.1103/PhysRevLett.50.346}
\end{barticle}
\endbibitem

\bibitem{Theiler1992}
\begin{barticle}
\bauthor{\bsnm{{Theiler}}, \binits{J.}},
\bauthor{\bsnm{{Eubank}}, \binits{S.}},
\bauthor{\bsnm{{Longtin}}, \binits{A.}},
\bauthor{\bsnm{{Galdrikian}}, \binits{B.}},
\bauthor{\bsnm{{Doyne Farmer}}, \binits{J.}}:
\batitle{{Testing for nonlinearity in time series: the method of surrogate
  data}}.
\bjtitle{Physica D Nonlinear Phenomena}
\bvolume{58}(\bissue{1-4}),
\bfpage{77}--\blpage{94}
(\byear{1992}).
\doiurl{10.1016/0167-2789(92)90102-S}
\end{barticle}
\endbibitem

\bibitem{Paluvs1999}
\begin{barticle}
\bauthor{\bsnm{{Palu{\v{s}} }}, \binits{M.}},
\bauthor{\bsnm{{Novotn{\'a}}}, \binits{D.}}:
\batitle{{Sunspot Cycle: A Driven Nonlinear Oscillator?}}
\bjtitle{Physical Review Letters}
\bvolume{83}(\bissue{17}),
\bfpage{3406}--\blpage{3409}
(\byear{1999}).
\doiurl{10.1103/PhysRevLett.83.3406}
\end{barticle}
\endbibitem

\bibitem{Price1992}
\begin{barticle}
\bauthor{\bsnm{{Price}}, \binits{C.P.}},
\bauthor{\bsnm{{Prichard}}, \binits{D.}},
\bauthor{\bsnm{{Hogenson}}, \binits{E.A.}}:
\batitle{{Do the sunspot numbers form a ``chaotic'' set?}}
\bjtitle{Journal of Geophysical Research}
\bvolume{97}(\bissue{A12}),
\bfpage{19113}--\blpage{19120}
(\byear{1992}).
\doiurl{10.1029/92JA01459}
\end{barticle}
\endbibitem

\bibitem{Carbonell1994}
\begin{barticle}
\bauthor{\bsnm{{Carbonell}}, \binits{M.}},
\bauthor{\bsnm{{Oliver}}, \binits{R.}},
\bauthor{\bsnm{{Ballester}}, \binits{J.L.}}:
\batitle{{A search for chaotic behaviour in solar activity.}}
\bjtitle{Astronomy and Astrophysics}
\bvolume{290},
\bfpage{983}--\blpage{994}
(\byear{1994})
\end{barticle}
\endbibitem

\bibitem{Hanslmeier2010}
\begin{barticle}
\bauthor{\bsnm{{Hanslmeier}}, \binits{A.}},
\bauthor{\bsnm{{Braj{\v{s}}a}}, \binits{R.}}:
\batitle{{The chaotic solar cycle. I. Analysis of cosmogenic $^{14}$C-data}}.
\bjtitle{Astronomy and Astrophysics}
\bvolume{509},
\bfpage{5}
(\byear{2010}).
\doiurl{10.1051/0004-6361/200913095}
\end{barticle}
\endbibitem

\bibitem{Deng2016}
\begin{barticle}
\bauthor{\bsnm{{Deng}}, \binits{L.H.}},
\bauthor{\bsnm{{Li}}, \binits{B.}},
\bauthor{\bsnm{{Xiang}}, \binits{Y.Y.}},
\bauthor{\bsnm{{Dun}}, \binits{G.T.}}:
\batitle{{Comparison of Chaotic and Fractal Properties of Polar Faculae with
  Sunspot Activity}}.
\bjtitle{The Astronomical Journal}
\bvolume{151}(\bissue{1}),
\bfpage{2}
(\byear{2016}).
\doiurl{10.3847/0004-6256/151/1/2}
\end{barticle}
\endbibitem

\bibitem{Mundt1991}
\begin{barticle}
\bauthor{\bsnm{{Mundt}}, \binits{M.D.}},
\bauthor{\bsnm{{Maguire}}, \binits{I.} \bsuffix{W.~Bruce}},
\bauthor{\bsnm{{Chase}}, \binits{R.R.P.}}:
\batitle{{Chaos in the sunspot cycle: Analysis and prediction}}.
\bjtitle{Journal of Geophysical Research}
\bvolume{96}(\bissue{A2}),
\bfpage{1705}--\blpage{1716}
(\byear{1991}).
\doiurl{10.1029/90JA02150}
\end{barticle}
\endbibitem

\bibitem{Rozelot1995}
\begin{barticle}
\bauthor{\bsnm{{Rozelot}}, \binits{J.P.}}:
\batitle{{On the chaotic behaviour of the solar activity.}}
\bjtitle{Astronomy and Astrophysics}
\bvolume{297},
\bfpage{45}
(\byear{1995})
\end{barticle}
\endbibitem

\bibitem{Mininni2000}
\begin{barticle}
\bauthor{\bsnm{{Mininni}}, \binits{P.D.}},
\bauthor{\bsnm{{G{\'o}mez}}, \binits{D.O.}},
\bauthor{\bsnm{{Mindlin}}, \binits{G.B.}}:
\batitle{{Stochastic Relaxation Oscillator Model for the Solar Cycle}}.
\bjtitle{Physical Review Letters}
\bvolume{85}(\bissue{25}),
\bfpage{5476}--\blpage{5479}
(\byear{2000}).
\doiurl{10.1103/PhysRevLett.85.5476}
\end{barticle}
\endbibitem

\bibitem{Lopes2014}
\begin{barticle}
\bauthor{\bsnm{{Lopes}}, \binits{I.}},
\bauthor{\bsnm{{Passos}}, \binits{D.}},
\bauthor{\bsnm{{Nagy}}, \binits{M.}},
\bauthor{\bsnm{{Petrovay}}, \binits{K.}}:
\batitle{{Oscillator Models of the Solar Cycle. Towards the Development of
  Inversion Methods}}.
\bjtitle{Space Science Review}
\bvolume{186}(\bissue{1-4}),
\bfpage{535}--\blpage{559}
(\byear{2014})
{\href{https://arxiv.org/abs/1407.4918}{{arXiv:1407.4918}}}
{[astro-ph.SR]}.
\doiurl{10.1007/s11214-014-0066-2}
\end{barticle}
\endbibitem

\bibitem{Tobias1995}
\begin{barticle}
\bauthor{\bsnm{{Tobias}}, \binits{S.M.}},
\bauthor{\bsnm{{Weiss}}, \binits{N.O.}},
\bauthor{\bsnm{{Kirk}}, \binits{V.}}:
\batitle{{Chaotically modulated stellar dynamos}}.
\bjtitle{Monthly Notices of the Royal Astronomical Society}
\bvolume{273}(\bissue{4}),
\bfpage{1150}--\blpage{1166}
(\byear{1995}).
\doiurl{10.1093/mnras/273.4.1150}
\end{barticle}
\endbibitem

\bibitem{Knobloch1998}
\begin{barticle}
\bauthor{\bsnm{{Knobloch}}, \binits{E.}},
\bauthor{\bsnm{{Tobias}}, \binits{S.M.}},
\bauthor{\bsnm{{Weiss}}, \binits{N.O.}}:
\batitle{{Modulation and symmetry changes in stellar dynamos}}.
\bjtitle{Monthly Notices of the Royal Astronomical Society}
\bvolume{297}(\bissue{4}),
\bfpage{1123}--\blpage{1138}
(\byear{1998}).
\doiurl{10.1046/j.1365-8711.1998.01572.x}
\end{barticle}
\endbibitem

\bibitem{Wilmot-Smith2005}
\begin{barticle}
\bauthor{\bsnm{{Wilmot-Smith}}, \binits{A.L.}},
\bauthor{\bsnm{{Martens}}, \binits{P.C.H.}},
\bauthor{\bsnm{{Nandy}}, \binits{D.}},
\bauthor{\bsnm{{Priest}}, \binits{E.R.}},
\bauthor{\bsnm{{Tobias}}, \binits{S.M.}}:
\batitle{{Low-order stellar dynamo models}}.
\bjtitle{Monthly Notices of the Royal Astronomical Society}
\bvolume{363}(\bissue{4}),
\bfpage{1167}--\blpage{1172}
(\byear{2005}).
\doiurl{10.1111/j.1365-2966.2005.09514.x}
\end{barticle}
\endbibitem

\bibitem{Jennings1991}
\begin{barticle}
\bauthor{\bsnm{{Jennings}}, \binits{R.L.}},
\bauthor{\bsnm{{Weiss}}, \binits{N.O.}}:
\batitle{{Symmetry breaking in stellar dynamos}}.
\bjtitle{Monthly Notices of the Royal Astronomical Society}
\bvolume{252},
\bfpage{249}--\blpage{260}
(\byear{1991}).
\doiurl{10.1093/mnras/252.2.249}
\end{barticle}
\endbibitem

\bibitem{Tobias1996}
\begin{barticle}
\bauthor{\bsnm{{Tobias}}, \binits{S.M.}}:
\batitle{{Grand minimia in nonlinear dynamos.}}
\bjtitle{Astronomy \& Astrophysics}
\bvolume{307},
\bfpage{21}
(\byear{1996})
\end{barticle}
\endbibitem

\bibitem{Yoshimura1978}
\begin{barticle}
\bauthor{\bsnm{{Yoshimura}}, \binits{H.}}:
\batitle{{Nonlinear astrophysical dynamos: multiple-period dynamo wave
  oscillations and long-term modulations of the 22 year solar cycle.}}
\bjtitle{The Astrophysical Journal}
\bvolume{226},
\bfpage{706}--\blpage{719}
(\byear{1978}).
\doiurl{10.1086/156653}
\end{barticle}
\endbibitem

\bibitem{Wilmot-Smith2006}
\begin{barticle}
\bauthor{\bsnm{{Wilmot-Smith}}, \binits{A.L.}},
\bauthor{\bsnm{{Nandy}}, \binits{D.}},
\bauthor{\bsnm{{Hornig}}, \binits{G.}},
\bauthor{\bsnm{{Martens}}, \binits{P.C.H.}}:
\batitle{{A Time Delay Model for Solar and Stellar Dynamos}}.
\bjtitle{The Astrophysical Journal}
\bvolume{652}(\bissue{1}),
\bfpage{696}--\blpage{708}
(\byear{2006}).
\doiurl{10.1086/508013}
\end{barticle}
\endbibitem

\bibitem{HazraS2014ApJ}
\begin{barticle}
\bauthor{\bsnm{{Hazra}}, \binits{S.}},
\bauthor{\bsnm{{Passos}}, \binits{D.}},
\bauthor{\bsnm{{Nandy}}, \binits{D.}}:
\batitle{{A Stochastically Forced Time Delay Solar Dynamo Model:
  Self-consistent Recovery from a Maunder-like Grand Minimum Necessitates a
  Mean-field Alpha Effect}}.
\bjtitle{The Astrophysical Journal}
\bvolume{789}(\bissue{1}),
\bfpage{5}
(\byear{2014})
{\href{https://arxiv.org/abs/1307.5751}{{arXiv:1307.5751}}}
{[astro-ph.SR]}.
\doiurl{10.1088/0004-637X/789/1/5}
\end{barticle}
\endbibitem

\bibitem{Tripathi2021}
\begin{barticle}
\bauthor{\bsnm{{Tripathi}}, \binits{B.}},
\bauthor{\bsnm{{Nandy}}, \binits{D.}},
\bauthor{\bsnm{{Banerjee}}, \binits{S.}}:
\batitle{{Stellar mid-life crisis: subcritical magnetic dynamos of solar-like
  stars and the breakdown of gyrochronology}}.
\bjtitle{Monthly Notices of the Royal Astronomical Society}
\bvolume{506}(\bissue{1}),
\bfpage{50}--\blpage{54}
(\byear{2021})
{\href{https://arxiv.org/abs/1812.05533}{{arXiv:1812.05533}}}
{[astro-ph.SR]}.
\doiurl{10.1093/mnrasl/slab035}
\end{barticle}
\endbibitem

\bibitem{vanSaders2016}
\begin{barticle}
\bauthor{\bsnm{{van Saders}}, \binits{J.L.}},
\bauthor{\bsnm{{Ceillier}}, \binits{T.}},
\bauthor{\bsnm{{Metcalfe}}, \binits{T.S.}},
\bauthor{\bsnm{{Silva Aguirre}}, \binits{V.}},
\bauthor{\bsnm{{Pinsonneault}}, \binits{M.H.}},
\bauthor{\bsnm{{Garc{\'\i}a}}, \binits{R.A.}},
\bauthor{\bsnm{{Mathur}}, \binits{S.}},
\bauthor{\bsnm{{Davies}}, \binits{G.R.}}:
\batitle{{Weakened magnetic braking as the origin of anomalously rapid rotation
  in old field stars}}.
\bjtitle{Nature}
\bvolume{529}(\bissue{7585}),
\bfpage{181}--\blpage{184}
(\byear{2016})
{\href{https://arxiv.org/abs/1601.02631}{{arXiv:1601.02631}}}
{[astro-ph.SR]}.
\doiurl{10.1038/nature16168}
\end{barticle}
\endbibitem

\bibitem{Durney2000}
\begin{barticle}
\bauthor{\bsnm{{Durney}}, \binits{B.R.}}:
\batitle{{On the Differences Between Odd and Even Solar Cycles}}.
\bjtitle{Solar Physics}
\bvolume{196}(\bissue{2}),
\bfpage{421}--\blpage{426}
(\byear{2000}).
\doiurl{10.1023/A:1005285315323}
\end{barticle}
\endbibitem

\bibitem{Charbonneau2001}
\begin{barticle}
\bauthor{\bsnm{{Charbonneau}}, \binits{P.}}:
\batitle{{Multiperiodicity, Chaos, and Intermittency in a Reduced Model of the
  Solar Cycle}}.
\bjtitle{Solar Physics}
\bvolume{199}(\bissue{2}),
\bfpage{385}--\blpage{404}
(\byear{2001}).
\doiurl{10.1023/A:1010387509792}
\end{barticle}
\endbibitem

\bibitem{Barnes1980}
\begin{bchapter}
\bauthor{\bsnm{{Barnes}}, \binits{J.A.}},
\bauthor{\bsnm{{Tryon}}, \binits{P.V.}},
\bauthor{\bsnm{{Sargent}}, \binits{I.} \bsuffix{H.~H.}}:
\bctitle{{Sunspot cycle simulation using random noise}}.
In: \beditor{\bsnm{{Pepin}}, \binits{R.O.}},
\beditor{\bsnm{{Eddy}}, \binits{J.A.}},
\beditor{\bsnm{{Merrill}}, \binits{R.B.}} (eds.)
\bbtitle{The Ancient Sun: Fossil Record in the Earth, Moon and Meteorites},
pp. \bfpage{159}--\blpage{163}
(\byear{1980})
\end{bchapter}
\endbibitem

\bibitem{Mininni2001}
\begin{barticle}
\bauthor{\bsnm{{Mininni}}, \binits{P.D.}},
\bauthor{\bsnm{{Gomez}}, \binits{D.O.}},
\bauthor{\bsnm{{Mindlin}}, \binits{G.B.}}:
\batitle{{Simple Model of a Stochastically Excited Solar Dynamo}}.
\bjtitle{Solar Physics}
\bvolume{201}(\bissue{2}),
\bfpage{203}--\blpage{223}
(\byear{2001}).
\doiurl{10.1023/A:1017515709106}
\end{barticle}
\endbibitem

\bibitem{Passos2011}
\begin{barticle}
\bauthor{\bsnm{{Passos}}, \binits{D.}},
\bauthor{\bsnm{{Lopes}}, \binits{I.}}:
\batitle{{Grand minima under the light of a low order dynamo model}}.
\bjtitle{Journal of Atmospheric and Solar-Terrestrial Physics}
\bvolume{73}(\bissue{2-3}),
\bfpage{191}--\blpage{197}
(\byear{2011}).
\doiurl{10.1016/j.jastp.2009.12.019}
\end{barticle}
\endbibitem

\bibitem{Cameron2017}
\begin{barticle}
\bauthor{\bsnm{{Cameron}}, \binits{R.H.}},
\bauthor{\bsnm{{Sch{\"u}ssler}}, \binits{M.}}:
\batitle{{Understanding Solar Cycle Variability}}.
\bjtitle{The Astrophysical JournalKalnay2003}
\bvolume{843}(\bissue{2}),
\bfpage{111}
(\byear{2017})
{\href{https://arxiv.org/abs/1705.10746}{{arXiv:1705.10746}}}
{[astro-ph.SR]}.
\doiurl{10.3847/1538-4357/aa767a}
\end{barticle}
\endbibitem

\bibitem{Karak2010}
\begin{barticle}
\bauthor{\bsnm{{Karak}}, \binits{B.B.}}:
\batitle{{Importance of Meridional Circulation in Flux Transport Dynamo: The
  Possibility of a Maunder-like Grand Minimum}}.
\bjtitle{The Astrophysical Journal}
\bvolume{724}(\bissue{2}),
\bfpage{1021}--\blpage{1029}
(\byear{2010})
{\href{https://arxiv.org/abs/1009.2479}{{arXiv:1009.2479}}}
{[astro-ph.SR]}.
\doiurl{10.1088/0004-637X/724/2/1021}
\end{barticle}
\endbibitem

\bibitem{Nandy2011}
\begin{barticle}
\bauthor{\bsnm{{Nandy}}, \binits{D.}},
\bauthor{\bsnm{{Mu{\~n}oz-Jaramillo}}, \binits{A.}},
\bauthor{\bsnm{{Martens}}, \binits{P.C.H.}}:
\batitle{{The unusual minimum of sunspot cycle 23 caused by meridional plasma
  flow variations}}.
\bjtitle{Nature}
\bvolume{471}(\bissue{7336}),
\bfpage{80}--\blpage{82}
(\byear{2011})
{\href{https://arxiv.org/abs/1303.0349}{{arXiv:1303.0349}}}
{[astro-ph.SR]}.
\doiurl{10.1038/nature09786}
\end{barticle}
\endbibitem

\bibitem{Karak2013}
\begin{barticle}
\bauthor{\bsnm{{Karak}}, \binits{B.B.}},
\bauthor{\bsnm{{Choudhuri}}, \binits{A.R.}}:
\batitle{{Studies of grand minima in sunspot cycles by using a flux transport
  solar dynamo model}}.
\bjtitle{Research in Astronomy and Astrophysics}
\bvolume{13}(\bissue{11}),
\bfpage{1339}--\blpage{1357}
(\byear{2013})
{\href{https://arxiv.org/abs/1306.5438}{{arXiv:1306.5438}}}
{[astro-ph.SR]}.
\doiurl{10.1088/1674-4527/13/11/005}
\end{barticle}
\endbibitem

\bibitem{Bushby2007}
\begin{barticle}
\bauthor{\bsnm{{Bushby}}, \binits{P.J.}},
\bauthor{\bsnm{{Tobias}}, \binits{S.M.}}:
\batitle{{On Predicting the Solar Cycle Using Mean-Field Models}}.
\bjtitle{The Astrophysical Journal}
\bvolume{661}(\bissue{2}),
\bfpage{1289}--\blpage{1296}
(\byear{2007})
{\href{https://arxiv.org/abs/0704.2345}{{arXiv:0704.2345}}}
{[astro-ph]}.
\doiurl{10.1086/516628}
\end{barticle}
\endbibitem

\bibitem{HazraS2020A&A}
\begin{barticle}
\bauthor{\bsnm{{Hazra}}, \binits{S.}},
\bauthor{\bsnm{{Brun}}, \binits{A.S.}},
\bauthor{\bsnm{{Nandy}}, \binits{D.}}:
\batitle{{Does the mean-field {\ensuremath{\alpha}} effect have any impact on
  the memory of the solar cycle?}}
\bjtitle{Astronomy \& Astrophysics}
\bvolume{642},
\bfpage{51}
(\byear{2020})
{\href{https://arxiv.org/abs/2003.02776}{{arXiv:2003.02776}}}
{[astro-ph.SR]}.
\doiurl{10.1051/0004-6361/201937287}
\end{barticle}
\endbibitem

\bibitem{Zhang2022}
\begin{barticle}
\bauthor{\bsnm{{Zhang}}, \binits{Z.}},
\bauthor{\bsnm{{Jiang}}, \binits{J.}}:
\batitle{{A Babcock-Leighton-type Solar Dynamo Operating in the Bulk of the
  Convection Zone}}.
\bjtitle{The Astrophysical Journal}
\bvolume{930}(\bissue{1}),
\bfpage{30}
(\byear{2022})
{\href{https://arxiv.org/abs/2204.14077}{{arXiv:2204.14077}}}
{[astro-ph.SR]}.
\doiurl{10.3847/1538-4357/ac6177}
\end{barticle}
\endbibitem

\bibitem{Charbonneau2005}
\begin{barticle}
\bauthor{\bsnm{{Charbonneau}}, \binits{P.}},
\bauthor{\bsnm{{St-Jean}}, \binits{C.}},
\bauthor{\bsnm{{Zacharias}}, \binits{P.}}:
\batitle{{Fluctuations in Babcock-Leighton Dynamos. I. Period Doubling and
  Transition to Chaos}}.
\bjtitle{The Astrophysical Journal}
\bvolume{619}(\bissue{1}),
\bfpage{613}--\blpage{622}
(\byear{2005}).
\doiurl{10.1086/426385}
\end{barticle}
\endbibitem

\bibitem{Charbonneau2007}
\begin{barticle}
\bauthor{\bsnm{{Charbonneau}}, \binits{P.}},
\bauthor{\bsnm{{Beaubien}}, \binits{G.}},
\bauthor{\bsnm{{St-Jean}}, \binits{C.}}:
\batitle{{Fluctuations in Babcock-Leighton Dynamos. II. Revisiting the
  Gnevyshev-Ohl Rule}}.
\bjtitle{The Astrophysical Journal}
\bvolume{658}(\bissue{1}),
\bfpage{657}--\blpage{662}
(\byear{2007}).
\doiurl{10.1086/511177}
\end{barticle}
\endbibitem

\bibitem{Charbonneau2000}
\begin{barticle}
\bauthor{\bsnm{{Charbonneau}}, \binits{P.}},
\bauthor{\bsnm{{Dikpati}}, \binits{M.}}:
\batitle{{Stochastic Fluctuations in a Babcock-Leighton Model of the Solar
  Cycle}}.
\bjtitle{The Astrophysical Journal}
\bvolume{543}(\bissue{2}),
\bfpage{1027}--\blpage{1043}
(\byear{2000}).
\doiurl{10.1086/317142}
\end{barticle}
\endbibitem

\bibitem{Choudhuri2012}
\begin{barticle}
\bauthor{\bsnm{{Choudhuri}}, \binits{A.R.}},
\bauthor{\bsnm{{Karak}}, \binits{B.B.}}:
\batitle{{Origin of Grand Minima in Sunspot Cycles}}.
\bjtitle{Physical Review Letters}
\bvolume{109}(\bissue{17}),
\bfpage{171103}
(\byear{2012})
{\href{https://arxiv.org/abs/1208.3947}{{arXiv:1208.3947}}}
{[astro-ph.SR]}.
\doiurl{10.1103/PhysRevLett.109.171103}
\end{barticle}
\endbibitem

\bibitem{Karak2017}
\begin{barticle}
\bauthor{\bsnm{{Karak}}, \binits{B.B.}},
\bauthor{\bsnm{{Miesch}}, \binits{M.}}:
\batitle{{Solar Cycle Variability Induced by Tilt Angle Scatter in a
  Babcock-Leighton Solar Dynamo Model}}.
\bjtitle{The Astrophysical Journal}
\bvolume{847}(\bissue{1}),
\bfpage{69}
(\byear{2017})
{\href{https://arxiv.org/abs/1706.08933}{{arXiv:1706.08933}}}
{[astro-ph.SR]}.
\doiurl{10.3847/1538-4357/aa8636}
\end{barticle}
\endbibitem

\bibitem{Kitchatinov2018}
\begin{barticle}
\bauthor{\bsnm{{Kitchatinov}}, \binits{L.L.}},
\bauthor{\bsnm{{Mordvinov}}, \binits{A.V.}},
\bauthor{\bsnm{{Nepomnyashchikh}}, \binits{A.A.}}:
\batitle{{Modelling variability of solar activity cycles}}.
\bjtitle{Astronomy \& Astrophysics}
\bvolume{615},
\bfpage{38}
(\byear{2018})
{\href{https://arxiv.org/abs/1804.02833}{{arXiv:1804.02833}}}
{[astro-ph.SR]}.
\doiurl{10.1051/0004-6361/201732549}
\end{barticle}
\endbibitem

\bibitem{Hazra2019}
\begin{barticle}
\bauthor{\bsnm{{Hazra}}, \binits{S.}},
\bauthor{\bsnm{{Nandy}}, \binits{D.}}:
\batitle{{The origin of parity changes in the solar cycle}}.
\bjtitle{Monthly Notices of the Royal Astronomical Society}
\bvolume{489}(\bissue{3}),
\bfpage{4329}--\blpage{4337}
(\byear{2019})
{\href{https://arxiv.org/abs/1906.06780}{{arXiv:1906.06780}}}
{[astro-ph.SR]}.
\doiurl{10.1093/mnras/stz2476}
\end{barticle}
\endbibitem

\bibitem{saha2022}
\begin{barticle}
\bauthor{\bsnm{{Saha}}, \binits{C.}},
\bauthor{\bsnm{{Chandra}}, \binits{S.}},
\bauthor{\bsnm{{Nandy}}, \binits{D.}}:
\batitle{{Evidence of persistence of weak magnetic cycles driven by meridional
  plasma flows during solar grand minima phases}}.
\bjtitle{Monthly Notices of the Royal Astronomical Society}
\bvolume{517}(\bissue{1}),
\bfpage{36}--\blpage{40}
(\byear{2022})
{\href{https://arxiv.org/abs/2209.14651}{{arXiv:2209.14651}}}
{[astro-ph.SR]}.
\doiurl{10.1093/mnrasl/slac104}
\end{barticle}
\endbibitem

\bibitem{Sanchez2014}
\begin{barticle}
\bauthor{\bsnm{{Sanchez}}, \binits{S.}},
\bauthor{\bsnm{{Fournier}}, \binits{A.}},
\bauthor{\bsnm{{Aubert}}, \binits{J.}}:
\batitle{{The Predictability of Advection-dominated Flux-transport Solar Dynamo
  Models}}.
\bjtitle{The Astrophysical Journal}
\bvolume{781}(\bissue{1}),
\bfpage{8}
(\byear{2014}).
\doiurl{10.1088/0004-637X/781/1/8}
\end{barticle}
\endbibitem

\bibitem{Kitchatinov2011}
\begin{barticle}
\bauthor{\bsnm{{Kitchatinov}}, \binits{L.L.}},
\bauthor{\bsnm{{Olemskoy}}, \binits{S.V.}}:
\batitle{{Does the Babcock-Leighton mechanism operate on the Sun?}}
\bjtitle{Astronomy Letters}
\bvolume{37}(\bissue{9}),
\bfpage{656}--\blpage{658}
(\byear{2011})
{\href{https://arxiv.org/abs/1109.1351}{{arXiv:1109.1351}}}
{[astro-ph.SR]}.
\doiurl{10.1134/S0320010811080031}
\end{barticle}
\endbibitem

\bibitem{Olemskoy2013}
\begin{barticle}
\bauthor{\bsnm{{Olemskoy}}, \binits{S.V.}},
\bauthor{\bsnm{{Choudhuri}}, \binits{A.R.}},
\bauthor{\bsnm{{Kitchatinov}}, \binits{L.L.}}:
\batitle{{Fluctuations in the alpha-effect and grand solar minima}}.
\bjtitle{Astronomy Reports}
\bvolume{57}(\bissue{6}),
\bfpage{458}--\blpage{468}
(\byear{2013})
{\href{https://arxiv.org/abs/1305.2660}{{arXiv:1305.2660}}}
{[astro-ph.SR]}.
\doiurl{10.1134/S1063772913050065}
\end{barticle}
\endbibitem

\bibitem{Whitbread2018}
\begin{barticle}
\bauthor{\bsnm{{Whitbread}}, \binits{T.}},
\bauthor{\bsnm{{Yeates}}, \binits{A.R.}},
\bauthor{\bsnm{{Mu{\~n}oz-Jaramillo}}, \binits{A.}}:
\batitle{{How Many Active Regions Are Necessary to Predict the Solar Dipole
  Moment?}}
\bjtitle{The Astrophysical Journal}
\bvolume{863}(\bissue{2}),
\bfpage{116}
(\byear{2018})
{\href{https://arxiv.org/abs/1807.01617}{{arXiv:1807.01617}}}
{[astro-ph.SR]}.
\doiurl{10.3847/1538-4357/aad17e}
\end{barticle}
\endbibitem

\bibitem{Petrovay2020b}
\begin{barticle}
\bauthor{\bsnm{{Petrovay}}, \binits{K.}},
\bauthor{\bsnm{{Nagy}}, \binits{M.}},
\bauthor{\bsnm{{Yeates}}, \binits{A.R.}}:
\batitle{{Towards an algebraic method of solar cycle prediction. I. Calculating
  the ultimate dipole contributions of individual active regions}}.
\bjtitle{Journal of Space Weather and Space Climate}
\bvolume{10},
\bfpage{50}
(\byear{2020})
{\href{https://arxiv.org/abs/2009.02299}{{arXiv:2009.02299}}}
{[astro-ph.SR]}.
\doiurl{10.1051/swsc/2020050}
\end{barticle}
\endbibitem

\bibitem{Jiang2019}
\begin{barticle}
\bauthor{\bsnm{{Jiang}}, \binits{J.}},
\bauthor{\bsnm{{Song}}, \binits{Q.}},
\bauthor{\bsnm{{Wang}}, \binits{J.-X.}},
\bauthor{\bsnm{{Baranyi}}, \binits{T.}}:
\batitle{{Different Contributions to Space Weather and Space Climate from
  Different Big Solar Active Regions}}.
\bjtitle{The Astrophysical Journal}
\bvolume{871}(\bissue{1}),
\bfpage{16}
(\byear{2019})
{\href{https://arxiv.org/abs/1901.00116}{{arXiv:1901.00116}}}
{[astro-ph.SR]}.
\doiurl{10.3847/1538-4357/aaf64a}
\end{barticle}
\endbibitem

\bibitem{Yeates2020}
\begin{barticle}
\bauthor{\bsnm{{Yeates}}, \binits{A.R.}}:
\batitle{{How Good Is the Bipolar Approximation of Active Regions for Surface
  Flux Transport?}}
\bjtitle{Solar Physics}
\bvolume{295}(\bissue{9}),
\bfpage{119}
(\byear{2020})
{\href{https://arxiv.org/abs/2008.03203}{{arXiv:2008.03203}}}
{[astro-ph.SR]}.
\doiurl{10.1007/s11207-020-01688-y}
\end{barticle}
\endbibitem

\bibitem{Jiao2021}
\begin{barticle}
\bauthor{\bsnm{{Jiao}}, \binits{Q.}},
\bauthor{\bsnm{{Jiang}}, \binits{J.}},
\bauthor{\bsnm{{Wang}}, \binits{Z.-F.}}:
\batitle{{Sunspot tilt angles revisited: Dependence on the solar cycle
  strength}}.
\bjtitle{Astronomy \& Astrophysics}
\bvolume{653},
\bfpage{27}
(\byear{2021})
{\href{https://arxiv.org/abs/2106.11615}{{arXiv:2106.11615}}}
{[astro-ph.SR]}.
\doiurl{10.1051/0004-6361/202141215}
\end{barticle}
\endbibitem

\bibitem{Solanki2008}
\begin{barticle}
\bauthor{\bsnm{{Solanki}}, \binits{S.K.}},
\bauthor{\bsnm{{Wenzler}}, \binits{T.}},
\bauthor{\bsnm{{Schmitt}}, \binits{D.}}:
\batitle{{Moments of the latitudinal dependence of the sunspot cycle: a new
  diagnostic of dynamo models}}.
\bjtitle{Astronomy \& Astrophysics}
\bvolume{483}(\bissue{2}),
\bfpage{623}--\blpage{632}
(\byear{2008}).
\doiurl{10.1051/0004-6361:20054282}
\end{barticle}
\endbibitem

\bibitem{Karak2020}
\begin{barticle}
\bauthor{\bsnm{{Karak}}, \binits{B.B.}}:
\batitle{{Dynamo Saturation through the Latitudinal Variation of Bipolar
  Magnetic Regions in the Sun}}.
\bjtitle{The Astrophysical Journal Letter}
\bvolume{901}(\bissue{2}),
\bfpage{35}
(\byear{2020})
{\href{https://arxiv.org/abs/2009.06969}{{arXiv:2009.06969}}}
{[astro-ph.SR]}.
\doiurl{10.3847/2041-8213/abb93f}
\end{barticle}
\endbibitem

\bibitem{Talafha2022}
\begin{barticle}
\bauthor{\bsnm{{Talafha}}, \binits{M.}},
\bauthor{\bsnm{{Nagy}}, \binits{M.}},
\bauthor{\bsnm{{Lemerle}}, \binits{A.}},
\bauthor{\bsnm{{Petrovay}}, \binits{K.}}:
\batitle{{Role of observable nonlinearities in solar cycle modulation}}.
\bjtitle{Astronomy \& Astrophysics}
\bvolume{660},
\bfpage{92}
(\byear{2022})
{\href{https://arxiv.org/abs/2112.14465}{{arXiv:2112.14465}}}
{[astro-ph.SR]}.
\doiurl{10.1051/0004-6361/202142572}
\end{barticle}
\endbibitem

\bibitem{schatten1978}
\begin{barticle}
\bauthor{\bsnm{{Schatten}}, \binits{K.H.}},
\bauthor{\bsnm{{Scherrer}}, \binits{P.H.}},
\bauthor{\bsnm{{Svalgaard}}, \binits{L.}},
\bauthor{\bsnm{{Wilcox}}, \binits{J.M.}}:
\batitle{{Using Dynamo Theory to predict the sunspot number during Solar Cycle
  21}}.
\bjtitle{Geophysical Research Letters}
\bvolume{5}(\bissue{5}),
\bfpage{411}--\blpage{414}
(\byear{1978}).
\doiurl{10.1029/GL005i005p00411}
\end{barticle}
\endbibitem

\bibitem{nandy2002predict}
\begin{bchapter}
\bauthor{\bsnm{{Nandy}}, \binits{D.}}:
\bctitle{{Can theoretical solar dynamo models predict future solar activity?}}
In: \bbtitle{34th COSPAR Scientific Assembly},
vol. \bseriesno{34},
p. \bfpage{53}
(\byear{2002})
\end{bchapter}
\endbibitem

\bibitem{Munoz2013ApJ}
\begin{barticle}
\bauthor{\bsnm{{Mu{\~n}oz-Jaramillo}}, \binits{A.}},
\bauthor{\bsnm{{Dasi-Espuig}}, \binits{M.}},
\bauthor{\bsnm{{Balmaceda}}, \binits{L.A.}},
\bauthor{\bsnm{{DeLuca}}, \binits{E.E.}}:
\batitle{{Solar Cycle Propagation, Memory, and Prediction: Insights from a
  Century of Magnetic Proxies}}.
\bjtitle{The Astrophysical Journal Letter}
\bvolume{767}(\bissue{2}),
\bfpage{25}
(\byear{2013})
{\href{https://arxiv.org/abs/1304.3151}{{arXiv:1304.3151}}}
{[astro-ph.SR]}.
\doiurl{10.1088/2041-8205/767/2/L25}
\end{barticle}
\endbibitem

\bibitem{Hathaway2016JGRA}
\begin{barticle}
\bauthor{\bsnm{{Hathaway}}, \binits{D.H.}},
\bauthor{\bsnm{{Upton}}, \binits{L.A.}}:
\batitle{{Predicting the amplitude and hemispheric asymmetry of solar cycle 25
  with surface flux transport}}.
\bjtitle{Journal of Geophysical Research (Space Physics)}
\bvolume{121}(\bissue{11}),
\bfpage{10744}--\blpage{10753}
(\byear{2016})
{\href{https://arxiv.org/abs/1611.05106}{{arXiv:1611.05106}}}
{[astro-ph.SR]}.
\doiurl{10.1002/2016JA023190}
\end{barticle}
\endbibitem

\bibitem{Upton2018GeoRL}
\begin{barticle}
\bauthor{\bsnm{{Upton}}, \binits{L.A.}},
\bauthor{\bsnm{{Hathaway}}, \binits{D.H.}}:
\batitle{{An Updated Solar Cycle 25 Prediction With AFT: The Modern Minimum}}.
\bjtitle{Geophysical Research Letters}
\bvolume{45}(\bissue{16}),
\bfpage{8091}--\blpage{8095}
(\byear{2018})
{\href{https://arxiv.org/abs/1808.04868}{{arXiv:1808.04868}}}
{[astro-ph.SR]}.
\doiurl{10.1029/2018GL078387}
\end{barticle}
\endbibitem

\bibitem{Iijima2017A&A}
\begin{barticle}
\bauthor{\bsnm{{Iijima}}, \binits{H.}},
\bauthor{\bsnm{{Hotta}}, \binits{H.}},
\bauthor{\bsnm{{Imada}}, \binits{S.}},
\bauthor{\bsnm{{Kusano}}, \binits{K.}},
\bauthor{\bsnm{{Shiota}}, \binits{D.}}:
\batitle{{Improvement of solar-cycle prediction: Plateau of solar axial dipole
  moment}}.
\bjtitle{Astronomy \& Astrophysics}
\bvolume{607},
\bfpage{2}
(\byear{2017})
{\href{https://arxiv.org/abs/1710.06528}{{arXiv:1710.06528}}}
{[astro-ph.SR]}.
\doiurl{10.1051/0004-6361/201731813}
\end{barticle}
\endbibitem

\bibitem{Labonville2019SoPh}
\begin{barticle}
\bauthor{\bsnm{{Labonville}}, \binits{F.}},
\bauthor{\bsnm{{Charbonneau}}, \binits{P.}},
\bauthor{\bsnm{{Lemerle}}, \binits{A.}}:
\batitle{{A Dynamo-based Forecast of Solar Cycle 25}}.
\bjtitle{Solar Physics}
\bvolume{294}(\bissue{6}),
\bfpage{82}
(\byear{2019}).
\doiurl{10.1007/s11207-019-1480-0}
\end{barticle}
\endbibitem

\bibitem{Dikpati2006GeoRL}
\begin{barticle}
\bauthor{\bsnm{{Dikpati}}, \binits{M.}},
\bauthor{\bsnm{{de Toma}}, \binits{G.}},
\bauthor{\bsnm{{Gilman}}, \binits{P.A.}}:
\batitle{{Predicting the strength of solar cycle 24 using a flux-transport
  dynamo-based tool}}.
\bjtitle{Geophysical Research Letters}
\bvolume{33}(\bissue{5}),
\bfpage{05102}
(\byear{2006}).
\doiurl{10.1029/2005GL025221}
\end{barticle}
\endbibitem

\bibitem{Choudhuri2007PhRvL}
\begin{barticle}
\bauthor{\bsnm{{Choudhuri}}, \binits{A.R.}},
\bauthor{\bsnm{{Chatterjee}}, \binits{P.}},
\bauthor{\bsnm{{Jiang}}, \binits{J.}}:
\batitle{{Predicting Solar Cycle 24 With a Solar Dynamo Model}}.
\bjtitle{Physical Review Letters}
\bvolume{98}(\bissue{13}),
\bfpage{131103}
(\byear{2007})
{\href{https://arxiv.org/abs/astro-ph/0701527}{{arXiv:astro-ph/0701527}}}
{[astro-ph]}.
\doiurl{10.1103/PhysRevLett.98.131103}
\end{barticle}
\endbibitem

\bibitem{Dikpati1999ApJ}
\begin{barticle}
\bauthor{\bsnm{{Dikpati}}, \binits{M.}},
\bauthor{\bsnm{{Charbonneau}}, \binits{P.}}:
\batitle{{A Babcock-Leighton Flux Transport Dynamo with Solar-like Differential
  Rotation}}.
\bjtitle{The Astrophysical Journal}
\bvolume{518}(\bissue{1}),
\bfpage{508}--\blpage{520}
(\byear{1999}).
\doiurl{10.1086/307269}
\end{barticle}
\endbibitem

\bibitem{Chatterjee2004A&A}
\begin{barticle}
\bauthor{\bsnm{{Chatterjee}}, \binits{P.}},
\bauthor{\bsnm{{Nandy}}, \binits{D.}},
\bauthor{\bsnm{{Choudhuri}}, \binits{A.R.}}:
\batitle{{Full-sphere simulations of a circulation-dominated solar dynamo:
  Exploring the parity issue}}.
\bjtitle{Astronomy \& Astrophysics}
\bvolume{427},
\bfpage{1019}--\blpage{1030}
(\byear{2004})
{\href{https://arxiv.org/abs/astro-ph/0405027}{{arXiv:astro-ph/0405027}}}
{[astro-ph]}.
\doiurl{10.1051/0004-6361:20041199}
\end{barticle}
\endbibitem

\bibitem{Hazra2019MNRAS}
\begin{barticle}
\bauthor{\bsnm{{Hazra}}, \binits{S.}},
\bauthor{\bsnm{{Nandy}}, \binits{D.}}:
\batitle{{The origin of parity changes in the solar cycle}}.
\bjtitle{Monthly Notices of the Royal Astronomical Society}
\bvolume{489}(\bissue{3}),
\bfpage{4329}--\blpage{4337}
(\byear{2019})
{\href{https://arxiv.org/abs/1906.06780}{{arXiv:1906.06780}}}
{[astro-ph.SR]}.
\doiurl{10.1093/mnras/stz2476}
\end{barticle}
\endbibitem

\bibitem{Andres2011ApJ}
\begin{barticle}
\bauthor{\bsnm{{Mu{\~n}oz-Jaramillo}}, \binits{A.}},
\bauthor{\bsnm{{Nandy}}, \binits{D.}},
\bauthor{\bsnm{{Martens}}, \binits{P.C.H.}}:
\batitle{{Magnetic Quenching of Turbulent Diffusivity: Reconciling
  Mixing-length Theory Estimates with Kinematic Dynamo Models of the Solar
  Cycle}}.
\bjtitle{The Astrophysical Journal Letter}
\bvolume{727}(\bissue{1}),
\bfpage{23}
(\byear{2011})
{\href{https://arxiv.org/abs/1007.1262}{{arXiv:1007.1262}}}
{[astro-ph.SR]}.
\doiurl{10.1088/2041-8205/727/1/L23}
\end{barticle}
\endbibitem

\bibitem{Dikpati2006ApJ}
\begin{barticle}
\bauthor{\bsnm{{Dikpati}}, \binits{M.}},
\bauthor{\bsnm{{Gilman}}, \binits{P.A.}}:
\batitle{{Simulating and Predicting Solar Cycles Using a Flux-Transport
  Dynamo}}.
\bjtitle{The Astrophysical Journal}
\bvolume{649}(\bissue{1}),
\bfpage{498}--\blpage{514}
(\byear{2006}).
\doiurl{10.1086/506314}
\end{barticle}
\endbibitem

\bibitem{Jiang2007MNRAS}
\begin{barticle}
\bauthor{\bsnm{{Jiang}}, \binits{J.}},
\bauthor{\bsnm{{Chatterjee}}, \binits{P.}},
\bauthor{\bsnm{{Choudhuri}}, \binits{A.R.}}:
\batitle{{Solar activity forecast with a dynamo model}}.
\bjtitle{Monthly Notices of the Royal Astronomical Society}
\bvolume{381}(\bissue{4}),
\bfpage{1527}--\blpage{1542}
(\byear{2007})
{\href{https://arxiv.org/abs/0707.2258}{{arXiv:0707.2258}}}
{[astro-ph]}.
\doiurl{10.1111/j.1365-2966.2007.12267.x}
\end{barticle}
\endbibitem

\bibitem{Guo2021SoPh}
\begin{barticle}
\bauthor{\bsnm{{Guo}}, \binits{W.}},
\bauthor{\bsnm{{Jiang}}, \binits{J.}},
\bauthor{\bsnm{{Wang}}, \binits{J.-X.}}:
\batitle{{A Dynamo-Based Prediction of Solar Cycle 25}}.
\bjtitle{Solar Physics}
\bvolume{296}(\bissue{9}),
\bfpage{136}
(\byear{2021})
{\href{https://arxiv.org/abs/2108.01412}{{arXiv:2108.01412}}}
{[astro-ph.SR]}.
\doiurl{10.1007/s11207-021-01878-2}
\end{barticle}
\endbibitem

\bibitem{Yeatesetal2007}
\begin{barticle}
\bauthor{\bsnm{{Yeates}}, \binits{A.R.}},
\bauthor{\bsnm{{Mackay}}, \binits{D.H.}},
\bauthor{\bsnm{{van Ballegooijen}}, \binits{A.A.}}:
\batitle{{Modelling the Global Solar Corona: Filament Chirality Observations
  and Surface Simulations}}.
\bjtitle{Solar Physics}
\bvolume{245},
\bfpage{87}--\blpage{107}
(\byear{2007})
{\href{https://arxiv.org/abs/0707.3256}{{arXiv:0707.3256}}}.
\doiurl{10.1007/s11207-007-9013-7}
\end{barticle}
\endbibitem

\bibitem{HazraG2019ApJ}
\begin{barticle}
\bauthor{\bsnm{{Hazra}}, \binits{G.}},
\bauthor{\bsnm{{Choudhuri}}, \binits{A.R.}}:
\batitle{{A New Formula for Predicting Solar Cycles}}.
\bjtitle{The Astrophysical Journal}
\bvolume{880}(\bissue{2}),
\bfpage{113}
(\byear{2019})
{\href{https://arxiv.org/abs/1811.01363}{{arXiv:1811.01363}}}
{[astro-ph.SR]}.
\doiurl{10.3847/1538-4357/ab2718}
\end{barticle}
\endbibitem

\bibitem{Kumar2022MNRAS}
\begin{barticle}
\bauthor{\bsnm{{Kumar}}, \binits{P.}},
\bauthor{\bsnm{{Biswas}}, \binits{A.}},
\bauthor{\bsnm{{Karak}}, \binits{B.B.}}:
\batitle{{Physical link of the polar field buildup with the Waldmeier effect
  broadens the scope of early solar cycle prediction: Cycle 25 is likely to be
  slightly stronger than Cycle 24}}.
\bjtitle{Monthly Notices of the Royal Astronomical Society}
\bvolume{513}(\bissue{1}),
\bfpage{112}--\blpage{116}
(\byear{2022})
{\href{https://arxiv.org/abs/2203.11494}{{arXiv:2203.11494}}}
{[astro-ph.SR]}.
\doiurl{10.1093/mnrasl/slac043}
\end{barticle}
\endbibitem

\bibitem{Yeates2013MNRAS}
\begin{barticle}
\bauthor{\bsnm{{Yeates}}, \binits{A.R.}},
\bauthor{\bsnm{{Mu{\~n}oz-Jaramillo}}, \binits{A.}}:
\batitle{{Kinematic active region formation in a three-dimensional solar dynamo
  model}}.
\bjtitle{Monthly Notices of the Royal Astronomical Society}
\bvolume{436}(\bissue{4}),
\bfpage{3366}--\blpage{3379}
(\byear{2013})
{\href{https://arxiv.org/abs/1309.6342}{{arXiv:1309.6342}}}
{[astro-ph.SR]}.
\doiurl{10.1093/mnras/stt1818}
\end{barticle}
\endbibitem

\bibitem{Miesch2014ApJ}
\begin{barticle}
\bauthor{\bsnm{{Miesch}}, \binits{M.S.}},
\bauthor{\bsnm{{Dikpati}}, \binits{M.}}:
\batitle{{A Three-dimensional Babcock-Leighton Solar Dynamo Model}}.
\bjtitle{The Astrophysical Journal Letter}
\bvolume{785}(\bissue{1}),
\bfpage{8}
(\byear{2014})
{\href{https://arxiv.org/abs/1401.6557}{{arXiv:1401.6557}}}
{[astro-ph.SR]}.
\doiurl{10.1088/2041-8205/785/1/L8}
\end{barticle}
\endbibitem

\bibitem{Hazra2017ApJ}
\begin{barticle}
\bauthor{\bsnm{{Hazra}}, \binits{G.}},
\bauthor{\bsnm{{Choudhuri}}, \binits{A.R.}},
\bauthor{\bsnm{{Miesch}}, \binits{M.S.}}:
\batitle{{A Theoretical Study of the Build-up of the Sun{\textquoteright}s
  Polar Magnetic Field by using a 3D Kinematic Dynamo Model}}.
\bjtitle{The Astrophysical Journal}
\bvolume{835}(\bissue{1}),
\bfpage{39}
(\byear{2017})
{\href{https://arxiv.org/abs/1610.02726}{{arXiv:1610.02726}}}
{[astro-ph.SR]}.
\doiurl{10.3847/1538-4357/835/1/39}
\end{barticle}
\endbibitem

\bibitem{Kumar2019A&A}
\begin{barticle}
\bauthor{\bsnm{{Kumar}}, \binits{R.}},
\bauthor{\bsnm{{Jouve}}, \binits{L.}},
\bauthor{\bsnm{{Nandy}}, \binits{D.}}:
\batitle{{A 3D kinematic Babcock Leighton solar dynamo model sustained by
  dynamic magnetic buoyancy and flux transport processes}}.
\bjtitle{Astronomy \& Astrophysics}
\bvolume{623},
\bfpage{54}
(\byear{2019})
{\href{https://arxiv.org/abs/1901.04251}{{arXiv:1901.04251}}}
{[astro-ph.SR]}.
\doiurl{10.1051/0004-6361/201834705}
\end{barticle}
\endbibitem

\bibitem{Whitbread2019A&A}
\begin{barticle}
\bauthor{\bsnm{{Whitbread}}, \binits{T.}},
\bauthor{\bsnm{{Yeates}}, \binits{A.R.}},
\bauthor{\bsnm{{Mu{\~n}oz-Jaramillo}}, \binits{A.}}:
\batitle{{The need for active region disconnection in 3D kinematic dynamo
  simulations}}.
\bjtitle{Astronomy \& Astrophysics}
\bvolume{627},
\bfpage{168}
(\byear{2019})
{\href{https://arxiv.org/abs/1907.02762}{{arXiv:1907.02762}}}
{[astro-ph.SR]}.
\doiurl{10.1051/0004-6361/201935986}
\end{barticle}
\endbibitem

\bibitem{Nagyetal2020}
\begin{barticle}
\bauthor{\bsnm{{Nagy}}, \binits{M.}},
\bauthor{\bsnm{{Lemerle}}, \binits{A.}},
\bauthor{\bsnm{{Charbonneau}}, \binits{P.}}:
\batitle{{Impact of nonlinear surface inflows into activity belts on the solar
  dynamo}}.
\bjtitle{Journal of Space Weather and Space Climate}
\bvolume{10},
\bfpage{62}
(\byear{2020}).
\doiurl{10.1051/swsc/2020064}
\end{barticle}
\endbibitem

\end{thebibliography}


\end{document}